\documentclass[preprints,article,accept,moreauthors,pdftex]{mdpi} 

\usepackage{amsfonts}

\renewcommand{\BibitemShut}[1]{}

\newcommand\cA{{\mathcal A}}

\newcommand\cL{{\mathcal L}}
\newcommand\cM{{\mathcal M}}
\newcommand\cN{{\mathcal N}}

\usepackage{verbatim}
\newcommand{\HCd}{\mathcal{H}}

\def\HCdt0{\tilde{\HCd}_{0}}

\newcommand\rf[1]{(\ref{eq:#1})}
\newcommand\lab[1]{\label{eq:#1}}
\newcommand\nonu{\nonumber}
\newcommand\br{\begin{eqnarray}}
\newcommand\er{\end{eqnarray}}
\newcommand\be{\begin{equation}}
\newcommand\ee{\end{equation}}

\newcommand\lb{\lbrack}
\newcommand\rb{\rbrack}

\renewcommand\({\left(}
\renewcommand\){\right)}
\newcommand\bgv{\bigg\vert}              %%

\newcommand\bc{\begin{center}}
\newcommand\ec{\end{center}}

\newcommand\partder[2]{\frac{{\partial {#1}}}{{\partial {#2}}}}
\renewcommand\d{\delta}
\newcommand\eps{\epsilon}
\newcommand\vareps{\varepsilon}

\renewcommand\G{\Gamma}

\renewcommand\k{\kappa}
\renewcommand\l{\lambda}
\renewcommand\L{\Lambda}
\newcommand\m{\mu}
\newcommand\n{\nu}
\newcommand\om{\omega}

\newcommand\vp{\varphi}
\renewcommand\P{\Phi}
\newcommand\pa{\partial}

\newcommand\pr{\prime}

\renewcommand\S{\Sigma}

\newcommand\wti{\widetilde}
\newcommand\twomat[4]{\left(\begD{array}{cc}  %%   2x2 matrix  %ESA
{#1} & {#2} \\ {#3} & {#4} \end{array} \right)}

\newcommand{\ct}[1]{\cite{#1}}

\newcommand\vpdot{\stackrel{.}{\varphi}}

\newcommand\udot{\stackrel{.}{u}}
\newcommand\uddot{\stackrel{..}{u}}

\newcommand\adot{\stackrel{.}{a}}

\newcommand\Hdot{\stackrel{.}{H}}
\newcommand\Hddot{\stackrel{..}{H}}

\firstpage{1} 
\pubvolume{xx}
\issuenum{1}
\articlenumber{5}
\pubyear{2019}
\copyrightyear{2019}
\history{Received: date; Accepted: date; Published: date}
\Title{Dynamically Generated Inflationary $\Lambda$CDM}

\Author{D. Benisty$^{1,2}$, E. I. Guendelman$^{1,2,3}$, E. Nissimov*$^{4}$ and S. Pacheva$^{4}$}

\address{%
$^{1}$ \quad Physics Department, Ben-Gurion University of the Negev, Beer-Sheva 84105, Israel\\
$^{2}$ \quad Frankfurt Institute for Advanced Studies (FIAS), Ruth-Moufang-Strasse~1, 60438 Frankfurt am Main, Germany\\ 
$^{3}$ \quad Bahamas Advanced Study Institute and Conferences, 4A Ocean Heights, Hill View Circle, Stella Maris, Long Island, The Bahamas\\
$^{4}$ Institute for Nuclear Research and Nuclear Energy, Bulgarian Academy of Sciences, Sofia, Bulgaria}

\corres{Correspondence: nissimov@inrne.bas.bg}

\abstract{Our primary objective is to construct a plausible unified model of % quintessential 
inflation, dark energy and dark matter from a fundamental Lagrangian action 
first principle, where all fundamental ingredients are systematically 
dynamically generated starting from a very simple model of modified gravity 
interacting with a single scalar field employing the formalism of non-Riemannian 
spacetime volume-elements. The non-Riemannian volume element in the initial scalar field action leads to a 
hidden nonlinear Noether 
symmetry which produces energy-momentum tensor identified as a sum of a 
dynamically generated cosmological constant and a dust-like dark matter. 
The non-Riemannian volume-element in the initial Einstein-Hilbert action 
upon passage to the physical Einstein-frame creates dynamically a second scalar 
field with a non-trivial inflationary potential and with an additional interaction 
with the dynamically generated dark matter. The resulting Einstein-frame action 
describes a fully dynamically generated inflationary model coupled to dark matter.
Numerical results for observables such as the scalar power spectral index and the 
tensor-to-scalar ratio conform to the latest $2018$ \textsl{PLANCK} data.}


\begin{document}
%%%%%%%%%%%%%%%%%%%%%%%%%%%%%%%%%%%%%%%%%%
\section{Introduction}

In the last decade or so a groundbreaking concept emerged about the intrinsic 
necessity to modify (extend) gravity theories beyond the framework of standard 
Einstein's general relativity. The main motivation for these developments is to 
overcome the limitations of the latter coming from: (i) Cosmology -- 
for solving the problems of dark energy and dark matter and explaining the 
large scale structure of the Universe  \ct{Perlmutter:1998np,Copeland:2006wr,Novikov:2016fzd};
(ii) Quantum field theory in curved spacetime -- because of the non-renormalizabilty of ultraviolet divergences in higher loops \ct{Benitez:2020szx,Budge:2020oyl,Bell:2020qus,Frohlich:2020igy,DAmbrosio:2020yfa,Novikov:2016hrc}; 
(iii) Modern string theory -- because of the natural appearance of higher-order 
curvature invariants and
scalar-tensor couplings in low-energy effective field theories \ct{Dekens:2019ept,Ma:2019wbc,Jenkins:2017jig,Brandyshev:2017ywi,Gomez:2020xdb}.

Another parallel crucial development is the emergence of 
the theoretical framework based on the concept of ``inflation'', which is a 
necessary part of the standard model of  cosmology, since it provides  
the solution to the fundamental puzzles of the old Big Bang theory, 
such as the horizon, the flatness, and the monopole problems
\cite{Guth:1980zm,Starobinsky:1979ty,Kazanas:1980tx,Starobinsky:1980te,Linde:1981mu,Albrecht:1982wi,Barrow:1983rx,Blau:1986cw}. It can be 
achieved through various mechanisms, for instance through the introduction of 
primordial scalar field(s) 
\cite{CervantesCota:1995tz,Berera:1995ie,ArmendarizPicon:1999rj,Kanti:1999ie,Garriga:1999vw,Gordon:2000hv,Bassett:2005xm,Chen:2009zp,Germani:2010gm,Kobayashi:2010cm,Feng:2010ya,Burrage:2010cu,Kobayashi:2011nu,Ohashi:2012wf,Paliathanasis:2014yfa,Dimakis:2020tzc,Dimakis:2019qfs,Benisty:2017lmt,Barrow:2016qkh,Barrow:2016wiy,Olive:1989nu,Linde:1993cn,Liddle:1994dx,Lidsey:1995np,Hossain:2014xha,Hossain:2014zma,Cai:2014uka,Geng:2015fla,Kamali:2016frd,Geng:2017mic,Dalianis:2018frf,Dalianis:2019asr,Benisty:2019pxb,Benisty:2018gzx,Benisty:2019vej,Gerbino:2016sgw,Giovannini:2020xeg,Brahma:2020cpy,Domcke:2020zez,Tenkanen:2020cvw,Martin:2020fgl,Cheon:2020vnj,Saleem:2020dzo,Giacintucci:2020glv,Aalsma:2020aib,Kogut:2020add,Arciniega:2020pcy,Rasheed:2020syk,Aldabergenov:2020pry,Tenkanen:2020dge,Shaposhnikov:2020geh,Garcia:2020mwi,Hirano:2019iie,Gialamas:2019nly}, or through correction terms into the modified gravitational action \cite{Kawasaki:2000yn,Bojowald:2002nz,Nojiri:2003ft,Kachru:2003sx,Nojiri:2005pu,Ferraro:2006jd,Cognola:2007zu,Cai:2010kp,Ashtekar:2011rm,Qiu:2011zr,Briscese:2012ys,Ellis:2013xoa,Basilakos:2013xpa,Sebastiani:2013eqa,Baumann:2014nda,Dalianis:2015fpa,Kanti:2015pda,DeLaurentis:2015fea,Basilakos:2015yoa,Bonanno:2015fga,Koshelev:2016xqb,Bamba:2016wjm,Motohashi:2017vdc,Oikonomou:2017ppp,Benisty:2018ywz,Benisty:2018fja,Antoniadis:2018ywb,Karam:2019dlv,Nojiri:2019kkp,Benisty:2019jqz,Benisty:2019tno,Benisty:2019bmi,Kinney:2018nny,Brustein:2017iet,Sherf:2018uth,Capozziello:2014hia,Gorbunov:2013dqa,Myrzakulov:2014hca,Bamba:2014jia,Benisty:2018ufz,Aashish:2020ufe,Rashidi:2020wwg,Odintsov:2020nwm,Antoniadis:2019jnz,Benisty:2018fgu,Chakraborty:2018scm}. 

Additionally, inflation was proved crucial in  
providing a framework for the generation of primordial density perturbations 
\cite{Mukhanov:1981xt,Guth:1982ec}. Since these perturbations affect  the 
Cosmic Background Radiation (CMB), the inflationary effect on observations can be 
investigated through the prediction for the  scalar spectral index of the curvature 
perturbations and its running, for the tensor spectral index, and for the tensor-to-scalar ratio.

Various classes of modified gravity theories have been employed to construct viable 
inflationary models: $f(R)$-gravity; scalar-tensor gravity; 
Gauss-Bonnet gravity (see \ct{Capozziello:2010zz,Nojiri:2017ncd} for a detailed accounts); recently also based on non-local gravity (\ct{Dimitrijevic:2019pct} and references therein) or based on brane-world scenarios (\ct{Bilic:2018uqx} and references therein). The first early successful cosmological model based on the extended $f(R)= R + R^2$-gravity produces the  classical Starobinsky inflationary scalar field potential \ct{Starobinsky:1979ty}.

Dynamically generated models of inflation from modified/extended gravity such as the Starobinsky model \ct{Starobinsky:1980te,Nojiri:2010wj,Berti:2015itd,Nojiri:2017ncd} still remain viable and produce some of the best fits to existing observational 
data compared to other inflationary models \ct{Akrami:2018odb}. 

Unification of inflation with dark energy and dark matter have been widely discussed
\ct{Nojiri:2003ft,Nojiri:2005pu,Nojiri:2006gh,Lozano:2019gck,Chamings:2019kcl,Liu:2019xhn,Cheng:2019bkh,Cahill:2019lwc,Bandyopadhyay:2019vdd,Kase:2019veo,Ketov:2019rzg,Mukhopadhyay:2019jla,Yang:2019nhz}. 
It is indeed challenging to describe both phases of acceleration using a single 
scalar field minimally couple to gravity, without affecting the thermal history 
of the universe which has been verified to a good accuracy. 
In order to enable  slow-roll behavior, the scalar field potential should exhibit 
shallow behaviour at early times followed by a steep region for most of the universe 
history turning shallow once again at late times. 
Although a simple exponential potential does not comply with the above picture,
% observational constraints related to inflation, 
here we present a simple modified gravity model naturally providing a
dynamically generated scalar potential, whose inflationary dynamics is compatible 
with the recent observational data. 
On the other hand, the task of describing particle creation 
will be discussed in our future work. 

Another specific broad class of modified (extended) gravitational theories is based 
on the formalism of 
{\em non-Riemannian spacetime volume-elements}. It was originally proposed in 
\ct{Guendelman:1996qy,Gronwald:1997ei,Guendelman:1999tb,Guendelman:1999qt,Guendelman:2007ph}, with a subsequent concise geometric 
formulation in \ct{Vasihoun:2014hpa,Guendelman:2015qva,Guendelman:2015jii}. This formalism was used as a 
basis for constructing a series of extended gravity-matter models describing 
unified dark energy and dark matter scenario \ct{Guendelman:2012gg,Guendelman:2015rea},
quintessential cosmological models with gravity-assisted and inflaton-assisted
dynamical suppression (in the ``early'' universe) or dynamical generation (in the
post-inflationary universe) of electroweak spontaneous symmetry
breaking and charge confinement \ct{Guendelman:2016lea,Guendelman:2018zcb}, as well as a novel mechanism for dynamical supersymmetric Brout-Englert-Higgs effect in supergravity 
\ct{Vasihoun:2014hpa}.

In the present paper our principal aim is to construct a plausible unified model,
\textsl{i.e.}, describing (most of the) principal physical manifestations of a
unification of inflation and dark energy interacting with dark matter, where
the formalism of the non-Riemannian spacetime volume-elements will play a fundamental
role. To this end we will consider a simple modified gravity interacting 
with a single scalar field where the Einstein-Hilbert part and the scalar field part
of the action are constructed 
within the formalism of the non-Riemannian volume-elements -- 
alternatives to the canonical Riemannian one $\sqrt{-g}$. 
The non-Riemannian volume element in the initial scalar field action leads to a 
hidden nonlinear Noether 
symmetry which produces energy-momentum tensor identified as a sum of a 
dynamically generated cosmological constant and a dynamically generated
dust-like dark matter. 
The non-Riemannian volume-element in the initial Einstein-Hilbert action 
upon passage to the physical Einstein-frame creates dynamically a second scalar 
field with a non-trivial inflationary potential and with an additional interaction 
with the dynamically generated dark matter. The resulting Einstein-frame action 
describes a fully dynamically generated unified model of inflation, dark
energy and dark matter.
Numerical results for observables such as the scalar power spectral index and the 
tensor-to-scalar ratio conform to the latest $2018$ \textsl{PLANCK} data.
%%%%%%%%%%%%%%%%%%

Let us briefly recall the essence of the non-Riemannian volume-form (volume-element) 
formalism. In integrals over differentiable manifolds (not necessarily 
Riemannian one, so {\em no} metric is needed) volume-forms are given by 
nonsingular maximal rank differential forms $\om$:
\br
\int_{\cM} \om \bigl(\ldots\bigr) = \int_{\cM} dx^D\, \Omega \bigl(\ldots\bigr)
\;,
\lab{omega-1}   
\er
where
\be
\om = \frac{1}{D!}\om_{\m_1 \ldots \m_D} dx^{\m_1}\wedge \ldots \wedge dx^{\m_D}
\quad ,\quad
\om_{\m_1 \ldots \m_D} = - \vareps_{\m_1 \ldots \m_D} \Omega \;.
\ee
Our conventions for the alternating symbols $\vareps^{\m_1,\ldots,\m_D}$ and
$\vareps_{\m_1,\ldots,\m_D}$ are: $\vareps^{01\ldots D-1}=1$ and 
$\vareps_{01\ldots D-1}=-1$). 
The volume element $\Omega$ transforms as scalar density under general 
coordinate reparametrizations.

In Riemannian $D$-dimensional spacetime manifolds a standard generally-covariant 
volume-form is defined through the ``D-bein'' (frame-bundle) canonical one-forms 
$e^A = e^A_\m dx^\m$ ($A=0,\ldots ,D-1$):
\be
\om = e^0 \wedge \ldots \wedge e^{D-1} = \det\Vert e^A_\m \Vert\,
dx^{\m_1}\wedge \ldots \wedge dx^{\m_D},
\lab{omega-riemannian}
\ee
yields:
\be
\Omega = \det\Vert e^A_\m \Vert = \sqrt{-\det\Vert g_{\m\n}\Vert} \; .
\ee
To construct modified gravitational theories as alternatives to ordinary
standard theories in Einstein's general relativity, instead of $\sqrt{-g}$ 
we can employ one or more alternative {\em non-Riemannian} 
volume element(s) as in \rf{omega-1} given by non-singular {\em exact} $D$-forms
$\om = d A$ where:
\be
A = \frac{1}{(D-1)!} A_{\m_1\ldots\m_{D-1}}
dx^{\m_1}\wedge\ldots\wedge dx^{\m_{-1}} 
\ee
so that the {\em non-Riemannian} volume element reads:
\be
\Omega \equiv \Phi(A) =
\frac{1}{(D-1)!}\vareps^{\m_1\ldots\m_D}\, \pa_{\m_1} A_{\m_2\ldots\m_D} \; .
\lab{Phi-D}
\ee
Thus, a non-Riemannian volume element is defined in terms of
the (scalar density of the) dual field-strength of an auxiliary rank 
$D-1$ tensor gauge field $A_{\m_1\ldots\m_{D-1}}$.
%%%%%%%%%%%%%%%%%%%%%%%%%

The modified gravity Lagrangain actions based on the non-Riemannian
volume-elements formalism are of the following generic form (here and in
what follows we will use units with $16\pi G_{\rm Newton}=1$):
\be
S = \int d^4 x \,\P_1 (B)\bigl( R + \cL_1\bigr) + \int d^4 x \,\P_0 (A)\cL_0 
+ \int d^4 x \,\sqrt{-g}\,\cL_2 \; ,
\lab{NRVF-generic}
\ee
where $\P_0 (A)$ and $\P_1 (B)$ are of the form \rf{Phi-D} (for $D=4$),
$R$ is the scalar curvature,  and
the Lagrangian densities $\cL_{0,1,2}$ contain the matter fields (and possibly
higher curvature terms, \textsl{e.g.} $R^2$).

%%%%%%%%%%%%%
A basic property of the class of actions \rf{NRVF-generic} is that the equations
of motion w.r.t. auxiliary gauge fields, defining the non-Riemannian
volume-elements $\P_0 (A)$ and $\P_1 (B)$ as in \rf{Phi-D}, produce 
dynamically generated {\em free integration constants} $M_1$, $M_0$:
\be
\begin{split}
\pa_\m \bigl(R + \cL_1\bigr) = 0 \; \to \; R + \cL_1 = - M_1 \\
\pa_\m \cL_0 = 0 \; \to \; \cL_0 = -2 M_0 \; ,    
\end{split}
\lab{integr-const}
\ee
(cf. Eqs.\rf{L-const} and \rf{B-eq} below) whose appearance will play an instrumental
role in the sequel.
%%%%%%%%%%%%

Further, let us stress on the following important characteric feature of the 
modified gravity-matter actions \rf{NRVF-generic}. When considering the gravity part 
in the first order (Palatini) framework (\textsl{i.e.}, $R = g^{\m\n} R_{\m\n}(\G)$ 
with a priori indepedent metric $g_{\m\n}$ and affine connection $\G_{\m\n}^\l$),
then the auxiliary rank 3 tensor gauge fields defining the non-Riemannian 
volume-elements in \rf{NRVF-generic} are almost {\em pure-gauge} 
degrees of freedom, \textsl{i.e.} they {\em do not} introduce any 
additional propagating gravitational degrees of 
freedom when passing to the physical Einstein-frame except for few discrete degrees 
of freedom with conserved canonical momenta appearing as arbitrary integration 
constants. This has been explicitly shown within the canonical Hamiltonian treatment 
\ct{Guendelman:2015qva,Guendelman:2016lea}. 

On the other hand, when we treat \rf{NRVF-generic} in the second order
(metric) formalism (the affine connection $\G_{\m\n}^\l$ is the canonical
Levi-Civitta connection in terms of $g_{\m\n}$), while passing to the physical 
Einstein-frame via conformal transformation (cf. Eq.\rf{g-bar} below):
\be
g_{\m\n} \to {\bar g}_{\m\n}= \chi_1 \, g_{\m\n},\quad
\chi_1 \equiv \frac{\P_1 (A)}{\sqrt{-g}} \; ,
\lab{g-bar-0}
\ee
the first non-Riemannian volume element $\P_1 (A)$ in \rf{NRVF-generic} 
is not any more (almost) ``pure gauge'', but creates a new dynamical canonical 
scalar field $u$ via $\chi_1 = \exp{\frac{u}{\sqrt{3}}}$, which will play
the role both of an inflaton field at early times, as well as
driving late-time de Sitter expansion (see Section 3 below).

%%%%%%%%%  PLAN of the paper  %%%%%%%%%%%
In Section 2 we briefly review our construction in \ct{Guendelman:2015rea} 
of a simple gravity-scalar-field model -- specific member of the class of
modified gravitational models \rf{NRVF-generic} of the form \rf{NRVF-0} below, 
which  yields an explicit dynamical generation of independent (non-interacting 
among each other)
dark energy and dark matter components in an unified description as a 
manifestation of a single material entity (``darkon'' scalar field) -- 
a simplest realization of a $\L$CDM model. 

In section 3 we extend the previous 
construction to dynamically generate, apart from dark matter, also
early-time inflation and late-time de Sitter expansion  -- via 
dynamical creation of an additional canonical scalar field $u$
(``inflaton'') out of a non-Riemannian volume-element with the following
properties: 
(i) $u$ aquires dynamically a non-trivial inflationary type scalar field 
potential driving inflation at early times of the universe' evolution;
(ii) At late times the same evolving $u$ flows towards a stable critical point of
the pertinent dynamical system describing the cosmological evolution,
driving a late-time de Sitter expansion in a dark energy dominated epoch; 
(iii) In this case the field $u$, 
% manifesting itself as a quintessence-like inflationary field, 
induces a specific interaction between the dark energy and dark matter. 

In Section 4 we study
the cosmological implications of the latter dynamically generated % quintessence-like 
inflationary model with interacting dark energy and dark matter. 
In Section 5 several plots of the numerical solutions for the
evolution of the dynamical % quintessence-like 
inflationary field and for the
behavior of the relevant inflationary slow-roll parameters and the
corresponding observables are presented.
Section 6 contains our conclusions and outlook.

%%%%%%%%%%%%%%%%%%%%%%%%%%%%%%%%%%%%%%%%%%%%%%%%%%%%%%%%%%%%%%%%%%%%%%%%%%%%%%%%%%%%
%%%%%%%%%%%%%%%%%%%%%%%%%%%%%%%%%%%%%%%%%%%%%%%%%%%%%%%%%%%%%%%%%%%%%%%%%%%%%%%%%%%%
\section{A Simple Model of Unification of Dark Energy and Dark Matter}

% Short review of E.I. Guendelman, E. Nissimov and S. Pacheva, "Dark Energy and 
% Dark Matter From Hidden Symmetry of Gravity Model with a Non-Riemannian Volume 
% Form", European Physics Journal C75 (2015) 472-479

In \ct{Guendelman:2015rea} we started with the following non-conventional 
gravity-scalar-field action  -- a simple particular case of the class
\rf{NRVF-generic} --
containing one metric-independent non-Riemannian
volume-element alongside with the standard Riemannian one:
% (using units with $16\pi G_{\rm Newton}=1$):
\be
S = \int d^4 x \sqrt{-g}\, R(g) + 
\int d^4 x \bigl(\sqrt{-g}+\P_0 (A)\bigr) L(\vp,X) \; ,
\lab{NRVF-0}
\ee
with the following notations:

\begin{itemize}
\item
The first term in \rf{NRVF-0} is the standard Einstein-Hilbert action with $R(g)$ 
denoting the scalar curvature w.r.t. metric $g_{\m\n}$ in the second order (metric) 
formalism;
\item
$\P_0 (A)$ is particular representative of a $D=4$ non-Riemannian
volume-element density \rf{Phi-D}:
\be
\P_0 (A) = \frac{1}{3!}\vareps^{\m\n\k\l} \pa_\m A_{\n\k\l} \; .
\lab{Phi-A}
\ee
\item
$L(\vp,X)$ is general-coordinate invariant Lagrangian of a single scalar field 
$\vp (x)$:
\be
L(\vp,X)=X-V(\vp),\quad X \equiv - \frac{1}{2} g^{\m\n}\pa_\m \vp \pa_\n \vp \; .
\lab{L-darkon}
\ee
\end{itemize}

Varying \rf{NRVF-0} w.r.t. $g^{\m\n}$, $\vp$ and $A_{\m\n\l}$ yield the
following equations of motion, respectively:
\be
R_{\m\n}(g) -\frac{1}{2} g_{\m\n} R(g) = \frac{1}{2} T_{\m\n} \quad ,\quad
% \lab{einstein-eqs} \\
T_{\m\n} = g_{\m\n} L(\vp,X) + 
\Bigl( 1+\frac{\P_0 (A)}{\sqrt{-g}}\Bigr) \pa_\m \vp\, \pa_\n \vp \; ;
\lab{EM-tensor}
\ee
\be
-\partder{V}{\vp} + \bigl(\P_0 (A)+\sqrt{-g}\bigr)^{-1}
\pa_\m \Bigl\lb\bigl(\P_0 (A)+\sqrt{-g}\bigr) 
g^{\m\n}\pa_\n \vp \Bigr\rb = 0 \; ;
\lab{vp-eqs}
\ee
\be
\pa_\m L (\vp,X) = 0 \quad \longrightarrow \quad
L(\vp,X) \equiv X - V(\vp) = - 2M_0 = {\rm const} \; ,
\lab{L-const}
\ee
where $M_0$ is arbitrary integration constant (the factor $2$ is for later 
convenience).

As stressed in \ct{Guendelman:2015rea}, the scalar field dynamics 
is determined entirely by the first-order differential equation -- the 
dynamical constraint Eq.\rf{L-const}.
The usual second order differential equation \rf{vp-eqs} for $\vp$ is in fact a
consequence of \rf{L-const} together with the energy-momentum conservation:
\begin{equation}
    \nabla^\m T_{\m\n} = 0.
\end{equation}
Also, as exhibited in \ct{Guendelman:2015rea}, the specific form of the scalar field
potential $V(\vp)$ does not affect the dynamics of the system \rf{NRVF-0},
see the remark below following \rf{hidden-sym}.
The same phenomenon occurs in the extension of \rf{NRVF-0} to the model
\rf{NRVF-1} in Section 3 and 4 below.

The canonical Hamiltonian analysis in \ct{Guendelman:2015rea} of the action \rf{NRVF-0}
reveals that the auxiliary gauge field $A_{\m\n\l}$ is in fact an almost
pure-gauge, \textsl{i.e.}, it is a non-propagating field-theoretic degree of
freedom with the integration constant $(-2M_0)$ identified with the
conserved Dirac-constrained canonical momentum conjugated to the ``pure gauge''
``magnetic'' component of $A_{\m\n\l}$. For a general canonical Hamiltonian
treatment of Lagrangian action with one or more non-Riemannian
volume-elements, we refer to \ct{Guendelman:2014waa}.

A crucial property of the model \rf{NRVF-0} is the existence of a hidden
nonlinear Noether symmetry revealed in \ct{Guendelman:2015rea}. Indeed, both
Eqs.\rf{vp-eqs}-\rf{L-const} can be equivalently rewritten in the following
current-conservation law form:
\be
\nabla_\m J^\m = 0 \quad ,\quad
J^\m \equiv \Bigl(1+\frac{\P_0(A)}{\sqrt{-g}}\Bigr)\sqrt{2X} g^{\m\n}\pa_\n \vp \; .
\lab{J-conserv}
\ee
The covariantly conserved current $J^\m$ \rf{J-conserv} is the Noether
current corresponding to the invariance (modulo total derivative) of the 
action \rf{NRVF-0} w.r.t following hidden nonlinear symmetry transformations:
\be
\d_\eps \vp = \eps \sqrt{X} \quad ,\quad \d_\eps g_{\m\n} = 0 \quad ,\quad
% \nonu \\
\d_\eps \cA^\m = - \eps \frac{1}{2\sqrt{X}} g^{\m\n}\pa_\n \vp 
\bigl(\P_0(A) + \sqrt{-g}\bigr)  \; ,
\lab{hidden-sym}
\ee
with $\cA^\m = \bigl(\cA^0 \equiv \frac{1}{3!} \vareps^{mkl} A_{mkl}\; ,
\cA^i \equiv -\frac{1}{2} \vareps^{ikl} A_{0kl}\bigr)$ -- 
``dual'' components of the auxiliary gauge field $A_{\m\n\l}$ \rf{Phi-A}.
% \be
% \cA^0 = \frac{1}{3!} \vareps^{mkl} A_{mkl} \;\; ,\;\;
% \cA^i \equiv -\frac{1}{2} \vareps^{ikl} A_{0kl} \; .
% \lab{A-short}
% \ee

\textbf{Remark.} We notice that the existence of the hidden nonlinear 
symmetry \rf{hidden-sym} of the action \rf{NRVF-0} {\em does not} depend on 
the specific form of the scalar field potential $V(\vp)$. 
% The only requirement is that the kinetic term $X$ must be positive.

The next important step is to rewrite $T_{\m\n}$ \rf{EM-tensor} and 
$J^\m$ \rf{J-conserv} in the relativistic hydrodynamical form 
(again taking into account \rf{L-const}):
\be
T_{\m\n} = \rho_0 u_\m u_\n - 2M_0 g_{\m\n} \quad ,\quad J^\m = \rho_0 u^\m \; .
\lab{T-J-hydro}
\ee
Here the integration constant % $M \equiv - p/2$ ($p$ -- pressure) 
$M_0$ appears as dynamically generated cosmological constant and:
\be
\rho_0 \equiv \Bigl(1+\frac{\P_0(A)}{\sqrt{-g}}\Bigr)\, 2X ,\quad
u_\m \equiv - \frac{\pa_\m \vp}{\sqrt{2X}} \quad 
({\rm note} \; u^\m u_\m = -1\;) \; .
\lab{rho-u-def}
\ee
We now find that the covariant conservation laws for the energy-momentum tensor 
\rf{T-J-hydro} $\nabla^\m T_{\m\n} = 0$ and the $J$-current \rf{J-conserv} 
acquire the form:
\be
\nabla^\m \bigl(\rho_0 u_\m u_\n\bigr) = 0 \quad ,\quad
\nabla^\m \bigl(\rho_0 u_\m\bigr) = 0 \; .
\lab{dust-hydro}
\ee
Eqs.\rf{dust-hydro} imply in turn the geodesic equation for the ``fluid'' 
4-velocity $u_\m$:
\be
u_\m \nabla^\m u_\n = 0 \; .
\lab{geodesic-eq}
\ee

Therefore, comparing \rf{T-J-hydro} with the standard expression for a perfect fluid
stress-energy tensor $T_{\m\n} = \bigl(\rho + p) u_\m u_\n + p g_{\m\n}$,
we see that
% \be
% p = - 2M \quad ,\quad \rho = \rho_0 + 2M \quad {\rm with} \;\;
% \rho_0 \; {\rm as ~in ~\rf{rho-u-def}} \; ,
% \lab{p-rho-def}
% \ee
$T_{\m\n}$ \rf{T-J-hydro} consists of two additive parts which have the
following interpretation according to the standard $\L$-CDM model
\ct{Frieman:2008sn,Mathews:2017xht,Liddle:2009zz,Liddle:1998ew,Dodelson:2003ft,Dodelson:2009kq,Baumann:2008aj,Dodelson:1999hm} (using notations
$p = p_{\rm DM} + p_{\rm DE}$ and $\rho = \rho_{\rm DM} + \rho_{\rm DE}$):

\begin{itemize}
\item
{\em Dynamically generated dark energy} part given by the second cosmological constant 
term in $T_{\m\n}$ \rf{T-J-hydro} due to \rf{L-const}, where
$p_{\rm DE} = -2M\, ,\, \rho_{\rm DE} = 2M$;
\item
{\em Dynamically generated dark matter} part given by the first term in \rf{T-J-hydro}, 
where $p_{\rm DM} = 0\, ,\, \rho_{\rm DM} = \rho_0$ with 
$\rho_0$ as in \rf{rho-u-def}, which in fact according to \rf{dust-hydro} and
\rf{geodesic-eq} describes a {\em dust fluid} with fluid density $\rho_0$
flowing along geodesics. Thus, we will call the $\vp$ scalar field by the alias
``darkon''.
\end{itemize}

The conservation laws \rf{dust-hydro} due to the hidden nonlinear Noether
symmetry \rf{hidden-sym} imply that in the model \rf{NRVF-0} there is 
{\em no interaction} between dark energy and dark matter -- 
they are separately conserved.

%%%%%%%%%%%%%%%%%%%%%%%%%%%%%%%%%%%%%%%%%%%%%%%%%%%%%%%%%%%%%%%%%%%%%%%%%%%%%%%%%%%%
%%%%%%%%%%%%%%%%%%%%%%%%%%%%%%%%%%%%%%%%%%%%%%%%%%%%%%%%%%%%%%%%%%%%%%%%%%%%%%%%%%%%
\section{Inflation and Unified Dark Energy and Dark Matter}
Now we will extend the simple model \rf{NRVF-0} of unified dark energy and dark matter by introducing another metric-independent non-Riemannian volume-element:
\be
\P_1 (B) = \frac{1}{3!}\vareps^{\m\n\k\l} \pa_\m B_{\n\k\l}
\lab{Phi-B}
\ee
inside the gravity (Einstein-Hilbert) part of the action (using again units with 
$16\pi G_{\rm Newton} = 1$):
\be
S = \int d^4 x\,\Bigl\{\P_1(B) \Bigl\lb R(g)
- 2 \L_0 \frac{\P_1(B)}{\sqrt{-g}}\Bigr\rb
+ \Bigl(\sqrt{-g}+\P_0 (A)\Bigr) \Bigl\lb -\frac{1}{2} g^{\m\n} \pa_\m\vp \pa_\n\vp
- V(\vp)\Bigr\rb\Bigr\} \; .
\lab{NRVF-1}
\ee
Here $\L_0$ is a dimensionful parameter to be identified later on as energy
scale of the inflationary universe' epoch.

The specific form of the action \rf{NRVF-1} may be justified by the requirement
about global Weyl-scale invariance under the transformations:
\be
g_{\m\n} \to \l g_{\m\n} \quad,\quad A_{\m\n\k} \to \l^2 A_{\m\n\k} \quad ,\quad 
B_{\m\n\k} \to \l B_{\m\n\k} \quad ,\quad \vp \to \l^{-\frac{1}{2}} \vp \; ,
\lab{scale-transf}
\ee
and provided we choose $V(\vp) = \vp^4$. Concerning global Weyl-scale invariance
let us note that it played an important role already since the first original papers on 
the non-canonical volume-form formalism \ct{Guendelman:1999qt}. In particular, 
models with spontaneously broken dilatation symmetry have been constructed 
along these lines, which are free of the Fifth Force Problem \ct{Guendelman:2007ph}.

The equations of motion of the action \rf{NRVF-1} w.r.t. $\vp$ and 
$A_{\m\n\l}$ are the same as in \rf{vp-eqs}-\rf{L-const}, therefore once
again \rf{NRVF-1} is invariant under the hidden nonlinear Noether symmetry 
\rf{hidden-sym} with the associated Noether conserved current \rf{J-conserv},
which we rewrite here for later convenience taking into account \rf{L-const}:
\be
\nabla_\m J^\m = 0 \quad ,\quad
J^\m =
\bigl(1+\chi_0\bigr)\sqrt{2\bigl(V(\vp)-2M_0\bigr)}\, g^{\m\n}\pa_\n \vp 
\quad ,\quad \chi_0 \equiv \frac{\P_0(A)}{\sqrt{-g}} \; .
\lab{J-conserv-1}
\ee
On the other hand, the equations of motion w.r.t. $g^{\m\n}$ and $B_{\m\n\l}$ 
now read:
\br
R_{\m\n}(g) + 
\frac{1}{\chi_1}\bigl(g_{\m\n}\Box\chi_1 -\nabla_\m \nabla_\n \chi_1\Bigr)
-\L_0 \chi_1 g_{\m\n} = \frac{1}{2\chi_1} T_{\m\n} \; ,
\lab{einstein-like} \\
R(g) - 4 \L_0 \chi_1 = - M_1 \quad ,\quad 
\chi_1 \equiv \frac{\P_1(B)}{\sqrt{-g}} \; ,
\lab{B-eq}
\er
where $T_{\m\n}$ is the same energy-momentum tensor as in \rf{EM-tensor}
or \rf{T-J-hydro}, which taking into account \rf{L-const} and using short-hand 
notation $\chi_0$ in \rf{J-conserv-1}) reads
$T_{\m\n} = -2M_0 g_{\m\n} + \bigl(1+\chi_0\bigr)\pa_\m \vp \pa_\n \vp \; ,$ 
% \be
% T_{\m\n} = -2M_0 g_{\m\n} + \bigl(1+\chi_0\bigr)\pa_\m \vp \pa_\n \vp \; ,
% \lab{}
% \ee
and $M_1$ is another free integration constant similar to $M_0$ in \rf{L-const}.
Taking trace of \rf{einstein-like} together with \rf{B-eq} imply a
dynamical equation for $\chi_1$ ($\chi_0$ and $\chi_1$ as defined in \rf{J-conserv-1}
and \rf{B-eq}, respectively):
\be
\Box\chi_1 - \frac{1}{3}M_1 \chi_1 - \frac{1}{6}T = 0 \quad ,\quad
T \equiv g^{\m\n} T_{\m\n} = -8M_0 - 2(1+\chi_0)\bigl(V(\vp) - 2M_0\bigr) \; ,
\lab{chi1-eq}
\ee

The passage to the Einstein-frame is accomplished via the conformal
transformation:
\be
g_{\m\n} \;\; \longrightarrow \;\; {\bar g}_{\m\n} = \chi_1 g_{\m\n} \; ,
\lab{g-bar}
\ee
on Eqs.\rf{einstein-like} and \rf{chi1-eq}, 
and upon using the known formulae for conformal transformations of Ricci
curvature tensor and covariant Dalambertian (see \textsl{e.g.} \ct{Dabrowski:2008kx}; bars indicate magnitudes in the ${\bar g}_{\m\n}$-frame):
\br
R_{\m\n}(g) = R_{\m\n}(\bar{g}) - 3 \frac{{\bar g}_{\m\n}}{\chi_1}
{\bar g}^{\k\l} \pa_\k \chi_1^{1/2} \pa_\l \chi_1^{1/2} 
% \nonu \\
+ \chi_1^{-1/2}\bigl({\bar \nabla}_\m {\bar \nabla}_\n \chi_1^{1/2} +
{\bar g}_{\m\n} {\bar{\Box}}\chi_1^{1/2}\bigr) \; ,
\lab{dabrowski-1} \\
\Box \chi_1 = \chi_1 \Bigl({\bar{\Box}}\chi_1 
- 2{\bar g}^{\m\n} \frac{\pa_\m \chi_1^{1/2} \pa_\n \chi_1}{\chi_1^{1/2}}\Bigr)
\; .
\lab{dabrowski-2} 
\er
In the process we introduce the field redefinition $\chi_1 \; \to \; u$:
\be
\chi_1 = \exp\Bigl\{\frac{u}{\sqrt{3}}\Bigr\} \; ,
\lab{u-def}
\ee
so that $u$ appears as a canonical scalar field in the Einstein-frame 
transformed equations \rf{einstein-like}, \rf{chi1-eq} and \rf{L-const}:
\br
{\bar R}_{\m\n} - \frac{1}{2} {\bar g}_{\m\n} {\bar R} = \frac{1}{2} {\bar T}_{\m\n} \; ,
\nonu \\
{\bar T}_{\m\n} = \pa_\m u\, \pa_\n u + {\bar g}_{\m\n}\Bigl\lb
-\frac{1}{2} {\bar g}^{\k\l}\pa_\k u\, \pa_\l u - U_{\rm eff}(u)\Bigr\rb
+ e^{-u/\sqrt{3}}(1+\chi_0)\pa_\m \vp\, \pa_\n \vp \; ,
\lab{einstein-EF}\\
{\bar\Box}u - \partder{U_{\rm eff}(u)}{u} 
+ \frac{1}{\sqrt{3}} e^{-2u/\sqrt{3}}(1+\chi_0)\bigl( V(\vp) - 2M_0\bigr) = 0 \; ,
\lab{u-eq} \\
\frac{1}{2} {\bar g}^{\m\n}\pa_\m \vp \pa_\n \vp 
+ e^{-u/\sqrt{3}}\bigl( V(\vp) - 2M_0\bigr) = 0 \; ,
\lab{L-const-EF}
\er
and most importantly $u$ acquires a non-trivial dynamically denerated potential:
\be
U_{\rm eff}(u) = 2\L_0 - M_1 e^{-u/\sqrt{3}} + 2M_0 e^{-2u/\sqrt{3}} 
\lab{U-eff}
\ee
due to the appearance of the free integration constants from the equations
of motion of the original-frame non-Riemannian spacetime volume-elements.
The hidden nonlinear Noether symmetry current conservation \rf{J-conserv},
equivalent to the $\vp$-equation of motion, becomes in the Einstein-frame:
\be
{\bar\nabla}_\m {\bar J}^\m = 0 \quad ,\quad  {\bar J}^\m = 
(1+\chi_0) e^{-u/\sqrt{3}}\sqrt{V(\vp)-2M_0}\, {\bar g}^{\m\n}\pa_\n \vp \; .
\lab{J-conserv-EF}
\ee

Thus, the Einstein-frame Lagrangian action producing the Einstein-frame
equations of motion \rf{einstein-EF}-\rf{J-conserv-EF} reads:
\br
S_{\rm EF} = \int d^4 x \sqrt{-{\bar g}} \Bigl\lb {\bar R} 
-\frac{1}{2} {\bar g}^{\m\n}\pa_\m u\,\pa_\n u - U_{\rm eff}(u)\Bigr\rb 
\nonu \\
+ \int d^4 x \sqrt{-{\bar g}} \bigl(1+\chi_0\bigr) e^{-u/\sqrt{3}}
\Bigl\lb-\frac{1}{2} {\bar g}^{\m\n}\pa_\m \vp\,\pa_\n \vp 
- e^{-u/\sqrt{3}}\bigl(V(\vp)-2M_0\bigr)\Bigr\rb \; ,
\lab{EF-action}
\er
with $U_{\rm eff}(u)$ as in \rf{U-eff} and where now $\chi_0$ (from 
\rf{J-conserv-1}) becomes a simple Lagrange multiplier.

The upper line in $S_{\rm EF}$ \rf{EF-action} represents an
inflationary Lagrangian action with dynamically generated inflationary
potential $U_{\rm eff}(u)$ \rf{U-eff} obtained in \ct{Benisty:2019tno} from a 
pure gravity 
initial action (without any matter fields) in terms of non-Riemannian volume-elements:
\be
S_0 = \int d^4 x\,\Bigl\{\P_1(B) \Bigl\lb R(g)
- 2 \L_0 \P_1(B)/\sqrt{-g}\Bigr\rb + \bigl(\P_0(A)\bigr)^2/\sqrt{-g}\Bigr\}
\ee
which is graphically depicted on Fig.1, is a generalization of the classic 
Starobinsky inflationary potential \ct{Starobinsky:1979ty}. In fact, the latter 
is a special case of \rf{U-eff} for the particular values of the parameters: 
$\L_0 = M_0=\frac{1}{4}M_1$.
\begin{figure}[h]
\centering
\includegraphics[width=0.7\textwidth]{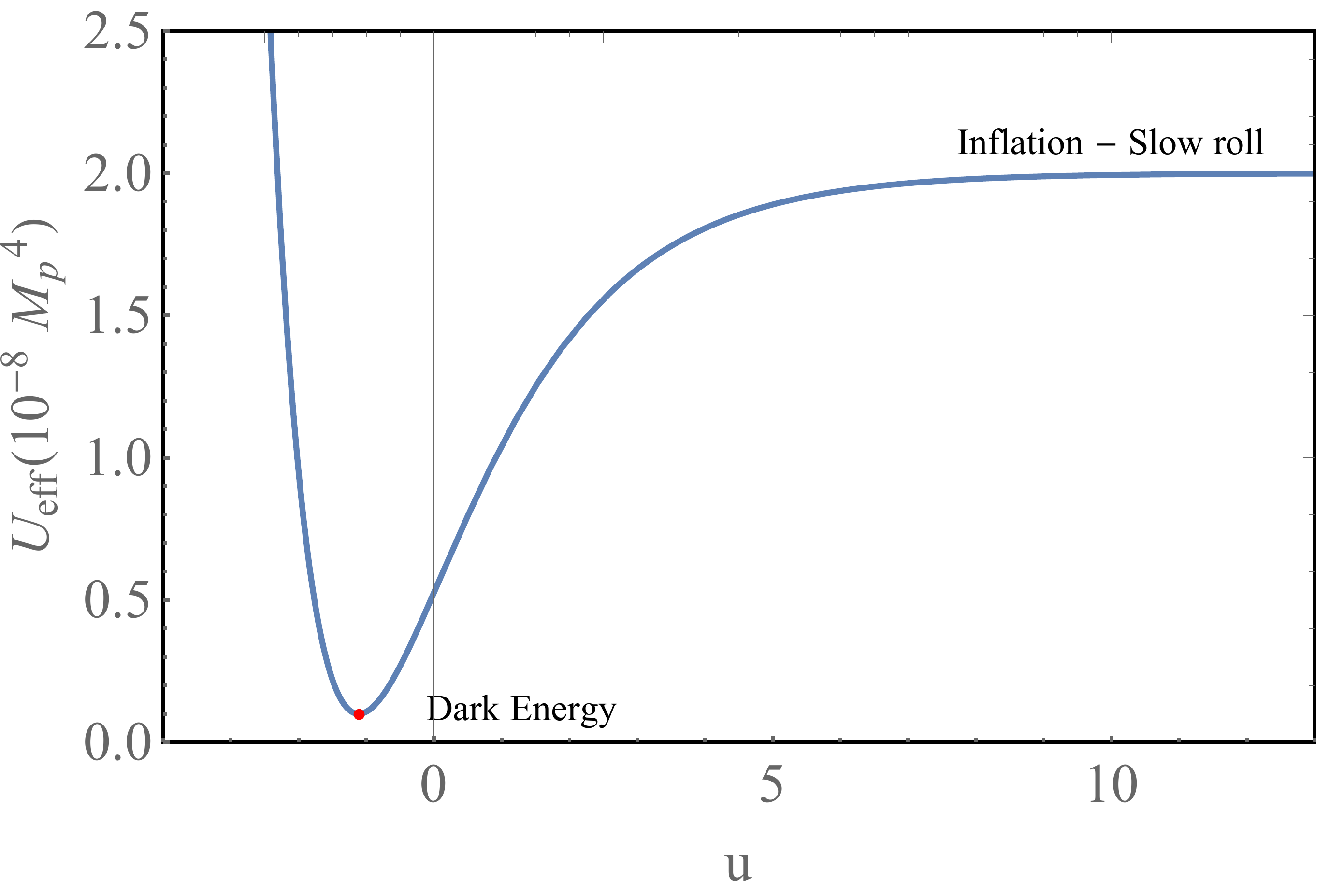}
\caption{Shape of the effective potential $U_{\rm eff} (u)$ in 
the Einstein-frame \rf{U-eff}. The physical unit for $u$ is $M_{Pl}/\sqrt{2}$.}
\label{fig1}
\end{figure}

$U_{\rm eff} (u)$ \rf{U-eff} possesses two main features relevant for
cosmological applications: 
% First, $U_{\rm eff} (u)$ \rf{U-eff} has an almost
% flat region for large positive $u$ and, second, it has a stable minimum for 
% a small finite value $u=u_{*}$:
\begin{itemize}
\item
(i) $U_{\rm eff} (u)$ \rf{U-eff} has an almost flat region for large
positive $u$: $U_{\rm eff} (u) \simeq 2\L_0$ for large $u$. This almost flat
region correspond to ``early'' universe' inflationary evolution with energy scale
$2\L_0$ as it will be evident from the autonomous dynamical system analysis of the
cosmological dynamics in Section 4;
\item
(ii) $U_{\rm eff} (u)$ \rf{U-eff} has has a stable minimum for 
a small finite value $u=u_{*}$:   
$\partder{U_{\rm eff}}{u} = 0\;$ for $u\equiv u_{*}$, where:
\be
\exp\bigl(-\frac{u_{*}}{\sqrt{3}}\bigr) = \frac{M_1}{4 M_0}  \quad ,
\frac{\pa^2 U_{\rm eff}}{\pa u^2}\bgv_{u=u_{*}} = \frac{M_1^2}{12 M_0} >0 \; . 
\lab{stable-min}
\ee
\item
(iii) As it will be explicitly exhibited in the dynamical system analysis in 
Section 4, the region of $u$ around the stable 
minimum at $u=u_{*}$ \rf{stable-min} correspond to late-time de Sitter expansion 
of the universe with slightly varied late-time Hubble parameter (dark energy
dominated epoch), where the minimum value of the potential:
\be
U_{\rm eff} (u_{*})= 2\L_0 - \frac{M_1^2}{8 M_0} \equiv 2 \L_{\rm DE}
\lab{DE-value}
\ee
is the asymptotic value at $t \to \infty$ of the dynamical dark energy density
\ct{Angus:2018tko,Zhang:2018gbq}. 
\end{itemize}

The lower line in $S_{\rm EF}$ \rf{EF-action} represents the interaction
between the dynamical % quintessence-like 
inflaton field $u$ and the ``darkon'' field $\vp$, 
in other words here we have unification of inflation, dark energy and dark-matter.
This is reflected in the structure of the Einstein-frame energy-momentum
tensor ${\bar T}_{\m\n}$ \rf{einstein-EF} -- the first two terms being the
stress-energy tensor of $u$ and the last term being the ``darkon''
stress-energy tensor coupled to $u$.

%%%%%%%%%%%%%%%%%%%%%%%%%%%%%%%%%%%%%%%%%%%%%%%%%%%%%%%%%%%%%%%%%%%%%%%%%%%%%%%%%%%%
%%%%%%%%%%%%%%%%%%%%%%%%%%%%%%%%%%%%%%%%%%%%%%%%%%%%%%%%%%%%%%%%%%%%%%%%%%%%%%%%%%%%
\section{Cosmological Implications}

Let us now consider reduction of the Einstein-frame action \rf{EF-action} to the
Friedmann-Lemaitre-Robertson-Walker (FLRW) framework with metric
$ds^2 = - N^2 dt^2 + a(t)^2 d{\vec x}^2$,  where $u=u(t)$ and $\vp=\vp(t)$:
\br
S_{\rm FLRW} = \int d^4 x \Bigl\{ - 6\frac{a\adot^2}{N}
+ N a^3 \Bigl\lb\frac{1}{2}\frac{\udot^2}{N^2} + M_1 e^{-u/\sqrt{3}} - 2 M_0 e^{-2u/\sqrt{3}} 
- 2\L_0\Bigr\rb 
\nonu \\
+ N a^3 (1+\chi_0) e^{-u/\sqrt{3}}\Bigl\lb \frac{1}{2} \frac{\vpdot^2}{N^2}
- e^{-u/\sqrt{3}}\bigl(V(\vp) - 2M_0\bigr)\Bigr\rb \Bigr\} \; .
\lab{EF-action-FLRW}
\er
The equations of motion w.r.t. $\chi_0$ and $\vp$ from \rf{EF-action-FLRW} 
are equivalent to the FLRW reduction of the dynamical constraint \rf{L-const-EF} 
and of the Noether current conservation \rf{J-conserv-EF}, respectively:
\be
\vpdot^2 = 2 e^{-u/\sqrt{3}}\bigl(V(\vp) - 2M_0\bigr) \quad ,\quad
\frac{d}{dt}\Bigl\lb a^3(1+\chi_0)e^{-u/\sqrt{3}}\sqrt{V(\vp)-2M_0}\,\vpdot\Bigr\rb
= 0 \; ,
\lab{FLRW-Lconst-Jconserv}
\ee
which imply the relation:
\be
(1+\chi_0)e^{-u/\sqrt{3}}\bigl(V(\vp)-2M_0\bigr) = 
\frac{c_0}{a^3} e^{u/2\sqrt{3}} \; ,
\lab{J-conserv-FLRW}
\ee
with $c_0$ a free integration constant. Taking into account \rf{J-conserv-FLRW},
the FLRW reduction of the Einstein-frame energy-momentum tensor \rf{einstein-EF}
becomes:
\br
{\bar T}_{00}\equiv \rho \quad, \quad {\bar T}_{ij}\equiv a^2 \d_{ij} p 
\quad, \quad {\bar T}_{0i}=0 \; ,
\lab{EM-tensor-FLRW} \\ 
% \rho = \frac{1}{2} \udot^2 + 2\L_0 - M_1 e^{-u/\sqrt{3}} 
% + 2 M_0 e^{-2u/\sqrt{3}} + 2 \frac{c_0}{a^3} e^{-u/2\sqrt{3}} \; ,
\rho = \frac{1}{2} \udot^2 + U_{\rm eff}(u) + 2 \frac{c_0}{a^3} e^{-u/2\sqrt{3}}
\quad ,\quad
% \lab{rho-FLRW} \\
% p = \frac{1}{2} \udot^2 - 2\L_0 + M_1 e^{-u/\sqrt{3}} - 2 M_0 e^{-2u/\sqrt{3}}
p = \frac{1}{2} \udot^2 - U_{\rm eff}(u) \; .
\lab{rho-p-FLRW}
\er
Relations \rf{rho-p-FLRW} explicitly show that the last term in $\rho$:
\be
\rho_{\rm DM} \equiv 2 \frac{c_0}{a^3} e^{-u/2\sqrt{3}}
\lab{rho-DM}
\ee
represents the ``dust'' dark matter part of the total energy denisty -- it
is ``dust'' because of absence ot corresponding contribution for the
pressure $p$ in \rf{rho-p-FLRW}.

The equation of motion from \rf{EF-action-FLRW} w.r.t. $u$ is ($H=\adot/a$
being the Hubble parameter):
\be
\uddot + 3H\udot + \partder{U_{\rm eff}}{u} 
- \frac{1}{\sqrt{3}}\frac{c_0}{a^3} e^{-u/2\sqrt{3}} = 0
\lab{u-eq-FLRW}
\ee
and, finally, the two Friedmann equations (varying \rf{EF-action-FLRW} w.r.t 
lapse $N$ and $a$) read:
\br
6H^2 = \frac{1}{2} \udot^2 + U_{\rm eff}(u) + 2 \frac{c_0}{a^3} e^{-u/2\sqrt{3}} \; ,
\lab{fried-1} \\
\Hdot = - \frac{1}{4}\Bigl(\udot^2 + 2 \frac{c_0}{a^3} e^{-u/2\sqrt{3}}\Bigr)\; .
\lab{fried-2} 
\er

\textbf{Remark.} We observe that due to the hidden nonlinear Noether symmetry 
current conservation \rf{J-conserv-FLRW}, the FLRW dynamics given by 
\rf{u-eq-FLRW}-\rf{fried-2} does not depend on the explicit form the ``darkon'' 
part of the FLRW action \rf{EF-action-FLRW} -- 
the only trace of the ``darkon'' is embodied in the integration constant $c_0$.

It is instructive to analyze the system FLRW equations \rf{u-eq-FLRW}-\rf{fried-2} 
as an automous dynamical system. To this end it is useful to rewrite the system
\rf{u-eq-FLRW}-\rf{fried-2} in terms of a set of dimensionless coordinates
(following the approach in \ct{Bahamonde:2017ize}): 
\be
x := \frac{\dot{u}}{\sqrt{12} H},\quad 
y := \frac{\sqrt{U_{\rm eff}(u) - 2\L_{\rm DE}}}{\sqrt{6} H}, \quad 
z := \frac{\sqrt{\L_{\rm DE}+\rho_{\rm DM}}}{\sqrt{3}H} \; ,
\lab{xyz-def}
\ee
with $L_{\rm DE}$ as in \rf{DE-value} and $\rho_{\rm DM}$ as in \rf{rho-DM}. 
In these coordinates the system defines a closed orbit:
\be
x^2 + y^2 + z^2 = 1 \; ,
% + \Omega_\text{stiff} = 1 \; ,
\lab{orbit}
\ee
which is equivalent to the first Friedmann equation \rf{fried-1}. Then Eqs.\rf{u-eq-FLRW}
and \rf{fried-2} can be represented as a 3-dimensional autonomous dynamical system for the
$(x,y, H$ variables (cf. \rf{xyz-def}):
\br
x^\pr = \frac{3}{2}x\Bigl\lb x^2 -1 - y^2 - \frac{\L_{\rm DE}}{3H^2}\Bigr\rb
+\frac{1}{2} \Bigl(1 - x^2 - y^2 - \frac{\L_{\rm DE}}{3H^2}\Bigr) 
\nonu \\
-\frac{2y}{H}\sqrt{\frac{M_0}{3}}\Bigl(\frac{M_1}{4M_0} - \sqrt{\frac{3}{M_0}}\,Hy\Bigr)
\; ,
\lab{x-eq} \\
y^\pr = 
\frac{2x}{H}\sqrt{\frac{M_0}{3}}\Bigl(\frac{M_1}{4M_0} - \sqrt{\frac{3}{M_0}}\,Hy\Bigr)
+\frac{3}{2}y\Bigl\lb 1 + x^2 - y^2 - \frac{\L_{\rm DE}}{3H^2}\Bigr\rb \; ,
\lab{y-eq} \\
H^\pr = -\frac{3}{2} H \Bigl\lb 1 + x^2 - y^2 - \frac{\L_{\rm DE}}{3H^2}\Bigr\rb \; , 
\lab{H-eq}
\er
where the primes indicate derivatives w.r.t. number of e-folds $\cN = \log(a)$ 
(meaning $\frac{d}{d\cN} = \frac{1}{H} \frac{d}{dt}$).

The dynamical system \rf{x-eq}-\rf{H-eq} possesses two critical points:

\begin{itemize}
\item
(A) Stable critical point:
\be
x_{*} = 0 \quad ,\quad y_{*} = 0 \quad ,\quad H_{*} =\sqrt{\frac{\L_{\rm DE}}{3}} \; ,
\lab{stable}
\ee
where all three eigenvalues of the stability matrix are negative or with negative
real parts ($\l_1 = -3 \;  ,\; 
\l_{2,3} = \frac{1}{2} \Bigl\lb - 3 + \sqrt{9 - \frac{M_1^2}{M_0 \L_{\rm DE}}}\Bigr\rb$).
The stable critical point \rf{stable} corresponds to the late-time asymptotics of the
universe' evolution where
according to the definitions \rf{xyz-def} $u(t) \to u_{*}$ -- the stable
minimum of the effective potential $U_{\rm eff}(u)$ \rf{U-eff}, so that 
$U_{\rm eff}(u) \to 2\L_{\rm DE}$, the dark matter energy density \rf{rho-DM} 
$\rho_{\rm DM} \to 0$, and $\Hdot \to 0$ accordin to \rf{H-eq},
\textsl{i.e.}, late-time accelerated expansion with $H_{*} =\sqrt{\frac{\L_{\rm DE}}{3}}$.
\item
(B) Unstable critical point:
\be
x_{**} = 0 \quad ,\quad y_{**} = \sqrt{1-\frac{\L_{\rm DE}}{\L_0}}
= \frac{M_1}{4 \sqrt{M_0 \L_0}} \quad ,\quad H_{**} =\sqrt{\frac{\L_0}{3}} \; ,
\lab{unstable}
\ee
where one of the three eigenvalues of the stability matrix is zero ($\l_1 = 0 \; , \;
\l_2= -3 \;,\; \l_3 = -3(1-\L_{\rm DE}/\L_0)= - \frac{3M_1^2}{16M_0\L_0}$).
According to the definitions \rf{xyz-def}, in the vicinity of the unstable
critical point \rf{unstable} $u(t)$ is very large positive ($u \to \infty$), so that 
$U_{\rm eff}(u) \simeq 2\L_0$, $\rho_{\rm DM}$ is vanishing 
$\rho_{\rm DM}\approx 0$, and we have
there a slow-roll inflationary evolution with inflationary scale $\L_0$ 
where the standard slow-roll parameters are very small:
\br
\epsilon = -\frac{\Hdot}{H^2} \approx 
\(\frac{\partder{U_{\rm eff}}{u} - \frac{1}{2\sqrt{3}}\,\rho_{\rm DM}}{U_{\rm eff} +
\rho_{\rm DM}}\)^2
+ \frac{3}{2} \frac{\rho_{\rm DM}}{U_{\rm eff} + \rho_{\rm DM}} \; ,
\lab{epsilon} \\
\eta = -\frac{\Hdot}{H^2} - \frac{\Hddot}{2H\Hdot} \approx
- 2 \frac{\frac{\pa^2 U_{\rm eff}}{\pa u^2} + \frac{1}{12}\,\rho_{\rm DM}}{
U_{\rm eff} + \rho_{\rm DM}} + {\rm O}(\rho_{\rm DM}) \; .
\lab{eta}
\er
\end{itemize}

%%%%%%%%%%%%%%%%%%%%%%%%%%%%%%%%%%%%%%%%%%%%%%%%%%%%%%%%%%%%%%%%%%%%%%%%%%%%%%%%%%%%
%%%%%%%%%%%%%%%%%%%%%%%%%%%%%%%%%%%%%%%%%%%%%%%%%%%%%%%%%%%%%%%%%%%%%%%%%%%%%%%%%%%%
\section{Numerical Solutions}

Going back to the system of equations \rf{u-eq-FLRW}-\rf{fried-2} we can use
\rf{fried-1} to replace the term 
$\rho_{\rm DM} \equiv 2 \frac{c_0}{a^3} e^{-u/2\sqrt{3}}$ in \rf{u-eq-FLRW} and 
\rf{fried-2} so that we will obtain a closed system of two coupled nonlinear
diffential equations for $\bigl(u(t), H(t)\bigr)$ of second and first order, 
respectively:
\br
\uddot + 3H\udot + \partder{U_{\rm eff}}{u} 
- \frac{1}{2\sqrt{3}}\Bigl\lb 6H^2 -\frac{1}{2} \udot^2 - U_{\rm eff}(u)\Bigr\rb = 0 \; ,
\lab{u-eq-final} \\
\Hdot = - \frac{1}{4}\Bigl\lb 6H^2 +\frac{1}{2} \udot^2 - U_{\rm eff}(u)\Bigr\rb \; ,
\lab{H-eq-final}
\er
where $U_{\rm eff}(u)$ is given by \rf{U-eff}:
$U_{\rm eff}(u) = 2\L_0 - M_1 e^{-u/\sqrt{3}} + 2M_0 e^{-2u/\sqrt{3}}$.

Below we present several plots qualitatively illustrating the evolutionary 
behavior of the numerical solution of the system \rf{u-eq-final}-\rf{H-eq-final} 
with initial conditions conforming to the unstable critical point $B$
\rf{unstable}: $\udot (0) = 0 \; ,\; H(0)=\sqrt{\L_0 /3}$ and $u(0)$ -- some
large inital value. As a numerical example, for the purpose of graphical
illustration,  we will take the following
numerical values of the paramaters (the physical units would be  
$10^{-9} M^4_{Pl}$):
\be
\L_0 = 50\;\; ,\;\; M_1=20\;\; ,\;\; M_0=0.501 \;\;
\longrightarrow \; \L_{\rm DE}=0.1 
\lab{num-example}
\ee
according to \rf{DE-value} (in reality $\L_{\rm DE}$ is much smaller than 1/500 part of $\L_0$: 
$\L_0 \sim 10^{-8} M^4_{Pl}$ \ct{Planck:2013jfk,Adam:2014bub}
and $\L_{DE} \sim 10^{-122} M^4_{Pl}$, cf. \ct{ArkaniHamed:2000tc}).

On Fig.2 below the plot represents the overall evolution of $u(t)$, whereas
on Fig.3 are the plots for the slow-roll parameters 
$\epsilon = -\frac{\Hdot}{H^2}$ and 
$\eta = -\frac{\Hdot}{H^2} - \frac{\Hddot}{2H\Hdot}$ clearly indicating the
end of inflation where their sharp grow-up starts. 

\begin{figure}[h]
\centering
\includegraphics[width=0.6\textwidth]{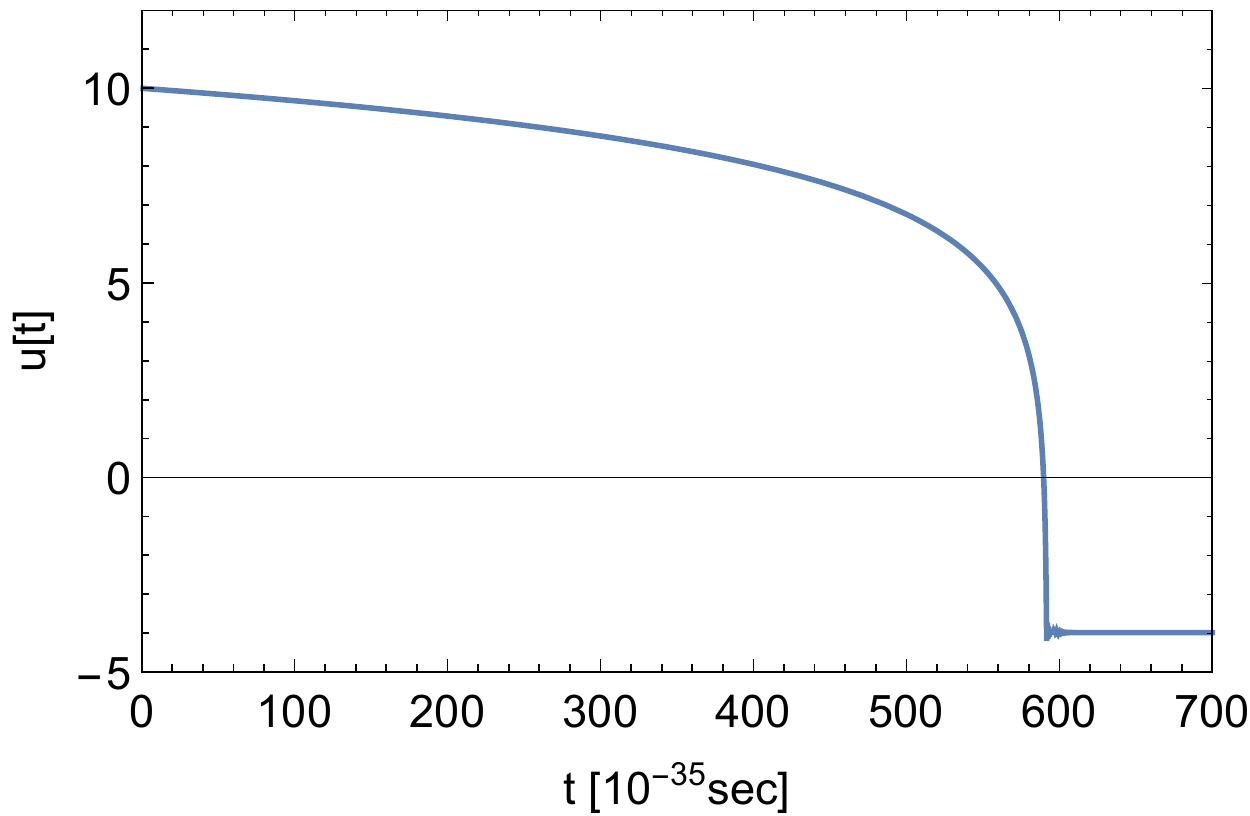}
\caption{Numerical shape of the evolution of $u(t)$. 
The physical unit for $u$ is $M_{Pl}/\sqrt{2}$.}
\label{fig2}
\end{figure}

\begin{figure}[h]
\centering
\includegraphics[width=0.49\textwidth]{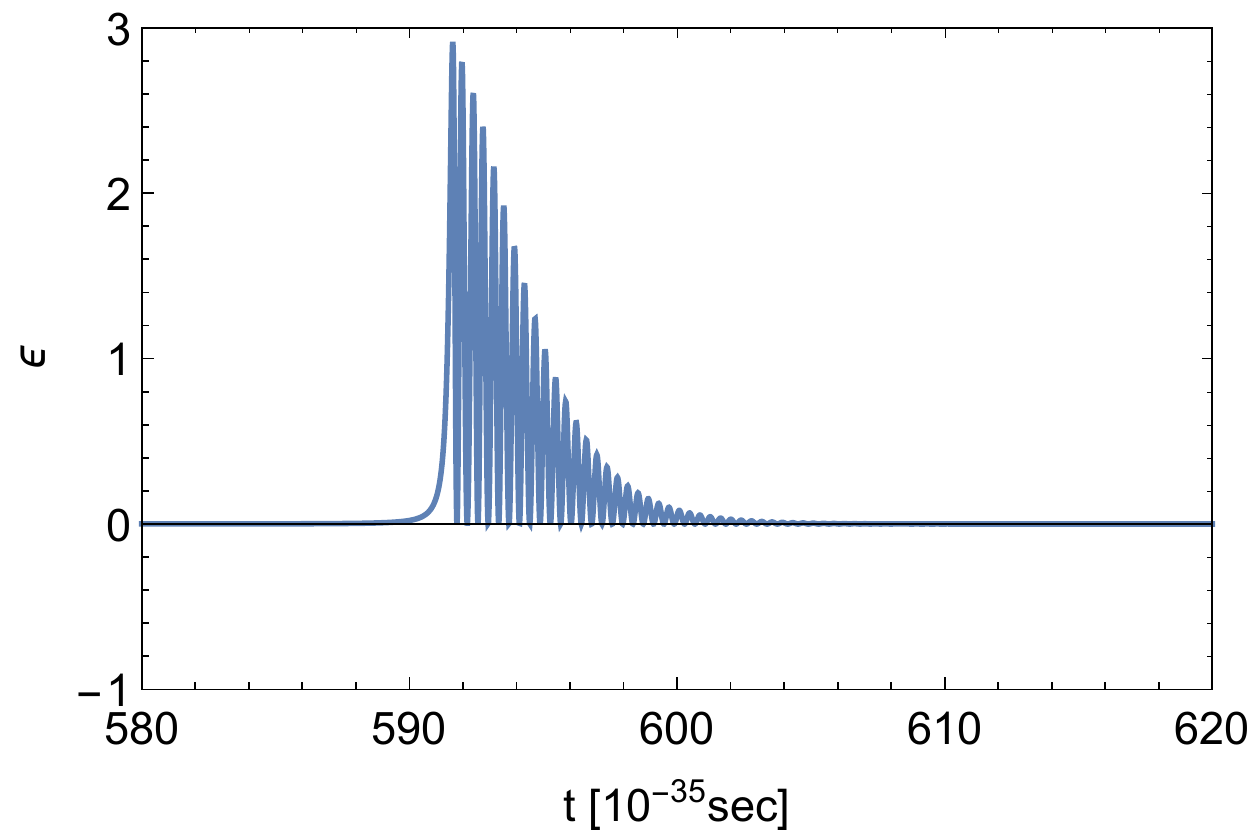}
\includegraphics[width=0.49\textwidth]{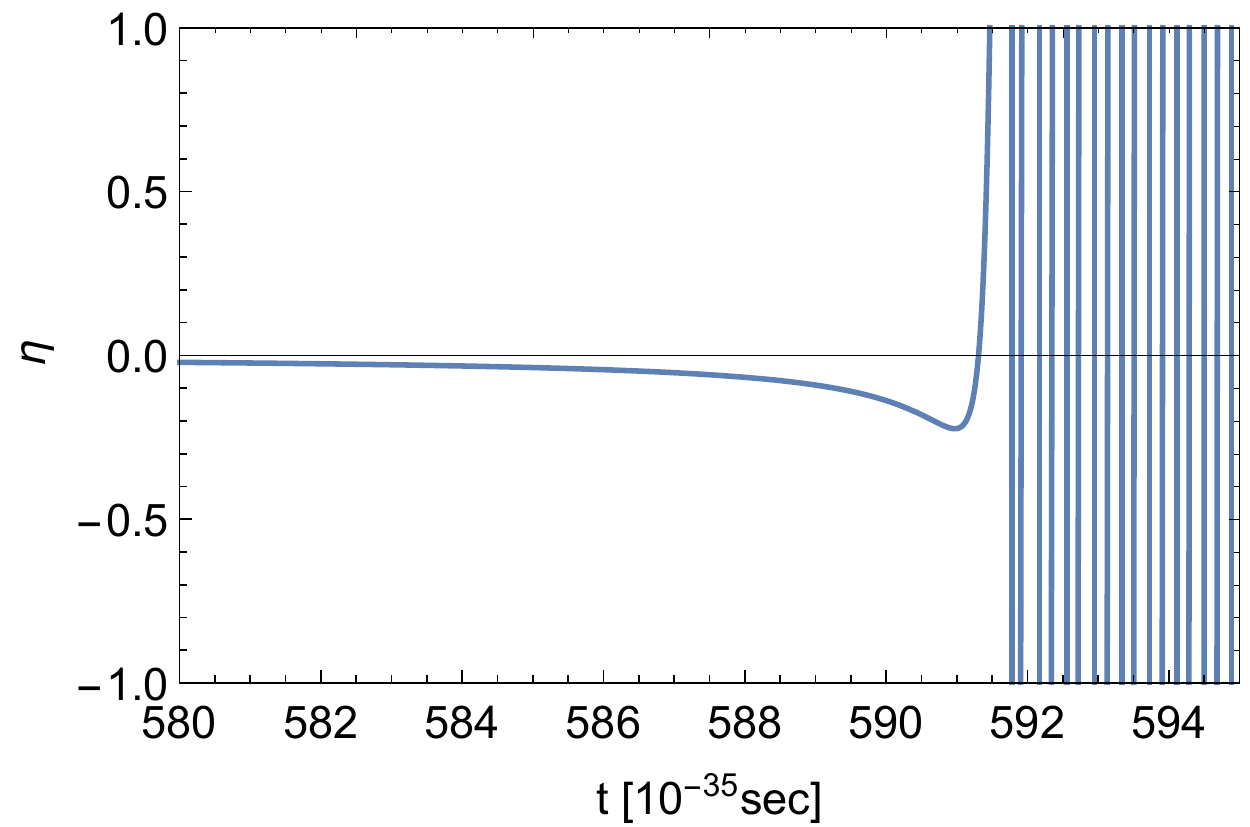}
\caption{Slow-roll parameters $\epsilon$ and $\eta$ before and around end of 
inflation. When $\epsilon = 1 $ the inflation ends.}
\label{fig3}
\end{figure}

%%%%%%%%%%%%%%%%%%%%%%%%
Fig.4 represents the plot of the evolution of the Hubble parameter $H(t)$ with a 
clear indication
of the two (quasi-)de Sitter epochs -- during early-times inflation with much higher
value of $H \simeq \sqrt{\frac{\L_0}{3}}$, and in late-times with much smaller value
of $H \simeq \sqrt{\frac{\L_{\rm DE}}{3}}$.

\begin{figure}[h]
\centering
\includegraphics[width=0.6\textwidth]{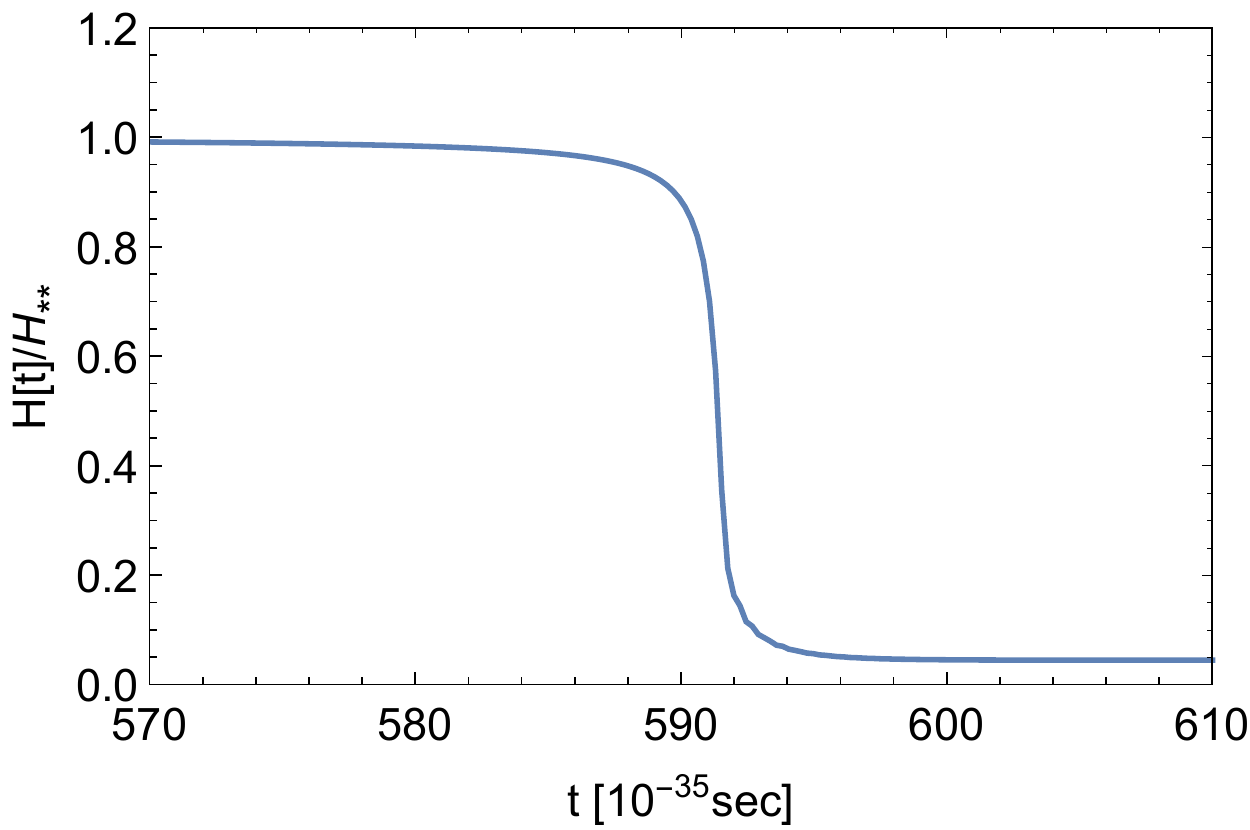}
\caption{Numerical shape of the evolution of $H(t)$. 
Here $H_{**} \equiv \sqrt{\L_0/3}$ as in \rf{unstable}.} 
\label{fig4}
\end{figure}
%%%%%%%%%%%%%%%%%%%%%%%

The plots on Fig.5 depict the oscillations of $u(t)$ and $\udot (t)$ 
%%%  REHEATING
occuring after the end of inflation.  %,\textsl{i.e.}, during reheating.

\begin{figure}[h]
\centering
\includegraphics[width=0.49\textwidth]{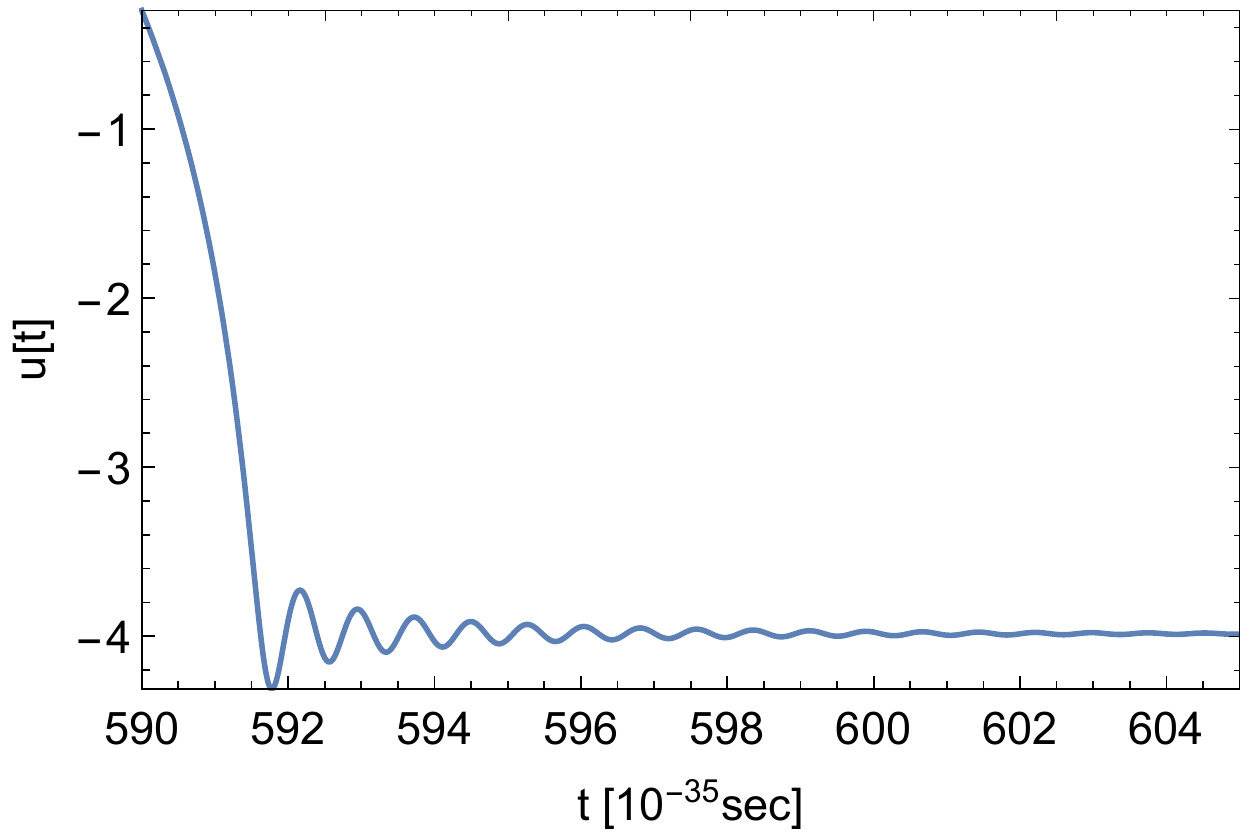}
\includegraphics[width=0.49\textwidth]{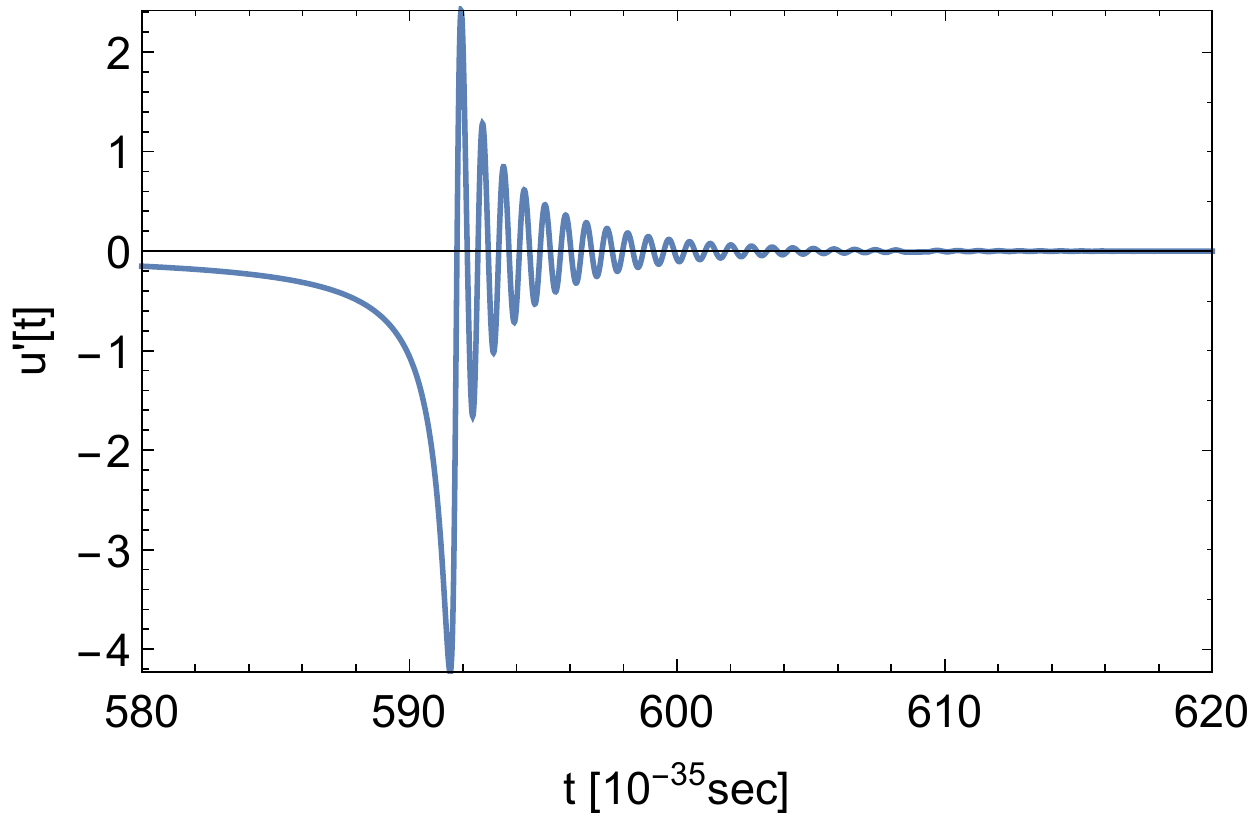}
\caption{On the left panel -- blown-up portion of the plot on Fig.2 around and 
after end of inflation depicting the oscillations of $u(t)$ after end of inflation.
%%%  REHEATING
% during reheating. 
On the right panel -- oscillations of $\udot(t)$ after end of inflation.}
%%%  REHEATING
% during reheating.}
\label{fig5}
\end{figure}

% A numerical illustration of the effect of the dust dark matter is provided on 
Fig.6 contains the plots of the evolution of $w=p/\rho$ -- 
the parameter of the equation of state with a clear indication of a 
%%%  REHEATING
short time epoch of matter domination after end of inflation. % during reheating.

%%%%%%%%%%
\begin{figure}[h]
\centering
\includegraphics[width=0.5\textwidth]{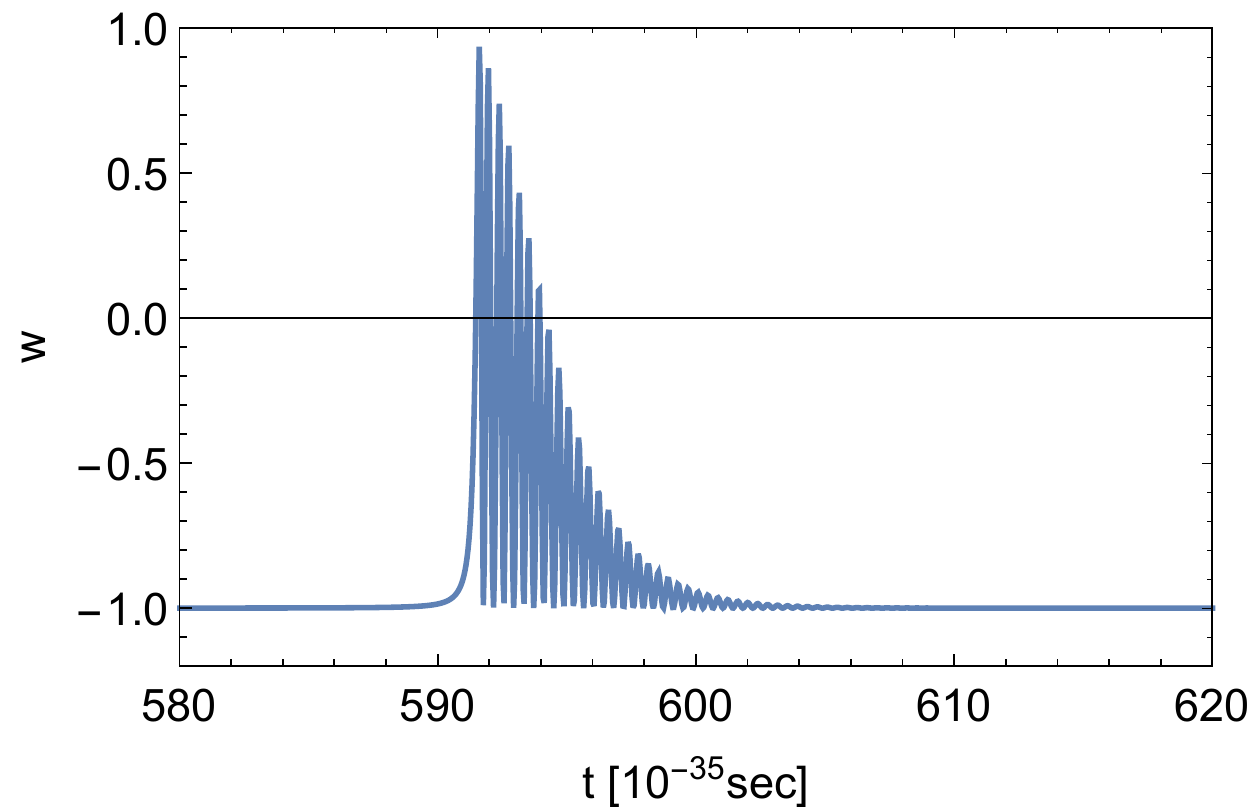}
\caption{Evolution of $w$ parameter of the equation of state with
%clearly indicating de Sitter behavior ($w \approx -1$) at early and late times.
sharp growth above $w \approx -1$ 
for a short time interval after end of inflation -- matter domination.}
%%%  REHEATING
% during reheating.}
\label{fig6}
\end{figure}
%%%%%%%%%%

To obtain plausible values for the observables -- the scalar power spectral index
$n_s$ and the tensor to scalar ratio $r$ 
\ct{Nojiri:2019kkp,Dalianis:2018frf,Martin:2013tda}. we need the functional
dependence of the slow-roll parameters $\epsilon$ and $\eta$ w.r.t. $\cN=\log(a)$ 
-- number of e-folds. 
More specifically, in $\cN_f$ is the number of e-folds at 
the end of inflation defined as $\epsilon (\cN_f) \approx 1$, then we need the 
values $\epsilon (\cN_i)$ and $\eta (\cN_i)$ at $\cN_i$ -- e-folds at the start 
of inflation, where it is assumed that $\cN_f - \cN_i \sim 60$. 
Then, according to \ct{Nojiri:2019kkp,Dalianis:2018frf}:
\br
r \approx 16 \epsilon (\cN_i) \quad ,\quad 
n_s \approx 1 - 6 \epsilon (\cN_i) + 2 \eta (\cN_i),
\lab{r-ns} 
\er
where the corresponding slow roll parameter read:
\br
\epsilon (\cN_i) = - \frac{H^\pr(\cN_i)}{H(\cN_i)}  \quad ,\quad
\eta(\cN_i) = - \frac{H^\pr(\cN_i)}{H(\cN_i)} 
- \frac{H^{\pr\pr}(\cN_i)}{2H(\cN_i) H^\pr(\cN_i)} \; ,
\lab{eps-eta-N}
\er
and where $H=H(\cN)$ is the functional dependence of Hubble parameter w.r.t.
the number of e-folds. To this end we employ numerical simulation of the 
autonomous dynamical system equations \rf{x-eq}-\rf{H-eq}.

From the inflationary scenario we know that the observed value of the 
inflationary scale $\L_0 \sim 10^{-8} M^4_{Pl}$ % cosmological constant 
is way larger than the current value ($\sim 10^{-122} M^4_{Pl}$)
of the cosmological constant $\L_{\rm DE}$ \rf{DE-value}. 
So, as in the numerical example above for the numerical solution of the system
for $u(t), H(t)$ \rf{u-eq-final}-\rf{H-eq-final}, we will take again the values
for the parameters according to \rf{num-example} meaning that we set
% $\L_0 = 50$, $M_1 = 20$, $\Lambda_{DE} = 0.1$, and set 
the initial condition for the Hubble parameter to be according to \rf{unstable}  
$H_{initial} = \sqrt{\frac{\L_0}{3}} = \sqrt{\frac{50}{3}}$. 
% with $\L_0 = 50$; $M_1 = 20$; $\Lambda_{DE} = 0.1$. 
With those numerical values we obtain for the observables \rf{r-ns} to be:
\be
r \approx 0.003683 \quad ,\quad n_s \approx 0.9638 \; , 
\lab{r-ns-1}
\ee
which are well inside the last \textsl{PLANCK} observed constraints 
\cite{Akrami:2018odb}:
\be
0.95 < n_s < 0.97 \quad, \quad r < 0.064 \; .
\lab{Planck-constr}
\ee

In order to see the pattern of the general behavior depending on the initial conditions,
we employ here Monte Carlo simulation with $10^4$ samples for the initial conditions using a normal distribution: $\Lambda_0 = 50 \pm 10$, $M_1 = 20 \pm 10$, 
while the error bar is taken to be $1 \, \sigma$.

Fig.7 shows how different values of initial 
conditions yield different number of $e$-folds until end of inflation
(where $\epsilon = 1$) and, accordingly, different values for the 
observables $r$ and $n_s$, whereas Fig.8 depicts the corresponding relation
between $r$ and $n_s$. Nevertheless, all the values of the latter fall within 
the constraint \rf{Planck-constr}.
%%%%%%%%%%
\begin{figure}[h]
\centering
\includegraphics[width=0.45\textwidth]{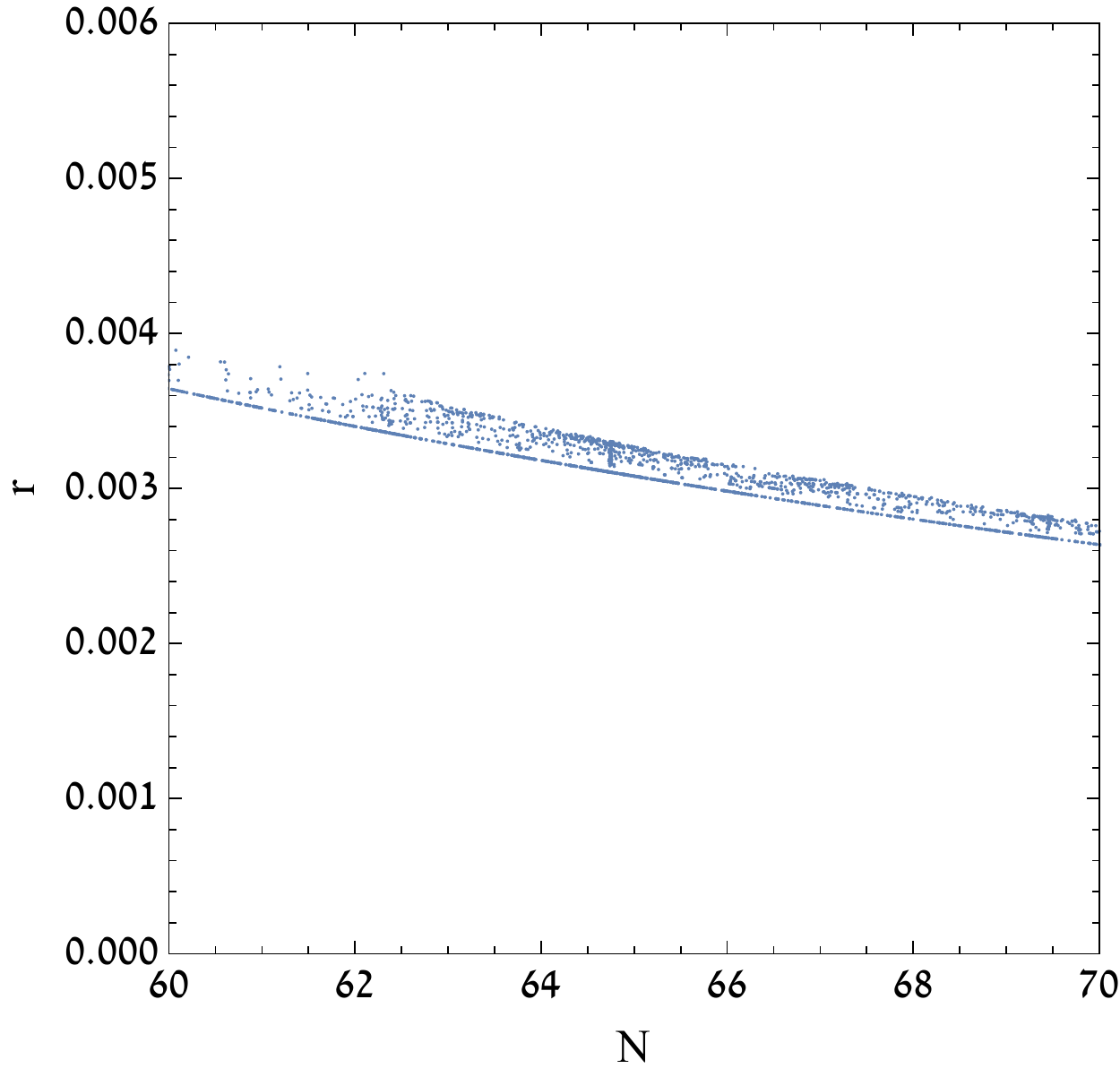}
\includegraphics[width=0.45\textwidth]{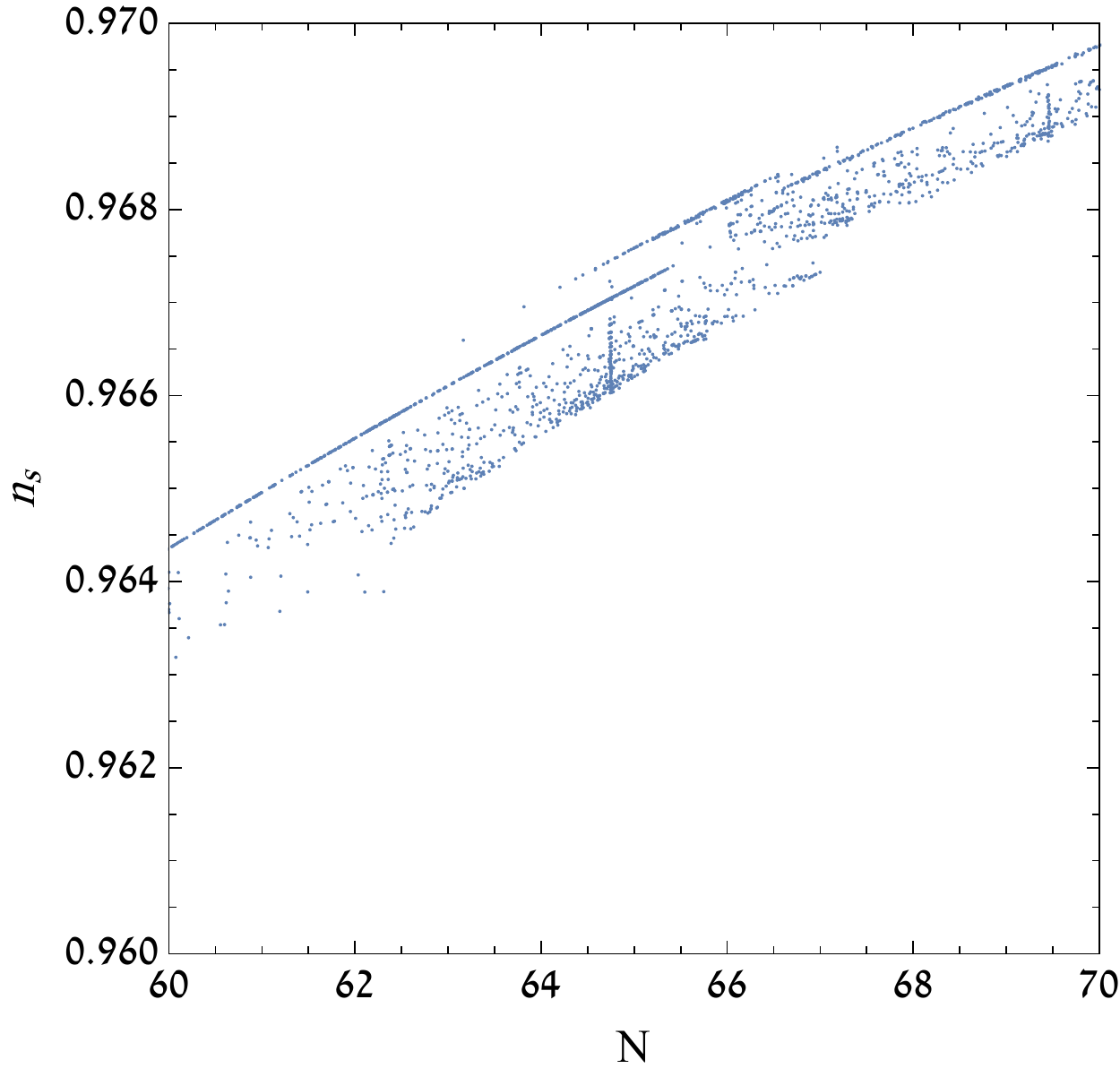}
\caption{The scalar to tensor ratio $r$ and the scalar spectral index $n_s$ vs. the 
number of $e$-folds for different values of the initial conditions. 
The sampling of the latter is done with a normal distribution 
$\Lambda_0 = 50 \pm 10$, $M_1 = 20 \pm 10$.}
\label{fig7}
\end{figure}
%%%%%%%%%%

%%%%%%%%%%
\begin{figure}[h]
\centering
\includegraphics[width=0.6\textwidth]{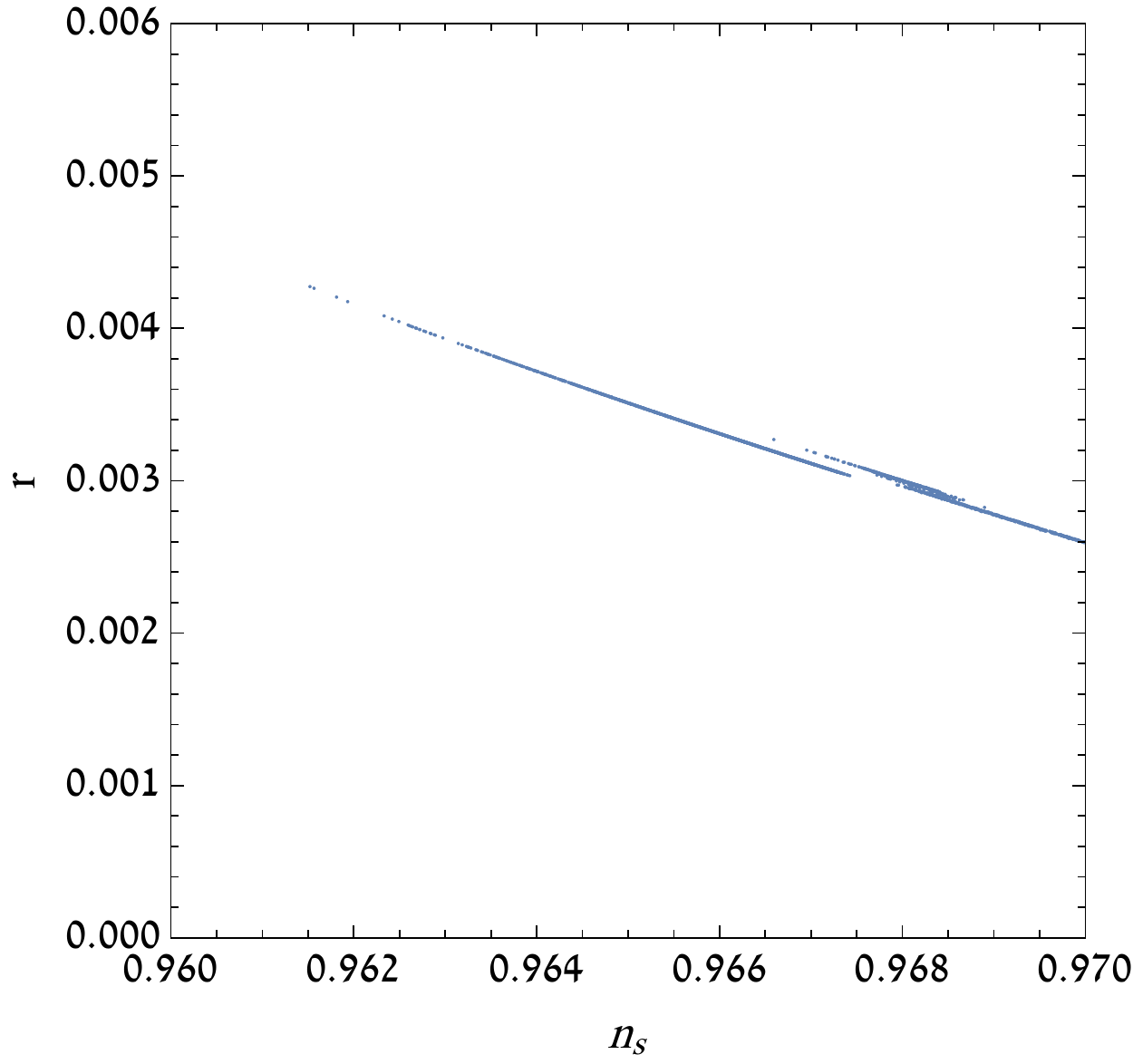}
\caption{The relation between the scalar to tensor ratio $r$ and the scalar spectral 
index $n_s$ via sampled initial conditions with a normal distribution 
$\Lambda_0 = 50 \pm 10$, $M_1 = 20 \pm 10$. All of the sampled values fall well 
inside the Planck data constraint \rf{Planck-constr}.}
\label{fig8}
\end{figure}
%%%%%%%%%x

%%%%%%%%%%%%%%%%%%%%%%%%%%%%%%%%%%%%%%%%%%%%%%%%%%%%%%%%%%%%%%%%%%%%%%%%%%%%%%%%%%%%
%%%%%%%%%%%%%%%%%%%%%%%%%%%%%%%%%%%%%%%%%%%%%%%%%%%%%%%%%%%%%%%%%%%%%%%%%%%%%%%%%%%%
\section{Conclusions and Outlook}
In the present paper, starting from the basic first
principle of Lagrangian field-theoretic actions combined with a non-canonical 
modification of gravity via employing non-Riemannian spacetime volume forms as
alternatives to the standard Riemannian one given by $\sqrt{-g}$, we have
constructed a unified model of dynamically generated inflation with dark energy
and dark matter coupled among themselves. Upon passage to the physical
Einstein frame our model captures the main properties of the slow-roll inflationary epoch
in early times, short period of matter domination after
end of inflation and late-time epoch of de Sitter expansion all driven by a
dynamically created scalar inflaton field. The numerical results for the observables
(scalar power spectral index and tensor to scalar ratio) conform to the 2018
\textsl{PLANCK} constraints.

In the present model dark matter in the form of a dust-like fluid is created 
already in the early universe inflationary epoch without a significant impact
on the inflationary dynamics.  After end of inflation the dust-like dark matter 
apart from a short period of matter domination still does not exert a sufficient 
impact, which means that one has to further extend the present formulation in order 
to take properly into account the full dark matter contribution to the evolution.  

One subject that has to be addressed is the ``reheating of the universe'', 
since of course we need temperature in the early universe to account for processes 
like Bing Bang Nucleosynthesis. There are many way to achieve this, due to the 
oscillating nature of inflaton solutions near the  minimum of the inflaton 
potential, which leads in general to particle creation.
%%%%%%%%%%%%%
For example, one possible way to complement the modified gravity-scalar field model
\rf{NRVF-1} in order to incorporate the effect of radiation after end of inflation 
is to include a coupling to the ``topological'' density of a electromagnetic
field $\cA_\m$ with field strength $F_{\m\n} = \pa_\m \cA_\n - \pa_\n \cA_\m$ 
in the following way:
\br
{\wti S} = \int d^4 x\,\Bigl\{\P_1(B) \Bigl\lb R(g)
- 2 \L_0 \frac{\P_1(B)}{\sqrt{-g}} \Bigr\rb
- \frac{\sqrt{-g}}{\P_1(B)} \vareps^{\m\n\k\l} F_{\m\n} F_{\k\l}
\nonu \\
+ \Bigl(\sqrt{-g}+\P_0 (A)\Bigr) \Bigl\lb -\frac{1}{2} g^{\m\n} \pa_\m\vp \pa_\n\vp
- V(\vp)\Bigr\rb\Bigr\} \; .
\lab{NRVF-2}
\er
Upon passage to the Einstein-frame via the conformal transformation \rf{g-bar}
the action \rf{NRVF-2} becomes (cf. \rf{EF-action}):
\br
{\wti S}_{\rm EF} = \int d^4 x \sqrt{-{\bar g}} \Bigl\lb {\bar R} 
-\frac{1}{2} {\bar g}^{\m\n}\pa_\m u\,\pa_\n u - U_{\rm eff}(u)
- e^{-u/\sqrt{3}}\vareps^{\m\n\k\l} F_{\m\n} F_{\k\l}\Bigr\rb 
\nonu \\
+ \int d^4 x \sqrt{-{\bar g}} \bigl(1+\chi_0\bigr) e^{-u/\sqrt{3}}
\Bigl\lb-\frac{1}{2} {\bar g}^{\m\n}\pa_\m \vp\,\pa_\n \vp 
- e^{-u/\sqrt{3}}\bigl(V(\vp)-2M_0\bigr)\Bigr\rb \; .
\lab{EF-action-FFdual}
\er
The coupling term $e^{-u/\sqrt{3}}\vareps^{\m\n\k\l} F_{\m\n} F_{\k\l}$ is
suppressed in the inflationary stage 
% where $u$ is large, whereas after end of inflation it may produce pairs of photons out of $u$. 
where the derivative of $u$ is small (because of the slow roll regime), whereas 
after end of inflation it may produce pairs of photons out of $u$ due to the 
appreciable time-derivative of $u$ resulting from the oscillations near the minimum of 
the effective potential. Of course,  many other possible interaction terms can be 
introduced. 

Finally, in the reheating stage  many particles can be produced, 
some of them could be no standard-model particles. If those are stable, they
could provide additional ``dark matter'' apart from the ``darkon'' dust-like
dark matter discussed here. Of course, if all created particles beyond those
of the standard models turn out to be unstable, then we will be left with the 
``darkon'' as the unique source of dark matter.

%%%%%%%%%%%%%%%%%%%%%%%%%%%%%%%%%%%%%%%%%%%%%%%%%%%%%%%%%%%%%%%%%%%%%%%%%%%%%%%%%%%%
\section{Acknowledgments}
We gratefully acknowledge support of our collaboration through the Exchange 
Agreement between Ben-Gurion University, Beer-Sheva, Israel and Bulgarian 
Academy of Sciences, Sofia, Bulgaria. E.N. and S.P. are thankful for support 
by Contract DN 18/1 from Bulgarian National Science Fund. D.B., E.G. and E.N. 
are also partially supported by COST Actions CA15117, CA16104 and CA18108. 
% Finally, we thank the referee for constructive remarks 
% contributing to improvement of the presentation.
\newpage
\externalbibliography{yes}
\bibliography{ref}

\begin{thebibliography}{-------}
\providecommand{\natexlab}[1]{#1}

\bibitem[Perlmutter \em{et~al.}(1999)Perlmutter et~al.]{Perlmutter:1998np}
Perlmutter, S.; others.
\newblock {Measurements of Omega and Lambda from 42 high redshift supernovae}.
\newblock {\em Astrophys. J.} {\bf 1999}, {\em 517},~565--586,
  \href{http://xxx.lanl.gov/abs/astro-ph/9812133}{{\normalfont
  [arXiv:astro-ph/astro-ph/9812133]}}.
\newblock
  doi:{\changeurlcolor{black}\href{https://doi.org/10.1086/307221}{\detokenize{10.1086/307221}}}.

\bibitem[Copeland \em{et~al.}(2006)Copeland, Sami, and
  Tsujikawa]{Copeland:2006wr}
Copeland, E.J.; Sami, M.; Tsujikawa, S.
\newblock {Dynamics of dark energy}.
\newblock {\em Int. J. Mod. Phys.} {\bf 2006}, {\em D15},~1753--1936,
  \href{http://xxx.lanl.gov/abs/hep-th/0603057}{{\normalfont
  [arXiv:hep-th/hep-th/0603057]}}.
\newblock
  doi:{\changeurlcolor{black}\href{https://doi.org/10.1142/S021827180600942X}{\detokenize{10.1142/S021827180600942X}}}.

\bibitem[Novikov(2016)]{Novikov:2016fzd}
Novikov, E.A.
\newblock {Quantum Modification of General Relativity}.
\newblock {\em Electron. J. Theor. Phys.} {\bf 2016}, {\em 13},~79--90.

\bibitem[Benitez \em{et~al.}(2020)Benitez, Gambini, Lehner, Liebling, and
  Pullin]{Benitez:2020szx}
Benitez, F.; Gambini, R.; Lehner, L.; Liebling, S.; Pullin, J.
\newblock {Critical collapse of a scalar field in semiclassical loop quantum
  gravity} {\bf 2020}.
\newblock  \href{http://xxx.lanl.gov/abs/2002.04044}{{\normalfont
  [arXiv:gr-qc/2002.04044]}}.

\bibitem[Budge \em{et~al.}(2020)Budge, Campbell, De~Laurentis, Keith~Ellis, and
  Seth]{Budge:2020oyl}
Budge, L.; Campbell, J.M.; De~Laurentis, G.; Keith~Ellis, R.; Seth, S.
\newblock {The one-loop amplitude for Higgs + 4 gluons with full mass effects}
  {\bf 2020}.
\newblock  \href{http://xxx.lanl.gov/abs/2002.04018}{{\normalfont
  [arXiv:hep-ph/2002.04018]}}.

\bibitem[Bell \em{et~al.}(2020)Bell, Beneke, Huber, and Li]{Bell:2020qus}
Bell, G.; Beneke, M.; Huber, T.; Li, X.Q.
\newblock {Two-loop non-leptonic penguin amplitude in QCD factorization} {\bf
  2020}.
\newblock  \href{http://xxx.lanl.gov/abs/2002.03262}{{\normalfont
  [arXiv:hep-ph/2002.03262]}}.

\bibitem[Fröhlich \em{et~al.}(2020)Fröhlich, Knowles, Schlein, and
  Sohinger]{Frohlich:2020igy}
Fröhlich, J.; Knowles, A.; Schlein, B.; Sohinger, V.
\newblock {A path-integral analysis of interacting Bose gases and loop gases}
  {\bf 2020}.
\newblock  \href{http://xxx.lanl.gov/abs/2001.11714}{{\normalfont
  [arXiv:math-ph/2001.11714]}}.

\bibitem[D'Ambrosio(2020)]{DAmbrosio:2020yfa}
D'Ambrosio, F.
\newblock {Semi-Classical Holomorphic Transition Amplitudes in Covariant Loop
  Quantum Gravity}.
\newblock PhD thesis, Marseille, CPT,  2020,
  \href{http://xxx.lanl.gov/abs/2001.04651}{{\normalfont
  [arXiv:gr-qc/2001.04651]}}.

\bibitem[Novikov(2016)]{Novikov:2016hrc}
Novikov, E.A.
\newblock {Ultralight gravitons with tiny electric dipole moment are seeping
  from the vacuum}.
\newblock {\em Mod. Phys. Lett.} {\bf 2016}, {\em A31},~1650092.
\newblock
  doi:{\changeurlcolor{black}\href{https://doi.org/10.1142/S0217732316500929}{\detokenize{10.1142/S0217732316500929}}}.

\bibitem[Dekens and Stoffer(2019)]{Dekens:2019ept}
Dekens, W.; Stoffer, P.
\newblock {Low-energy effective field theory below the electroweak scale:
  matching at one loop}.
\newblock {\em JHEP} {\bf 2019}, {\em 10},~197,
  \href{http://xxx.lanl.gov/abs/1908.05295}{{\normalfont
  [arXiv:hep-ph/1908.05295]}}.
\newblock
  doi:{\changeurlcolor{black}\href{https://doi.org/10.1007/JHEP10(2019)197}{\detokenize{10.1007/JHEP10(2019)197}}}.

\bibitem[Ma and Pezzella(2019)]{Ma:2019wbc}
Ma, C.T.; Pezzella, F.
\newblock {Stringy Effects at Low-Energy Limit and Double Field Theory} {\bf
  2019}.
\newblock  \href{http://xxx.lanl.gov/abs/1909.00411}{{\normalfont
  [arXiv:hep-th/1909.00411]}}.

\bibitem[Jenkins \em{et~al.}(2018)Jenkins, Manohar, and
  Stoffer]{Jenkins:2017jig}
Jenkins, E.E.; Manohar, A.V.; Stoffer, P.
\newblock {Low-Energy Effective Field Theory below the Electroweak Scale:
  Operators and Matching}.
\newblock {\em JHEP} {\bf 2018}, {\em 03},~016,
  \href{http://xxx.lanl.gov/abs/1709.04486}{{\normalfont
  [arXiv:hep-ph/1709.04486]}}.
\newblock
  doi:{\changeurlcolor{black}\href{https://doi.org/10.1007/JHEP03(2018)016}{\detokenize{10.1007/JHEP03(2018)016}}}.

\bibitem[Brandyshev(2017)]{Brandyshev:2017ywi}
Brandyshev, P.E.
\newblock {Cosmological solutions in low-energy effective field theory for type
  IIA superstrings}.
\newblock {\em Grav. Cosmol.} {\bf 2017}, {\em 23},~15--19.
\newblock
  doi:{\changeurlcolor{black}\href{https://doi.org/10.1134/S0202289317010029}{\detokenize{10.1134/S0202289317010029}}}.

\bibitem[Gomez and Jimenez(2020)]{Gomez:2020xdb}
Gomez, C.; Jimenez, R.
\newblock {Cosmology from Quantum Information} {\bf 2020}.
\newblock  \href{http://xxx.lanl.gov/abs/2002.04294}{{\normalfont
  [arXiv:hep-th/2002.04294]}}.

\bibitem[Guth(1981)]{Guth:1980zm}
Guth, A.H.
\newblock {The Inflationary Universe: A Possible Solution to the Horizon and
  Flatness Problems}.
\newblock {\em Phys. Rev.} {\bf 1981}, {\em D23},~347--356.
\newblock [Adv. Ser. Astrophys. Cosmol.3,139(1987)],
  doi:{\changeurlcolor{black}\href{https://doi.org/10.1103/PhysRevD.23.347}{\detokenize{10.1103/PhysRevD.23.347}}}.

\bibitem[Starobinsky(1979)]{Starobinsky:1979ty}
Starobinsky, A.A.
\newblock {Spectrum of relict gravitational radiation and the early state of
  the universe}.
\newblock {\em JETP Lett.} {\bf 1979}, {\em 30},~682--685.
\newblock [,767(1979)].

\bibitem[Kazanas(1980)]{Kazanas:1980tx}
Kazanas, D.
\newblock {Dynamics of the Universe and Spontaneous Symmetry Breaking}.
\newblock {\em Astrophys. J.} {\bf 1980}, {\em 241},~L59--L63.
\newblock
  doi:{\changeurlcolor{black}\href{https://doi.org/10.1086/183361}{\detokenize{10.1086/183361}}}.

\bibitem[Starobinsky(1980)]{Starobinsky:1980te}
Starobinsky, A.A.
\newblock {A New Type of Isotropic Cosmological Models Without Singularity}.
\newblock {\em Phys. Lett.} {\bf 1980}, {\em 91B},~99--102.
\newblock [,771(1980)],
  doi:{\changeurlcolor{black}\href{https://doi.org/10.1016/0370-2693(80)90670-X}{\detokenize{10.1016/0370-2693(80)90670-X}}}.

\bibitem[Linde(1982)]{Linde:1981mu}
Linde, A.D.
\newblock {A New Inflationary Universe Scenario: A Possible Solution of the
  Horizon, Flatness, Homogeneity, Isotropy and Primordial Monopole Problems}.
\newblock {\em Phys. Lett.} {\bf 1982}, {\em 108B},~389--393.
\newblock [Adv. Ser. Astrophys. Cosmol.3,149(1987)],
  doi:{\changeurlcolor{black}\href{https://doi.org/10.1016/0370-2693(82)91219-9}{\detokenize{10.1016/0370-2693(82)91219-9}}}.

\bibitem[Albrecht and Steinhardt(1982)]{Albrecht:1982wi}
Albrecht, A.; Steinhardt, P.J.
\newblock {Cosmology for Grand Unified Theories with Radiatively Induced
  Symmetry Breaking}.
\newblock {\em Phys. Rev. Lett.} {\bf 1982}, {\em 48},~1220--1223.
\newblock [Adv. Ser. Astrophys. Cosmol.3,158(1987)],
  doi:{\changeurlcolor{black}\href{https://doi.org/10.1103/PhysRevLett.48.1220}{\detokenize{10.1103/PhysRevLett.48.1220}}}.

\bibitem[Barrow and Ottewill(1983)]{Barrow:1983rx}
Barrow, J.D.; Ottewill, A.C.
\newblock {The Stability of General Relativistic Cosmological Theory}.
\newblock {\em J. Phys.} {\bf 1983}, {\em A16},~2757.
\newblock
  doi:{\changeurlcolor{black}\href{https://doi.org/10.1088/0305-4470/16/12/022}{\detokenize{10.1088/0305-4470/16/12/022}}}.

\bibitem[Blau \em{et~al.}(1987)Blau, Guendelman, and Guth]{Blau:1986cw}
Blau, S.K.; Guendelman, E.I.; Guth, A.H.
\newblock {The Dynamics of False Vacuum Bubbles}.
\newblock {\em Phys. Rev.} {\bf 1987}, {\em D35},~1747.
\newblock
  doi:{\changeurlcolor{black}\href{https://doi.org/10.1103/PhysRevD.35.1747}{\detokenize{10.1103/PhysRevD.35.1747}}}.

\bibitem[Cervantes-Cota and Dehnen(1995)]{CervantesCota:1995tz}
Cervantes-Cota, J.L.; Dehnen, H.
\newblock {Induced gravity inflation in the standard model of particle
  physics}.
\newblock {\em Nucl. Phys.} {\bf 1995}, {\em B442},~391--412,
  \href{http://xxx.lanl.gov/abs/astro-ph/9505069}{{\normalfont
  [arXiv:astro-ph/astro-ph/9505069]}}.
\newblock
  doi:{\changeurlcolor{black}\href{https://doi.org/10.1016/0550-3213(95)00128-X}{\detokenize{10.1016/0550-3213(95)00128-X}}}.

\bibitem[Berera(1995)]{Berera:1995ie}
Berera, A.
\newblock {Warm inflation}.
\newblock {\em Phys. Rev. Lett.} {\bf 1995}, {\em 75},~3218--3221,
  \href{http://xxx.lanl.gov/abs/astro-ph/9509049}{{\normalfont
  [arXiv:astro-ph/astro-ph/9509049]}}.
\newblock
  doi:{\changeurlcolor{black}\href{https://doi.org/10.1103/PhysRevLett.75.3218}{\detokenize{10.1103/PhysRevLett.75.3218}}}.

\bibitem[Armendariz-Picon \em{et~al.}(1999)Armendariz-Picon, Damour, and
  Mukhanov]{ArmendarizPicon:1999rj}
Armendariz-Picon, C.; Damour, T.; Mukhanov, V.F.
\newblock {k - inflation}.
\newblock {\em Phys. Lett.} {\bf 1999}, {\em B458},~209--218,
  \href{http://xxx.lanl.gov/abs/hep-th/9904075}{{\normalfont
  [arXiv:hep-th/hep-th/9904075]}}.
\newblock
  doi:{\changeurlcolor{black}\href{https://doi.org/10.1016/S0370-2693(99)00603-6}{\detokenize{10.1016/S0370-2693(99)00603-6}}}.

\bibitem[Kanti and Olive(1999)]{Kanti:1999ie}
Kanti, P.; Olive, K.A.
\newblock {Assisted chaotic inflation in higher dimensional theories}.
\newblock {\em Phys. Lett.} {\bf 1999}, {\em B464},~192--198,
  \href{http://xxx.lanl.gov/abs/hep-ph/9906331}{{\normalfont
  [arXiv:hep-ph/hep-ph/9906331]}}.
\newblock
  doi:{\changeurlcolor{black}\href{https://doi.org/10.1016/S0370-2693(99)00982-X}{\detokenize{10.1016/S0370-2693(99)00982-X}}}.

\bibitem[Garriga and Mukhanov(1999)]{Garriga:1999vw}
Garriga, J.; Mukhanov, V.F.
\newblock {Perturbations in k-inflation}.
\newblock {\em Phys. Lett.} {\bf 1999}, {\em B458},~219--225,
  \href{http://xxx.lanl.gov/abs/hep-th/9904176}{{\normalfont
  [arXiv:hep-th/hep-th/9904176]}}.
\newblock
  doi:{\changeurlcolor{black}\href{https://doi.org/10.1016/S0370-2693(99)00602-4}{\detokenize{10.1016/S0370-2693(99)00602-4}}}.

\bibitem[Gordon \em{et~al.}(2000)Gordon, Wands, Bassett, and
  Maartens]{Gordon:2000hv}
Gordon, C.; Wands, D.; Bassett, B.A.; Maartens, R.
\newblock {Adiabatic and entropy perturbations from inflation}.
\newblock {\em Phys. Rev.} {\bf 2000}, {\em D63},~023506,
  \href{http://xxx.lanl.gov/abs/astro-ph/0009131}{{\normalfont
  [arXiv:astro-ph/astro-ph/0009131]}}.
\newblock
  doi:{\changeurlcolor{black}\href{https://doi.org/10.1103/PhysRevD.63.023506}{\detokenize{10.1103/PhysRevD.63.023506}}}.

\bibitem[Bassett \em{et~al.}(2006)Bassett, Tsujikawa, and
  Wands]{Bassett:2005xm}
Bassett, B.A.; Tsujikawa, S.; Wands, D.
\newblock {Inflation dynamics and reheating}.
\newblock {\em Rev. Mod. Phys.} {\bf 2006}, {\em 78},~537--589,
  \href{http://xxx.lanl.gov/abs/astro-ph/0507632}{{\normalfont
  [arXiv:astro-ph/astro-ph/0507632]}}.
\newblock
  doi:{\changeurlcolor{black}\href{https://doi.org/10.1103/RevModPhys.78.537}{\detokenize{10.1103/RevModPhys.78.537}}}.

\bibitem[Chen and Wang(2010)]{Chen:2009zp}
Chen, X.; Wang, Y.
\newblock {Quasi-Single Field Inflation and Non-Gaussianities}.
\newblock {\em JCAP} {\bf 2010}, {\em 1004},~027,
  \href{http://xxx.lanl.gov/abs/0911.3380}{{\normalfont
  [arXiv:hep-th/0911.3380]}}.
\newblock
  doi:{\changeurlcolor{black}\href{https://doi.org/10.1088/1475-7516/2010/04/027}{\detokenize{10.1088/1475-7516/2010/04/027}}}.

\bibitem[Germani and Kehagias(2010)]{Germani:2010gm}
Germani, C.; Kehagias, A.
\newblock {New Model of Inflation with Non-minimal Derivative Coupling of
  Standard Model Higgs Boson to Gravity}.
\newblock {\em Phys. Rev. Lett.} {\bf 2010}, {\em 105},~011302,
  \href{http://xxx.lanl.gov/abs/1003.2635}{{\normalfont
  [arXiv:hep-ph/1003.2635]}}.
\newblock
  doi:{\changeurlcolor{black}\href{https://doi.org/10.1103/PhysRevLett.105.011302}{\detokenize{10.1103/PhysRevLett.105.011302}}}.

\bibitem[Kobayashi \em{et~al.}(2010)Kobayashi, Yamaguchi, and
  Yokoyama]{Kobayashi:2010cm}
Kobayashi, T.; Yamaguchi, M.; Yokoyama, J.
\newblock {G-inflation: Inflation driven by the Galileon field}.
\newblock {\em Phys. Rev. Lett.} {\bf 2010}, {\em 105},~231302,
  \href{http://xxx.lanl.gov/abs/1008.0603}{{\normalfont
  [arXiv:hep-th/1008.0603]}}.
\newblock
  doi:{\changeurlcolor{black}\href{https://doi.org/10.1103/PhysRevLett.105.231302}{\detokenize{10.1103/PhysRevLett.105.231302}}}.

\bibitem[Feng \em{et~al.}(2010)Feng, Li, and Saridakis]{Feng:2010ya}
Feng, C.J.; Li, X.Z.; Saridakis, E.N.
\newblock {Preventing eternality in phantom inflation}.
\newblock {\em Phys. Rev.} {\bf 2010}, {\em D82},~023526,
  \href{http://xxx.lanl.gov/abs/1004.1874}{{\normalfont
  [arXiv:astro-ph.CO/1004.1874]}}.
\newblock
  doi:{\changeurlcolor{black}\href{https://doi.org/10.1103/PhysRevD.82.023526}{\detokenize{10.1103/PhysRevD.82.023526}}}.

\bibitem[Burrage \em{et~al.}(2011)Burrage, de~Rham, Seery, and
  Tolley]{Burrage:2010cu}
Burrage, C.; de~Rham, C.; Seery, D.; Tolley, A.J.
\newblock {Galileon inflation}.
\newblock {\em JCAP} {\bf 2011}, {\em 1101},~014,
  \href{http://xxx.lanl.gov/abs/1009.2497}{{\normalfont
  [arXiv:hep-th/1009.2497]}}.
\newblock
  doi:{\changeurlcolor{black}\href{https://doi.org/10.1088/1475-7516/2011/01/014}{\detokenize{10.1088/1475-7516/2011/01/014}}}.

\bibitem[Kobayashi \em{et~al.}(2011)Kobayashi, Yamaguchi, and
  Yokoyama]{Kobayashi:2011nu}
Kobayashi, T.; Yamaguchi, M.; Yokoyama, J.
\newblock {Generalized G-inflation: Inflation with the most general
  second-order field equations}.
\newblock {\em Prog. Theor. Phys.} {\bf 2011}, {\em 126},~511--529,
  \href{http://xxx.lanl.gov/abs/1105.5723}{{\normalfont
  [arXiv:hep-th/1105.5723]}}.
\newblock
  doi:{\changeurlcolor{black}\href{https://doi.org/10.1143/PTP.126.511}{\detokenize{10.1143/PTP.126.511}}}.

\bibitem[Ohashi and Tsujikawa(2012)]{Ohashi:2012wf}
Ohashi, J.; Tsujikawa, S.
\newblock {Potential-driven Galileon inflation}.
\newblock {\em JCAP} {\bf 2012}, {\em 1210},~035,
  \href{http://xxx.lanl.gov/abs/1207.4879}{{\normalfont
  [arXiv:gr-qc/1207.4879]}}.
\newblock
  doi:{\changeurlcolor{black}\href{https://doi.org/10.1088/1475-7516/2012/10/035}{\detokenize{10.1088/1475-7516/2012/10/035}}}.

\bibitem[Paliathanasis and Tsamparlis(2014)]{Paliathanasis:2014yfa}
Paliathanasis, A.; Tsamparlis, M.
\newblock {Two scalar field cosmology: Conservation laws and exact solutions}.
\newblock {\em Phys. Rev.} {\bf 2014}, {\em D90},~043529,
  \href{http://xxx.lanl.gov/abs/1408.1798}{{\normalfont
  [arXiv:gr-qc/1408.1798]}}.
\newblock
  doi:{\changeurlcolor{black}\href{https://doi.org/10.1103/PhysRevD.90.043529}{\detokenize{10.1103/PhysRevD.90.043529}}}.

\bibitem[Dimakis and Paliathanasis(2020)]{Dimakis:2020tzc}
Dimakis, N.; Paliathanasis, A.
\newblock {Crossing the phantom divide line as an effect of quantum
  transitions} {\bf 2020}.
\newblock  \href{http://xxx.lanl.gov/abs/2001.09687}{{\normalfont
  [arXiv:gr-qc/2001.09687]}}.

\bibitem[Dimakis \em{et~al.}(2019)Dimakis, Paliathanasis, Terzis, and
  Christodoulakis]{Dimakis:2019qfs}
Dimakis, N.; Paliathanasis, A.; Terzis, P.A.; Christodoulakis, T.
\newblock {Cosmological Solutions in Multiscalar Field Theory}.
\newblock {\em Eur. Phys. J.} {\bf 2019}, {\em C79},~618,
  \href{http://xxx.lanl.gov/abs/1904.09713}{{\normalfont
  [arXiv:gr-qc/1904.09713]}}.
\newblock
  doi:{\changeurlcolor{black}\href{https://doi.org/10.1140/epjc/s10052-019-7130-8}{\detokenize{10.1140/epjc/s10052-019-7130-8}}}.

\bibitem[Benisty and Guendelman(2018)]{Benisty:2017lmt}
Benisty, D.; Guendelman, E.I.
\newblock {A transition between bouncing hyper-inflation to $\Lambda$CDM from
  diffusive scalar fields}.
\newblock {\em Int. J. Mod. Phys.} {\bf 2018}, {\em A33},~1850119,
  \href{http://xxx.lanl.gov/abs/1710.10588}{{\normalfont
  [arXiv:gr-qc/1710.10588]}}.
\newblock
  doi:{\changeurlcolor{black}\href{https://doi.org/10.1142/S0217751X18501191}{\detokenize{10.1142/S0217751X18501191}}}.

\bibitem[Barrow and Paliathanasis(2016)]{Barrow:2016qkh}
Barrow, J.D.; Paliathanasis, A.
\newblock {Observational Constraints on New Exact Inflationary Scalar-field
  Solutions}.
\newblock {\em Phys. Rev.} {\bf 2016}, {\em D94},~083518,
  \href{http://xxx.lanl.gov/abs/1609.01126}{{\normalfont
  [arXiv:gr-qc/1609.01126]}}.
\newblock
  doi:{\changeurlcolor{black}\href{https://doi.org/10.1103/PhysRevD.94.083518}{\detokenize{10.1103/PhysRevD.94.083518}}}.

\bibitem[Barrow and Paliathanasis(2018)]{Barrow:2016wiy}
Barrow, J.D.; Paliathanasis, A.
\newblock {Reconstructions of the dark-energy equation of state and the
  inflationary potential}.
\newblock {\em Gen. Rel. Grav.} {\bf 2018}, {\em 50},~82,
  \href{http://xxx.lanl.gov/abs/1611.06680}{{\normalfont
  [arXiv:gr-qc/1611.06680]}}.
\newblock
  doi:{\changeurlcolor{black}\href{https://doi.org/10.1007/s10714-018-2402-4}{\detokenize{10.1007/s10714-018-2402-4}}}.

\bibitem[Olive(1990)]{Olive:1989nu}
Olive, K.A.
\newblock {Inflation}.
\newblock {\em Phys. Rept.} {\bf 1990}, {\em 190},~307--403.
\newblock
  doi:{\changeurlcolor{black}\href{https://doi.org/10.1016/0370-1573(90)90144-Q}{\detokenize{10.1016/0370-1573(90)90144-Q}}}.

\bibitem[Linde(1994)]{Linde:1993cn}
Linde, A.D.
\newblock {Hybrid inflation}.
\newblock {\em Phys. Rev.} {\bf 1994}, {\em D49},~748--754,
  \href{http://xxx.lanl.gov/abs/astro-ph/9307002}{{\normalfont
  [arXiv:astro-ph/astro-ph/9307002]}}.
\newblock
  doi:{\changeurlcolor{black}\href{https://doi.org/10.1103/PhysRevD.49.748}{\detokenize{10.1103/PhysRevD.49.748}}}.

\bibitem[Liddle \em{et~al.}(1994)Liddle, Parsons, and Barrow]{Liddle:1994dx}
Liddle, A.R.; Parsons, P.; Barrow, J.D.
\newblock {Formalizing the slow roll approximation in inflation}.
\newblock {\em Phys. Rev.} {\bf 1994}, {\em D50},~7222--7232,
  \href{http://xxx.lanl.gov/abs/astro-ph/9408015}{{\normalfont
  [arXiv:astro-ph/astro-ph/9408015]}}.
\newblock
  doi:{\changeurlcolor{black}\href{https://doi.org/10.1103/PhysRevD.50.7222}{\detokenize{10.1103/PhysRevD.50.7222}}}.

\bibitem[Lidsey \em{et~al.}(1997)Lidsey, Liddle, Kolb, Copeland, Barreiro, and
  Abney]{Lidsey:1995np}
Lidsey, J.E.; Liddle, A.R.; Kolb, E.W.; Copeland, E.J.; Barreiro, T.; Abney, M.
\newblock {Reconstructing the inflation potential : An overview}.
\newblock {\em Rev. Mod. Phys.} {\bf 1997}, {\em 69},~373--410,
  \href{http://xxx.lanl.gov/abs/astro-ph/9508078}{{\normalfont
  [arXiv:astro-ph/astro-ph/9508078]}}.
\newblock
  doi:{\changeurlcolor{black}\href{https://doi.org/10.1103/RevModPhys.69.373}{\detokenize{10.1103/RevModPhys.69.373}}}.

\bibitem[Hossain \em{et~al.}(2014)Hossain, Myrzakulov, Sami, and
  Saridakis]{Hossain:2014xha}
Hossain, M.W.; Myrzakulov, R.; Sami, M.; Saridakis, E.N.
\newblock {Variable gravity: A suitable framework for quintessential
  inflation}.
\newblock {\em Phys. Rev.} {\bf 2014}, {\em D90},~023512,
  \href{http://xxx.lanl.gov/abs/1402.6661}{{\normalfont
  [arXiv:gr-qc/1402.6661]}}.
\newblock
  doi:{\changeurlcolor{black}\href{https://doi.org/10.1103/PhysRevD.90.023512}{\detokenize{10.1103/PhysRevD.90.023512}}}.

\bibitem[Wali~Hossain \em{et~al.}(2015)Wali~Hossain, Myrzakulov, Sami, and
  Saridakis]{Hossain:2014zma}
Wali~Hossain, M.; Myrzakulov, R.; Sami, M.; Saridakis, E.N.
\newblock {Unification of inflation and dark energy à la quintessential
  inflation}.
\newblock {\em Int. J. Mod. Phys.} {\bf 2015}, {\em D24},~1530014,
  \href{http://xxx.lanl.gov/abs/1410.6100}{{\normalfont
  [arXiv:gr-qc/1410.6100]}}.
\newblock
  doi:{\changeurlcolor{black}\href{https://doi.org/10.1142/S0218271815300141}{\detokenize{10.1142/S0218271815300141}}}.

\bibitem[Cai \em{et~al.}(2015)Cai, Gong, Pi, Saridakis, and Wu]{Cai:2014uka}
Cai, Y.F.; Gong, J.O.; Pi, S.; Saridakis, E.N.; Wu, S.Y.
\newblock {On the possibility of blue tensor spectrum within single field
  inflation}.
\newblock {\em Nucl. Phys.} {\bf 2015}, {\em B900},~517--532,
  \href{http://xxx.lanl.gov/abs/1412.7241}{{\normalfont
  [arXiv:hep-th/1412.7241]}}.
\newblock
  doi:{\changeurlcolor{black}\href{https://doi.org/10.1016/j.nuclphysb.2015.09.025}{\detokenize{10.1016/j.nuclphysb.2015.09.025}}}.

\bibitem[Geng \em{et~al.}(2015)Geng, Hossain, Myrzakulov, Sami, and
  Saridakis]{Geng:2015fla}
Geng, C.Q.; Hossain, M.W.; Myrzakulov, R.; Sami, M.; Saridakis, E.N.
\newblock {Quintessential inflation with canonical and noncanonical scalar
  fields and Planck 2015 results}.
\newblock {\em Phys. Rev.} {\bf 2015}, {\em D92},~023522,
  \href{http://xxx.lanl.gov/abs/1502.03597}{{\normalfont
  [arXiv:gr-qc/1502.03597]}}.
\newblock
  doi:{\changeurlcolor{black}\href{https://doi.org/10.1103/PhysRevD.92.023522}{\detokenize{10.1103/PhysRevD.92.023522}}}.

\bibitem[Kamali \em{et~al.}(2016)Kamali, Basilakos, and
  Mehrabi]{Kamali:2016frd}
Kamali, V.; Basilakos, S.; Mehrabi, A.
\newblock {Tachyon warm-intermediate inflation in the light of Planck data}.
\newblock {\em Eur. Phys. J.} {\bf 2016}, {\em C76},~525,
  \href{http://xxx.lanl.gov/abs/1604.05434}{{\normalfont
  [arXiv:gr-qc/1604.05434]}}.
\newblock
  doi:{\changeurlcolor{black}\href{https://doi.org/10.1140/epjc/s10052-016-4380-6}{\detokenize{10.1140/epjc/s10052-016-4380-6}}}.

\bibitem[Geng \em{et~al.}(2017)Geng, Lee, Sami, Saridakis, and
  Starobinsky]{Geng:2017mic}
Geng, C.Q.; Lee, C.C.; Sami, M.; Saridakis, E.N.; Starobinsky, A.A.
\newblock {Observational constraints on successful model of quintessential
  Inflation}.
\newblock {\em JCAP} {\bf 2017}, {\em 1706},~011,
  \href{http://xxx.lanl.gov/abs/1705.01329}{{\normalfont
  [arXiv:gr-qc/1705.01329]}}.
\newblock
  doi:{\changeurlcolor{black}\href{https://doi.org/10.1088/1475-7516/2017/06/011}{\detokenize{10.1088/1475-7516/2017/06/011}}}.

\bibitem[Dalianis \em{et~al.}(2019)Dalianis, Kehagias, and
  Tringas]{Dalianis:2018frf}
Dalianis, I.; Kehagias, A.; Tringas, G.
\newblock {Primordial black holes from $\alpha$-attractors}.
\newblock {\em JCAP} {\bf 2019}, {\em 1901},~037,
  \href{http://xxx.lanl.gov/abs/1805.09483}{{\normalfont
  [arXiv:astro-ph.CO/1805.09483]}}.
\newblock
  doi:{\changeurlcolor{black}\href{https://doi.org/10.1088/1475-7516/2019/01/037}{\detokenize{10.1088/1475-7516/2019/01/037}}}.

\bibitem[Dalianis and Tringas(2019)]{Dalianis:2019asr}
Dalianis, I.; Tringas, G.
\newblock {Primordial black hole remnants as dark matter produced in thermal,
  matter, and runaway-quintessence postinflationary scenarios}.
\newblock {\em Phys. Rev.} {\bf 2019}, {\em D100},~083512,
  \href{http://xxx.lanl.gov/abs/1905.01741}{{\normalfont
  [arXiv:astro-ph.CO/1905.01741]}}.
\newblock
  doi:{\changeurlcolor{black}\href{https://doi.org/10.1103/PhysRevD.100.083512}{\detokenize{10.1103/PhysRevD.100.083512}}}.

\bibitem[Benisty(2019)]{Benisty:2019pxb}
Benisty, D.
\newblock {Inflation from Fermions} {\bf 2019}.
\newblock  \href{http://xxx.lanl.gov/abs/1912.11124}{{\normalfont
  [arXiv:gr-qc/1912.11124]}}.

\bibitem[Benisty and Guendelman(2018)]{Benisty:2018gzx}
Benisty, D.; Guendelman, E.I.
\newblock {Inflation compactification from dynamical spacetime}.
\newblock {\em Phys. Rev.} {\bf 2018}, {\em D98},~043522,
  \href{http://xxx.lanl.gov/abs/1805.09314}{{\normalfont
  [arXiv:gr-qc/1805.09314]}}.
\newblock
  doi:{\changeurlcolor{black}\href{https://doi.org/10.1103/PhysRevD.98.043522}{\detokenize{10.1103/PhysRevD.98.043522}}}.

\bibitem[Benisty \em{et~al.}(2019)Benisty, Guendelman, and
  Saridakis]{Benisty:2019vej}
Benisty, D.; Guendelman, E.I.; Saridakis, E.N.
\newblock {The Scale Factor Potential Approach to Inflation} {\bf 2019}.
\newblock  \href{http://xxx.lanl.gov/abs/1909.01982}{{\normalfont
  [arXiv:gr-qc/1909.01982]}}.

\bibitem[Gerbino \em{et~al.}(2017)Gerbino, Freese, Vagnozzi, Lattanzi, Mena,
  Giusarma, and Ho]{Gerbino:2016sgw}
Gerbino, M.; Freese, K.; Vagnozzi, S.; Lattanzi, M.; Mena, O.; Giusarma, E.;
  Ho, S.
\newblock {Impact of neutrino properties on the estimation of inflationary
  parameters from current and future observations}.
\newblock {\em Phys. Rev.} {\bf 2017}, {\em D95},~043512,
  \href{http://xxx.lanl.gov/abs/1610.08830}{{\normalfont
  [arXiv:astro-ph.CO/1610.08830]}}.
\newblock
  doi:{\changeurlcolor{black}\href{https://doi.org/10.1103/PhysRevD.95.043512}{\detokenize{10.1103/PhysRevD.95.043512}}}.

\bibitem[Giovannini(2020)]{Giovannini:2020xeg}
Giovannini, M.
\newblock {Planckian hypersurfaces, inflation and bounces} {\bf 2020}.
\newblock  \href{http://xxx.lanl.gov/abs/2001.11799}{{\normalfont
  [arXiv:hep-th/2001.11799]}}.

\bibitem[Brahma \em{et~al.}(2020)Brahma, Brandenberger, and
  Yeom]{Brahma:2020cpy}
Brahma, S.; Brandenberger, R.; Yeom, D.h.
\newblock {Swampland, Trans-Planckian Censorship and Fine-Tuning Problem for
  Inflation: Tunnelling Wavefunction to the Rescue} {\bf 2020}.
\newblock  \href{http://xxx.lanl.gov/abs/2002.02941}{{\normalfont
  [arXiv:hep-th/2002.02941]}}.

\bibitem[Domcke \em{et~al.}(2020)Domcke, Guidetti, Welling, and
  Westphal]{Domcke:2020zez}
Domcke, V.; Guidetti, V.; Welling, Y.; Westphal, A.
\newblock {Resonant backreaction in axion inflation} {\bf 2020}.
\newblock  \href{http://xxx.lanl.gov/abs/2002.02952}{{\normalfont
  [arXiv:astro-ph.CO/2002.02952]}}.

\bibitem[Tenkanen and Tomberg(2020)]{Tenkanen:2020cvw}
Tenkanen, T.; Tomberg, E.
\newblock {Initial conditions for plateau inflation} {\bf 2020}.
\newblock  \href{http://xxx.lanl.gov/abs/2002.02420}{{\normalfont
  [arXiv:astro-ph.CO/2002.02420]}}.

\bibitem[Martin \em{et~al.}(2020)Martin, Papanikolaou, Pinol, and
  Vennin]{Martin:2020fgl}
Martin, J.; Papanikolaou, T.; Pinol, L.; Vennin, V.
\newblock {Metric preheating and radiative decay in single-field inflation}
  {\bf 2020}.
\newblock  \href{http://xxx.lanl.gov/abs/2002.01820}{{\normalfont
  [arXiv:astro-ph.CO/2002.01820]}}.

\bibitem[Cheon and Lee(2020)]{Cheon:2020vnj}
Cheon, K.; Lee, J.
\newblock {N=2 PNGB Quintessence Dark Energy} {\bf 2020}.
\newblock  \href{http://xxx.lanl.gov/abs/2002.01756}{{\normalfont
  [arXiv:gr-qc/2002.01756]}}.

\bibitem[Saleem and Zubair(2020)]{Saleem:2020dzo}
Saleem, R.; Zubair, M.
\newblock {Inflationary solution of Hamilton Jacobi equations during weak
  dissipative regime}.
\newblock {\em Phys. Scripta} {\bf 2020}, {\em 95},~035214.
\newblock
  doi:{\changeurlcolor{black}\href{https://doi.org/10.1088/1402-4896/ab4954}{\detokenize{10.1088/1402-4896/ab4954}}}.

\bibitem[Giacintucci \em{et~al.}(2020)Giacintucci, Markevitch,
  Johnston-Hollitt, Wik, Wang, and Clarke]{Giacintucci:2020glv}
Giacintucci, S.; Markevitch, M.; Johnston-Hollitt, M.; Wik, D.R.; Wang, Q.H.S.;
  Clarke, T.E.
\newblock {Discovery of a giant radio fossil in the Ophiuchus galaxy cluster}
  {\bf 2020}.
\newblock  \href{http://xxx.lanl.gov/abs/2002.01291}{{\normalfont
  [arXiv:astro-ph.GA/2002.01291]}}.

\bibitem[Aalsma and Shiu(2020)]{Aalsma:2020aib}
Aalsma, L.; Shiu, G.
\newblock {Chaos and complementarity in de Sitter space} {\bf 2020}.
\newblock  \href{http://xxx.lanl.gov/abs/2002.01326}{{\normalfont
  [arXiv:hep-th/2002.01326]}}.

\bibitem[Kogut and Fixsen(2020)]{Kogut:2020add}
Kogut, A.; Fixsen, D.J.
\newblock {Calibration Method and Uncertainty for the Primordial Inflation
  Explorer (PIXIE)} {\bf 2020}.
\newblock  \href{http://xxx.lanl.gov/abs/2002.00976}{{\normalfont
  [arXiv:astro-ph.IM/2002.00976]}}.

\bibitem[Arciniega \em{et~al.}(2020)Arciniega, Jaime, and
  Piccinelli]{Arciniega:2020pcy}
Arciniega, G.; Jaime, L.; Piccinelli, G.
\newblock {Inflationary predictions of Geometric Inflation} {\bf 2020}.
\newblock  \href{http://xxx.lanl.gov/abs/2001.11094}{{\normalfont
  [arXiv:gr-qc/2001.11094]}}.

\bibitem[Rasheed \em{et~al.}(2020)Rasheed, Golanbari, Sayar, Akhtari,
  Sheikhahmadi, Mohammadi, and Saaidi]{Rasheed:2020syk}
Rasheed, M.A.; Golanbari, T.; Sayar, K.; Akhtari, L.; Sheikhahmadi, H.;
  Mohammadi, A.; Saaidi, K.
\newblock {Warm Tachyon Inflation and Swampland Criteria} {\bf 2020}.
\newblock  \href{http://xxx.lanl.gov/abs/2001.10042}{{\normalfont
  [arXiv:gr-qc/2001.10042]}}.

\bibitem[Aldabergenov \em{et~al.}(2020)Aldabergenov, Aoki, and
  Ketov]{Aldabergenov:2020pry}
Aldabergenov, Y.; Aoki, S.; Ketov, S.V.
\newblock {Minimal Starobinsky supergravity coupled to dilaton-axion
  superfield} {\bf 2020}.
\newblock  \href{http://xxx.lanl.gov/abs/2001.09574}{{\normalfont
  [arXiv:hep-th/2001.09574]}}.

\bibitem[Tenkanen(2020)]{Tenkanen:2020dge}
Tenkanen, T.
\newblock {Tracing the high energy theory of gravity: an introduction to
  Palatini inflation} {\bf 2020}.
\newblock  \href{http://xxx.lanl.gov/abs/2001.10135}{{\normalfont
  [arXiv:astro-ph.CO/2001.10135]}}.

\bibitem[Shaposhnikov \em{et~al.}(2020)Shaposhnikov, Shkerin, and
  Zell]{Shaposhnikov:2020geh}
Shaposhnikov, M.; Shkerin, A.; Zell, S.
\newblock {Standard Model Meets Gravity: Electroweak Symmetry Breaking and
  Inflation} {\bf 2020}.
\newblock  \href{http://xxx.lanl.gov/abs/2001.09088}{{\normalfont
  [arXiv:hep-th/2001.09088]}}.

\bibitem[Garcia \em{et~al.}(2020)Garcia, Amin, and Green]{Garcia:2020mwi}
Garcia, M.A.G.; Amin, M.A.; Green, D.
\newblock {Curvature Perturbations From Stochastic Particle Production During
  Inflation} {\bf 2020}.
\newblock  \href{http://xxx.lanl.gov/abs/2001.09158}{{\normalfont
  [arXiv:astro-ph.CO/2001.09158]}}.

\bibitem[Hirano(2019)]{Hirano:2019iie}
Hirano, K.
\newblock {Inflation with very small tensor-to-scalar ratio} {\bf 2019}.
\newblock  \href{http://xxx.lanl.gov/abs/1912.12515}{{\normalfont
  [arXiv:astro-ph.CO/1912.12515]}}.

\bibitem[Gialamas and Lahanas(2019)]{Gialamas:2019nly}
Gialamas, I.D.; Lahanas, A.B.
\newblock {Reheating in $R^2$ Palatini inflationary models} {\bf 2019}.
\newblock  \href{http://xxx.lanl.gov/abs/1911.11513}{{\normalfont
  [arXiv:gr-qc/1911.11513]}}.

\bibitem[Kawasaki \em{et~al.}(2000)Kawasaki, Yamaguchi, and
  Yanagida]{Kawasaki:2000yn}
Kawasaki, M.; Yamaguchi, M.; Yanagida, T.
\newblock {Natural chaotic inflation in supergravity}.
\newblock {\em Phys. Rev. Lett.} {\bf 2000}, {\em 85},~3572--3575,
  \href{http://xxx.lanl.gov/abs/hep-ph/0004243}{{\normalfont
  [arXiv:hep-ph/hep-ph/0004243]}}.
\newblock
  doi:{\changeurlcolor{black}\href{https://doi.org/10.1103/PhysRevLett.85.3572}{\detokenize{10.1103/PhysRevLett.85.3572}}}.

\bibitem[Bojowald(2002)]{Bojowald:2002nz}
Bojowald, M.
\newblock {Inflation from quantum geometry}.
\newblock {\em Phys. Rev. Lett.} {\bf 2002}, {\em 89},~261301,
  \href{http://xxx.lanl.gov/abs/gr-qc/0206054}{{\normalfont
  [arXiv:gr-qc/gr-qc/0206054]}}.
\newblock
  doi:{\changeurlcolor{black}\href{https://doi.org/10.1103/PhysRevLett.89.261301}{\detokenize{10.1103/PhysRevLett.89.261301}}}.

\bibitem[Nojiri and Odintsov(2003)]{Nojiri:2003ft}
Nojiri, S.; Odintsov, S.D.
\newblock {Modified gravity with negative and positive powers of the curvature:
  Unification of the inflation and of the cosmic acceleration}.
\newblock {\em Phys. Rev.} {\bf 2003}, {\em D68},~123512,
  \href{http://xxx.lanl.gov/abs/hep-th/0307288}{{\normalfont
  [arXiv:hep-th/hep-th/0307288]}}.
\newblock
  doi:{\changeurlcolor{black}\href{https://doi.org/10.1103/PhysRevD.68.123512}{\detokenize{10.1103/PhysRevD.68.123512}}}.

\bibitem[Kachru \em{et~al.}(2003)Kachru, Kallosh, Linde, Maldacena, McAllister,
  and Trivedi]{Kachru:2003sx}
Kachru, S.; Kallosh, R.; Linde, A.D.; Maldacena, J.M.; McAllister, L.P.;
  Trivedi, S.P.
\newblock {Towards inflation in string theory}.
\newblock {\em JCAP} {\bf 2003}, {\em 0310},~013,
  \href{http://xxx.lanl.gov/abs/hep-th/0308055}{{\normalfont
  [arXiv:hep-th/hep-th/0308055]}}.
\newblock
  doi:{\changeurlcolor{black}\href{https://doi.org/10.1088/1475-7516/2003/10/013}{\detokenize{10.1088/1475-7516/2003/10/013}}}.

\bibitem[Nojiri and Odintsov(2006)]{Nojiri:2005pu}
Nojiri, S.; Odintsov, S.D.
\newblock {Unifying phantom inflation with late-time acceleration: Scalar
  phantom-non-phantom transition model and generalized holographic dark
  energy}.
\newblock {\em Gen. Rel. Grav.} {\bf 2006}, {\em 38},~1285--1304,
  \href{http://xxx.lanl.gov/abs/hep-th/0506212}{{\normalfont
  [arXiv:hep-th/hep-th/0506212]}}.
\newblock
  doi:{\changeurlcolor{black}\href{https://doi.org/10.1007/s10714-006-0301-6}{\detokenize{10.1007/s10714-006-0301-6}}}.

\bibitem[Ferraro and Fiorini(2007)]{Ferraro:2006jd}
Ferraro, R.; Fiorini, F.
\newblock {Modified teleparallel gravity: Inflation without inflaton}.
\newblock {\em Phys. Rev.} {\bf 2007}, {\em D75},~084031,
  \href{http://xxx.lanl.gov/abs/gr-qc/0610067}{{\normalfont
  [arXiv:gr-qc/gr-qc/0610067]}}.
\newblock
  doi:{\changeurlcolor{black}\href{https://doi.org/10.1103/PhysRevD.75.084031}{\detokenize{10.1103/PhysRevD.75.084031}}}.

\bibitem[Cognola \em{et~al.}(2008)Cognola, Elizalde, Nojiri, Odintsov,
  Sebastiani, and Zerbini]{Cognola:2007zu}
Cognola, G.; Elizalde, E.; Nojiri, S.; Odintsov, S.D.; Sebastiani, L.; Zerbini,
  S.
\newblock {A Class of viable modified f(R) gravities describing inflation and
  the onset of accelerated expansion}.
\newblock {\em Phys. Rev.} {\bf 2008}, {\em D77},~046009,
  \href{http://xxx.lanl.gov/abs/0712.4017}{{\normalfont
  [arXiv:hep-th/0712.4017]}}.
\newblock
  doi:{\changeurlcolor{black}\href{https://doi.org/10.1103/PhysRevD.77.046009}{\detokenize{10.1103/PhysRevD.77.046009}}}.

\bibitem[Cai and Saridakis(2011)]{Cai:2010kp}
Cai, Y.F.; Saridakis, E.N.
\newblock {Inflation in Entropic Cosmology: Primordial Perturbations and
  non-Gaussianities}.
\newblock {\em Phys. Lett.} {\bf 2011}, {\em B697},~280--287,
  \href{http://xxx.lanl.gov/abs/1011.1245}{{\normalfont
  [arXiv:hep-th/1011.1245]}}.
\newblock
  doi:{\changeurlcolor{black}\href{https://doi.org/10.1016/j.physletb.2011.02.020}{\detokenize{10.1016/j.physletb.2011.02.020}}}.

\bibitem[Ashtekar and Sloan(2011)]{Ashtekar:2011rm}
Ashtekar, A.; Sloan, D.
\newblock {Probability of Inflation in Loop Quantum Cosmology}.
\newblock {\em Gen. Rel. Grav.} {\bf 2011}, {\em 43},~3619--3655,
  \href{http://xxx.lanl.gov/abs/1103.2475}{{\normalfont
  [arXiv:gr-qc/1103.2475]}}.
\newblock
  doi:{\changeurlcolor{black}\href{https://doi.org/10.1007/s10714-011-1246-y}{\detokenize{10.1007/s10714-011-1246-y}}}.

\bibitem[Qiu and Saridakis(2012)]{Qiu:2011zr}
Qiu, T.; Saridakis, E.N.
\newblock {Entropic Force Scenarios and Eternal Inflation}.
\newblock {\em Phys. Rev.} {\bf 2012}, {\em D85},~043504,
  \href{http://xxx.lanl.gov/abs/1107.1013}{{\normalfont
  [arXiv:hep-th/1107.1013]}}.
\newblock
  doi:{\changeurlcolor{black}\href{https://doi.org/10.1103/PhysRevD.85.043504}{\detokenize{10.1103/PhysRevD.85.043504}}}.

\bibitem[Briscese \em{et~al.}(2013)Briscese, Marcianò, Modesto, and
  Saridakis]{Briscese:2012ys}
Briscese, F.; Marcianò, A.; Modesto, L.; Saridakis, E.N.
\newblock {Inflation in (Super-)renormalizable Gravity}.
\newblock {\em Phys. Rev.} {\bf 2013}, {\em D87},~083507,
  \href{http://xxx.lanl.gov/abs/1212.3611}{{\normalfont
  [arXiv:hep-th/1212.3611]}}.
\newblock
  doi:{\changeurlcolor{black}\href{https://doi.org/10.1103/PhysRevD.87.083507}{\detokenize{10.1103/PhysRevD.87.083507}}}.

\bibitem[Ellis \em{et~al.}(2013)Ellis, Nanopoulos, and Olive]{Ellis:2013xoa}
Ellis, J.; Nanopoulos, D.V.; Olive, K.A.
\newblock {No-Scale Supergravity Realization of the Starobinsky Model of
  Inflation}.
\newblock {\em Phys. Rev. Lett.} {\bf 2013}, {\em 111},~111301,
  \href{http://xxx.lanl.gov/abs/1305.1247}{{\normalfont
  [arXiv:hep-th/1305.1247]}}.
\newblock [Erratum: Phys. Rev. Lett.111,no.12,129902(2013)],
  doi:{\changeurlcolor{black}\href{https://doi.org/10.1103/PhysRevLett.111.129902,
  10.1103/PhysRevLett.111.111301}{\detokenize{10.1103/PhysRevLett.111.129902,
  10.1103/PhysRevLett.111.111301}}}.

\bibitem[Basilakos \em{et~al.}(2013)Basilakos, Lima, and
  Sola]{Basilakos:2013xpa}
Basilakos, S.; Lima, J.A.S.; Sola, J.
\newblock {From inflation to dark energy through a dynamical Lambda: an attempt
  at alleviating fundamental cosmic puzzles}.
\newblock {\em Int. J. Mod. Phys.} {\bf 2013}, {\em D22},~1342008,
  \href{http://xxx.lanl.gov/abs/1307.6251}{{\normalfont
  [arXiv:astro-ph.CO/1307.6251]}}.
\newblock
  doi:{\changeurlcolor{black}\href{https://doi.org/10.1142/S021827181342008X}{\detokenize{10.1142/S021827181342008X}}}.

\bibitem[Sebastiani \em{et~al.}(2014)Sebastiani, Cognola, Myrzakulov, Odintsov,
  and Zerbini]{Sebastiani:2013eqa}
Sebastiani, L.; Cognola, G.; Myrzakulov, R.; Odintsov, S.D.; Zerbini, S.
\newblock {Nearly Starobinsky inflation from modified gravity}.
\newblock {\em Phys. Rev.} {\bf 2014}, {\em D89},~023518,
  \href{http://xxx.lanl.gov/abs/1311.0744}{{\normalfont
  [arXiv:gr-qc/1311.0744]}}.
\newblock
  doi:{\changeurlcolor{black}\href{https://doi.org/10.1103/PhysRevD.89.023518}{\detokenize{10.1103/PhysRevD.89.023518}}}.

\bibitem[Baumann and McAllister(2015)]{Baumann:2014nda}
Baumann, D.; McAllister, L.
\newblock {\em {Inflation and String Theory}}; Cambridge Monographs on
  Mathematical Physics, Cambridge University Press,  2015;
  \href{http://xxx.lanl.gov/abs/1404.2601}{{\normalfont
  [arXiv:hep-th/1404.2601]}}.
\newblock
  doi:{\changeurlcolor{black}\href{https://doi.org/10.1017/CBO9781316105733}{\detokenize{10.1017/CBO9781316105733}}}.

\bibitem[Dalianis and Farakos(2015)]{Dalianis:2015fpa}
Dalianis, I.; Farakos, F.
\newblock {On the initial conditions for inflation with plateau potentials: the
  $R+R^2$ (super)gravity case}.
\newblock {\em JCAP} {\bf 2015}, {\em 1507},~044,
  \href{http://xxx.lanl.gov/abs/1502.01246}{{\normalfont
  [arXiv:gr-qc/1502.01246]}}.
\newblock
  doi:{\changeurlcolor{black}\href{https://doi.org/10.1088/1475-7516/2015/07/044}{\detokenize{10.1088/1475-7516/2015/07/044}}}.

\bibitem[Kanti \em{et~al.}(2015)Kanti, Gannouji, and Dadhich]{Kanti:2015pda}
Kanti, P.; Gannouji, R.; Dadhich, N.
\newblock {Gauss-Bonnet Inflation}.
\newblock {\em Phys. Rev.} {\bf 2015}, {\em D92},~041302,
  \href{http://xxx.lanl.gov/abs/1503.01579}{{\normalfont
  [arXiv:hep-th/1503.01579]}}.
\newblock
  doi:{\changeurlcolor{black}\href{https://doi.org/10.1103/PhysRevD.92.041302}{\detokenize{10.1103/PhysRevD.92.041302}}}.

\bibitem[De~Laurentis \em{et~al.}(2015)De~Laurentis, Paolella, and
  Capozziello]{DeLaurentis:2015fea}
De~Laurentis, M.; Paolella, M.; Capozziello, S.
\newblock {Cosmological inflation in $F(R,\mathcal{G})$ gravity}.
\newblock {\em Phys. Rev.} {\bf 2015}, {\em D91},~083531,
  \href{http://xxx.lanl.gov/abs/1503.04659}{{\normalfont
  [arXiv:gr-qc/1503.04659]}}.
\newblock
  doi:{\changeurlcolor{black}\href{https://doi.org/10.1103/PhysRevD.91.083531}{\detokenize{10.1103/PhysRevD.91.083531}}}.

\bibitem[Basilakos \em{et~al.}(2016)Basilakos, Mavromatos, and
  Solà]{Basilakos:2015yoa}
Basilakos, S.; Mavromatos, N.E.; Solà, J.
\newblock {Starobinsky-like inflation and running vacuum in the context of
  Supergravity}.
\newblock {\em Universe} {\bf 2016}, {\em 2},~14,
  \href{http://xxx.lanl.gov/abs/1505.04434}{{\normalfont
  [arXiv:gr-qc/1505.04434]}}.
\newblock
  doi:{\changeurlcolor{black}\href{https://doi.org/10.3390/universe2030014}{\detokenize{10.3390/universe2030014}}}.

\bibitem[Bonanno and Platania(2015)]{Bonanno:2015fga}
Bonanno, A.; Platania, A.
\newblock {Asymptotically safe inflation from quadratic gravity}.
\newblock {\em Phys. Lett.} {\bf 2015}, {\em B750},~638--642,
  \href{http://xxx.lanl.gov/abs/1507.03375}{{\normalfont
  [arXiv:gr-qc/1507.03375]}}.
\newblock
  doi:{\changeurlcolor{black}\href{https://doi.org/10.1016/j.physletb.2015.10.005}{\detokenize{10.1016/j.physletb.2015.10.005}}}.

\bibitem[Koshelev \em{et~al.}(2016)Koshelev, Modesto, Rachwal, and
  Starobinsky]{Koshelev:2016xqb}
Koshelev, A.S.; Modesto, L.; Rachwal, L.; Starobinsky, A.A.
\newblock {Occurrence of exact $R^2$ inflation in non-local UV-complete
  gravity}.
\newblock {\em JHEP} {\bf 2016}, {\em 11},~067,
  \href{http://xxx.lanl.gov/abs/1604.03127}{{\normalfont
  [arXiv:hep-th/1604.03127]}}.
\newblock
  doi:{\changeurlcolor{black}\href{https://doi.org/10.1007/JHEP11(2016)067}{\detokenize{10.1007/JHEP11(2016)067}}}.

\bibitem[Bamba \em{et~al.}(2017)Bamba, Odintsov, and Saridakis]{Bamba:2016wjm}
Bamba, K.; Odintsov, S.D.; Saridakis, E.N.
\newblock {Inflationary cosmology in unimodular $F(T)$ gravity}.
\newblock {\em Mod. Phys. Lett.} {\bf 2017}, {\em A32},~1750114,
  \href{http://xxx.lanl.gov/abs/1605.02461}{{\normalfont
  [arXiv:gr-qc/1605.02461]}}.
\newblock
  doi:{\changeurlcolor{black}\href{https://doi.org/10.1142/S0217732317501140}{\detokenize{10.1142/S0217732317501140}}}.

\bibitem[Motohashi and Starobinsky(2017)]{Motohashi:2017vdc}
Motohashi, H.; Starobinsky, A.A.
\newblock {$f(R)$ constant-roll inflation}.
\newblock {\em Eur. Phys. J.} {\bf 2017}, {\em C77},~538,
  \href{http://xxx.lanl.gov/abs/1704.08188}{{\normalfont
  [arXiv:astro-ph.CO/1704.08188]}}.
\newblock
  doi:{\changeurlcolor{black}\href{https://doi.org/10.1140/epjc/s10052-017-5109-x}{\detokenize{10.1140/epjc/s10052-017-5109-x}}}.

\bibitem[Oikonomou(2018)]{Oikonomou:2017ppp}
Oikonomou, V.K.
\newblock {Autonomous dynamical system approach for inflationary Gauss–Bonnet
  modified gravity}.
\newblock {\em Int. J. Mod. Phys.} {\bf 2018}, {\em D27},~1850059,
  \href{http://xxx.lanl.gov/abs/1711.03389}{{\normalfont
  [arXiv:gr-qc/1711.03389]}}.
\newblock
  doi:{\changeurlcolor{black}\href{https://doi.org/10.1142/S0218271818500591}{\detokenize{10.1142/S0218271818500591}}}.

\bibitem[Benisty \em{et~al.}(2019)Benisty, Vasak, Guendelman, and
  Struckmeier]{Benisty:2018ywz}
Benisty, D.; Vasak, D.; Guendelman, E.; Struckmeier, J.
\newblock {Energy transfer from spacetime into matter and a bouncing inflation
  from covariant canonical gauge theory of gravity}.
\newblock {\em Mod. Phys. Lett.} {\bf 2019}, {\em A34},~1950164,
  \href{http://xxx.lanl.gov/abs/1807.03557}{{\normalfont
  [arXiv:gr-qc/1807.03557]}}.
\newblock
  doi:{\changeurlcolor{black}\href{https://doi.org/10.1142/S0217732319501645}{\detokenize{10.1142/S0217732319501645}}}.

\bibitem[Benisty and Guendelman(2019)]{Benisty:2018fja}
Benisty, D.; Guendelman, E.I.
\newblock {Two scalar fields inflation from scale-invariant gravity with
  modified measure}.
\newblock {\em Class. Quant. Grav.} {\bf 2019}, {\em 36},~095001,
  \href{http://xxx.lanl.gov/abs/1809.09866}{{\normalfont
  [arXiv:gr-qc/1809.09866]}}.
\newblock
  doi:{\changeurlcolor{black}\href{https://doi.org/10.1088/1361-6382/ab14af}{\detokenize{10.1088/1361-6382/ab14af}}}.

\bibitem[Antoniadis \em{et~al.}(2018)Antoniadis, Karam, Lykkas, and
  Tamvakis]{Antoniadis:2018ywb}
Antoniadis, I.; Karam, A.; Lykkas, A.; Tamvakis, K.
\newblock {Palatini inflation in models with an $R^2$ term}.
\newblock {\em JCAP} {\bf 2018}, {\em 1811},~028,
  \href{http://xxx.lanl.gov/abs/1810.10418}{{\normalfont
  [arXiv:gr-qc/1810.10418]}}.
\newblock
  doi:{\changeurlcolor{black}\href{https://doi.org/10.1088/1475-7516/2018/11/028}{\detokenize{10.1088/1475-7516/2018/11/028}}}.

\bibitem[Karam \em{et~al.}(2019)Karam, Pappas, and Tamvakis]{Karam:2019dlv}
Karam, A.; Pappas, T.; Tamvakis, K.
\newblock {Frame-dependence of inflationary observables in scalar-tensor
  gravity}.
\newblock {\em PoS} {\bf 2019}, {\em CORFU2018},~064,
  \href{http://xxx.lanl.gov/abs/1903.03548}{{\normalfont
  [arXiv:gr-qc/1903.03548]}}.
\newblock
  doi:{\changeurlcolor{black}\href{https://doi.org/10.22323/1.347.0064}{\detokenize{10.22323/1.347.0064}}}.

\bibitem[Nojiri \em{et~al.}(2019)Nojiri, Odintsov, and
  Saridakis]{Nojiri:2019kkp}
Nojiri, S.; Odintsov, S.D.; Saridakis, E.N.
\newblock {Holographic inflation}.
\newblock {\em Phys. Lett.} {\bf 2019}, {\em B797},~134829,
  \href{http://xxx.lanl.gov/abs/1904.01345}{{\normalfont
  [arXiv:gr-qc/1904.01345]}}.
\newblock
  doi:{\changeurlcolor{black}\href{https://doi.org/10.1016/j.physletb.2019.134829}{\detokenize{10.1016/j.physletb.2019.134829}}}.

\bibitem[Benisty \em{et~al.}(2019{\natexlab{a}})Benisty, Guendelman, Saridakis,
  Stoecker, Struckmeier, and Vasak]{Benisty:2019jqz}
Benisty, D.; Guendelman, E.I.; Saridakis, E.N.; Stoecker, H.; Struckmeier, J.;
  Vasak, D.
\newblock {Inflation from fermions with curvature-dependent mass} {\bf 2019}.
\newblock  \href{http://xxx.lanl.gov/abs/1905.03731}{{\normalfont
  [arXiv:gr-qc/1905.03731]}}.

\bibitem[Benisty \em{et~al.}(2019{\natexlab{b}})Benisty, Guendelman, Nissimov,
  and Pacheva]{Benisty:2019tno}
Benisty, D.; Guendelman, E.; Nissimov, E.; Pacheva, S.
\newblock {Dynamically Generated Inflation from Non-Riemannian Volume Forms}
  {\bf 2019}.
\newblock  \href{http://xxx.lanl.gov/abs/1906.06691}{{\normalfont
  [arXiv:gr-qc/1906.06691]}}.

\bibitem[Benisty \em{et~al.}(2019{\natexlab{c}})Benisty, Guendelman, Nissimov,
  and Pacheva]{Benisty:2019bmi}
Benisty, D.; Guendelman, E.I.; Nissimov, E.; Pacheva, S.
\newblock {Dynamically generated inflationary two-field potential via
  non-Riemannian volume forms} {\bf 2019}.
\newblock  \href{http://xxx.lanl.gov/abs/1907.07625}{{\normalfont
  [arXiv:astro-ph.CO/1907.07625]}}.

\bibitem[Kinney \em{et~al.}(2019)Kinney, Vagnozzi, and
  Visinelli]{Kinney:2018nny}
Kinney, W.H.; Vagnozzi, S.; Visinelli, L.
\newblock {The zoo plot meets the swampland: mutual (in)consistency of
  single-field inflation, string conjectures, and cosmological data}.
\newblock {\em Class. Quant. Grav.} {\bf 2019}, {\em 36},~117001,
  \href{http://xxx.lanl.gov/abs/1808.06424}{{\normalfont
  [arXiv:astro-ph.CO/1808.06424]}}.
\newblock
  doi:{\changeurlcolor{black}\href{https://doi.org/10.1088/1361-6382/ab1d87}{\detokenize{10.1088/1361-6382/ab1d87}}}.

\bibitem[Brustein and Sherf(2018)]{Brustein:2017iet}
Brustein, R.; Sherf, Y.
\newblock {Causality Violations in Lovelock Theories}.
\newblock {\em Phys. Rev.} {\bf 2018}, {\em D97},~084019,
  \href{http://xxx.lanl.gov/abs/1711.05140}{{\normalfont
  [arXiv:hep-th/1711.05140]}}.
\newblock
  doi:{\changeurlcolor{black}\href{https://doi.org/10.1103/PhysRevD.97.084019}{\detokenize{10.1103/PhysRevD.97.084019}}}.

\bibitem[Sherf(2019)]{Sherf:2018uth}
Sherf, Y.
\newblock {Hyperbolicity Constraints in Extended Gravity Theories}.
\newblock {\em Phys. Scripta} {\bf 2019}, {\em 94},~085005,
  \href{http://xxx.lanl.gov/abs/1806.09984}{{\normalfont
  [arXiv:gr-qc/1806.09984]}}.
\newblock
  doi:{\changeurlcolor{black}\href{https://doi.org/10.1088/1402-4896/ab1352}{\detokenize{10.1088/1402-4896/ab1352}}}.

\bibitem[Capozziello \em{et~al.}(2014)Capozziello, De~Laurentis, and
  Luongo]{Capozziello:2014hia}
Capozziello, S.; De~Laurentis, M.; Luongo, O.
\newblock {Connecting early and late universe by $f(R)$ gravity}.
\newblock {\em Int. J. Mod. Phys.} {\bf 2014}, {\em D24},~1541002,
  \href{http://xxx.lanl.gov/abs/1411.2822}{{\normalfont
  [arXiv:gr-qc/1411.2822]}}.
\newblock
  doi:{\changeurlcolor{black}\href{https://doi.org/10.1142/S0218271815410023}{\detokenize{10.1142/S0218271815410023}}}.

\bibitem[Gorbunov and Tokareva(2014)]{Gorbunov:2013dqa}
Gorbunov, D.; Tokareva, A.
\newblock {Scale-invariance as the origin of dark radiation?}
\newblock {\em Phys. Lett.} {\bf 2014}, {\em B739},~50--55,
  \href{http://xxx.lanl.gov/abs/1307.5298}{{\normalfont
  [arXiv:astro-ph.CO/1307.5298]}}.
\newblock
  doi:{\changeurlcolor{black}\href{https://doi.org/10.1016/j.physletb.2014.10.036}{\detokenize{10.1016/j.physletb.2014.10.036}}}.

\bibitem[Myrzakulov \em{et~al.}(2015)Myrzakulov, Odintsov, and
  Sebastiani]{Myrzakulov:2014hca}
Myrzakulov, R.; Odintsov, S.; Sebastiani, L.
\newblock {Inflationary universe from higher-derivative quantum gravity}.
\newblock {\em Phys. Rev.} {\bf 2015}, {\em D91},~083529,
  \href{http://xxx.lanl.gov/abs/1412.1073}{{\normalfont
  [arXiv:gr-qc/1412.1073]}}.
\newblock
  doi:{\changeurlcolor{black}\href{https://doi.org/10.1103/PhysRevD.91.083529}{\detokenize{10.1103/PhysRevD.91.083529}}}.

\bibitem[Bamba \em{et~al.}(2014)Bamba, Myrzakulov, Odintsov, and
  Sebastiani]{Bamba:2014jia}
Bamba, K.; Myrzakulov, R.; Odintsov, S.D.; Sebastiani, L.
\newblock {Trace-anomaly driven inflation in modified gravity and the BICEP2
  result}.
\newblock {\em Phys. Rev.} {\bf 2014}, {\em D90},~043505,
  \href{http://xxx.lanl.gov/abs/1403.6649}{{\normalfont
  [arXiv:hep-th/1403.6649]}}.
\newblock
  doi:{\changeurlcolor{black}\href{https://doi.org/10.1103/PhysRevD.90.043505}{\detokenize{10.1103/PhysRevD.90.043505}}}.

\bibitem[Benisty \em{et~al.}(2018)Benisty, Guendelman, Vasak, Struckmeier, and
  Stoecker]{Benisty:2018ufz}
Benisty, D.; Guendelman, E.I.; Vasak, D.; Struckmeier, J.; Stoecker, H.
\newblock {Quadratic curvature theories formulated as Covariant Canonical Gauge
  theories of Gravity}.
\newblock {\em Phys. Rev.} {\bf 2018}, {\em D98},~106021,
  \href{http://xxx.lanl.gov/abs/1809.10447}{{\normalfont
  [arXiv:gr-qc/1809.10447]}}.
\newblock
  doi:{\changeurlcolor{black}\href{https://doi.org/10.1103/PhysRevD.98.106021}{\detokenize{10.1103/PhysRevD.98.106021}}}.

\bibitem[Aashish and Panda(2020)]{Aashish:2020ufe}
Aashish, S.; Panda, S.
\newblock {Covariant quantum corrections to a scalar field model inspired by
  nonminimal natural inflation} {\bf 2020}.
\newblock  \href{http://xxx.lanl.gov/abs/2001.07350}{{\normalfont
  [arXiv:gr-qc/2001.07350]}}.

\bibitem[Rashidi and Nozari(2020)]{Rashidi:2020wwg}
Rashidi, N.; Nozari, K.
\newblock {Gauss-Bonnet Inflation after Planck2018} {\bf 2020}.
\newblock  \href{http://xxx.lanl.gov/abs/2001.07012}{{\normalfont
  [arXiv:astro-ph.CO/2001.07012]}}.

\bibitem[Odintsov and Oikonomou(2020)]{Odintsov:2020nwm}
Odintsov, S.D.; Oikonomou, V.K.
\newblock {Geometric Inflation and Dark Energy with Axion $F(R)$ Gravity}.
\newblock {\em Phys. Rev.} {\bf 2020}, {\em D101},~044009,
  \href{http://xxx.lanl.gov/abs/2001.06830}{{\normalfont
  [arXiv:gr-qc/2001.06830]}}.
\newblock
  doi:{\changeurlcolor{black}\href{https://doi.org/10.1103/PhysRevD.101.044009}{\detokenize{10.1103/PhysRevD.101.044009}}}.

\bibitem[Antoniadis \em{et~al.}(2019)Antoniadis, Karam, Lykkas, Pappas, and
  Tamvakis]{Antoniadis:2019jnz}
Antoniadis, I.; Karam, A.; Lykkas, A.; Pappas, T.; Tamvakis, K.
\newblock {Single-field inflation in models with an $R^2$ term}.
\newblock  {19th Hellenic School and Workshops on Elementary Particle Physics
  and Gravity (CORFU2019) Corfu, Greece, August 31-September 25, 2019},  2019,
  \href{http://xxx.lanl.gov/abs/1912.12757}{{\normalfont
  [arXiv:gr-qc/1912.12757]}}.

\bibitem[Benisty and Guendelman(2018)]{Benisty:2018fgu}
Benisty, D.; Guendelman, E.I.
\newblock {Correspondence between the first and second order formalism by a
  metricity constraint}.
\newblock {\em Phys. Rev.} {\bf 2018}, {\em D98},~044023,
  \href{http://xxx.lanl.gov/abs/1805.09667}{{\normalfont
  [arXiv:gr-qc/1805.09667]}}.
\newblock
  doi:{\changeurlcolor{black}\href{https://doi.org/10.1103/PhysRevD.98.044023}{\detokenize{10.1103/PhysRevD.98.044023}}}.

\bibitem[Chakraborty \em{et~al.}(2018)Chakraborty, Paul, and
  SenGupta]{Chakraborty:2018scm}
Chakraborty, S.; Paul, T.; SenGupta, S.
\newblock {Inflation driven by Einstein-Gauss-Bonnet gravity}.
\newblock {\em Phys. Rev.} {\bf 2018}, {\em D98},~083539,
  \href{http://xxx.lanl.gov/abs/1804.03004}{{\normalfont
  [arXiv:gr-qc/1804.03004]}}.
\newblock
  doi:{\changeurlcolor{black}\href{https://doi.org/10.1103/PhysRevD.98.083539}{\detokenize{10.1103/PhysRevD.98.083539}}}.

\bibitem[Mukhanov and Chibisov(1981)]{Mukhanov:1981xt}
Mukhanov, V.F.; Chibisov, G.V.
\newblock {Quantum Fluctuations and a Nonsingular Universe}.
\newblock {\em JETP Lett.} {\bf 1981}, {\em 33},~532--535.
\newblock [Pisma Zh. Eksp. Teor. Fiz.33,549(1981)].

\bibitem[Guth and Pi(1982)]{Guth:1982ec}
Guth, A.H.; Pi, S.Y.
\newblock {Fluctuations in the New Inflationary Universe}.
\newblock {\em Phys. Rev. Lett.} {\bf 1982}, {\em 49},~1110--1113.
\newblock
  doi:{\changeurlcolor{black}\href{https://doi.org/10.1103/PhysRevLett.49.1110}{\detokenize{10.1103/PhysRevLett.49.1110}}}.

\bibitem[Faraoni and Capozziello(2011)]{Capozziello:2010zz}
Faraoni, V.; Capozziello, S.
\newblock {\em {Beyond Einstein Gravity}}; Vol. 170, Springer: Dordrecht,
  2011.
\newblock
  doi:{\changeurlcolor{black}\href{https://doi.org/10.1007/978-94-007-0165-6}{\detokenize{10.1007/978-94-007-0165-6}}}.

\bibitem[Nojiri \em{et~al.}(2017)Nojiri, Odintsov, and
  Oikonomou]{Nojiri:2017ncd}
Nojiri, S.; Odintsov, S.D.; Oikonomou, V.K.
\newblock {Modified Gravity Theories on a Nutshell: Inflation, Bounce and
  Late-time Evolution}.
\newblock {\em Phys. Rept.} {\bf 2017}, {\em 692},~1--104,
  \href{http://xxx.lanl.gov/abs/1705.11098}{{\normalfont
  [arXiv:gr-qc/1705.11098]}}.
\newblock
  doi:{\changeurlcolor{black}\href{https://doi.org/10.1016/j.physrep.2017.06.001}{\detokenize{10.1016/j.physrep.2017.06.001}}}.

\bibitem[Dimitrijevic \em{et~al.}(2019)Dimitrijevic, Dragovich, Koshelev,
  Rakic, and Stankovic]{Dimitrijevic:2019pct}
Dimitrijevic, I.; Dragovich, B.; Koshelev, A.S.; Rakic, Z.; Stankovic, J.
\newblock {Cosmological Solutions of a Nonlocal Square Root Gravity}.
\newblock {\em Phys. Lett.} {\bf 2019}, {\em B797},~134848,
  \href{http://xxx.lanl.gov/abs/1906.07560}{{\normalfont
  [arXiv:gr-qc/1906.07560]}}.
\newblock
  doi:{\changeurlcolor{black}\href{https://doi.org/10.1016/j.physletb.2019.134848}{\detokenize{10.1016/j.physletb.2019.134848}}}.

\bibitem[Bilic \em{et~al.}(2019)Bilic, Dimitrijevic, Djordjevic, Milosevic, and
  Stojanovic]{Bilic:2018uqx}
Bilic, N.; Dimitrijevic, D.D.; Djordjevic, G.S.; Milosevic, M.; Stojanovic, M.
\newblock {Tachyon inflation in the holographic braneworld}.
\newblock {\em JCAP} {\bf 2019}, {\em 1908},~034,
  \href{http://xxx.lanl.gov/abs/1809.07216}{{\normalfont
  [arXiv:gr-qc/1809.07216]}}.
\newblock
  doi:{\changeurlcolor{black}\href{https://doi.org/10.1088/1475-7516/2019/08/034}{\detokenize{10.1088/1475-7516/2019/08/034}}}.

\bibitem[Nojiri and Odintsov(2011)]{Nojiri:2010wj}
Nojiri, S.; Odintsov, S.D.
\newblock {Unified cosmic history in modified gravity: from F(R) theory to
  Lorentz non-invariant models}.
\newblock {\em Phys. Rept.} {\bf 2011}, {\em 505},~59--144,
  \href{http://xxx.lanl.gov/abs/1011.0544}{{\normalfont
  [arXiv:gr-qc/1011.0544]}}.
\newblock
  doi:{\changeurlcolor{black}\href{https://doi.org/10.1016/j.physrep.2011.04.001}{\detokenize{10.1016/j.physrep.2011.04.001}}}.

\bibitem[Berti \em{et~al.}(2015)Berti et~al.]{Berti:2015itd}
Berti, E.; others.
\newblock {Testing General Relativity with Present and Future Astrophysical
  Observations}.
\newblock {\em Class. Quant. Grav.} {\bf 2015}, {\em 32},~243001,
  \href{http://xxx.lanl.gov/abs/1501.07274}{{\normalfont
  [arXiv:gr-qc/1501.07274]}}.
\newblock
  doi:{\changeurlcolor{black}\href{https://doi.org/10.1088/0264-9381/32/24/243001}{\detokenize{10.1088/0264-9381/32/24/243001}}}.

\bibitem[Akrami \em{et~al.}(2018)Akrami et~al.]{Akrami:2018odb}
Akrami, Y.; others.
\newblock {Planck 2018 results. X. Constraints on inflation} {\bf 2018}.
\newblock  \href{http://xxx.lanl.gov/abs/1807.06211}{{\normalfont
  [arXiv:astro-ph.CO/1807.06211]}}.

\bibitem[Nojiri and Odintsov(2006)]{Nojiri:2006gh}
Nojiri, S.; Odintsov, S.D.
\newblock {Modified f(R) gravity consistent with realistic cosmology: From
  matter dominated epoch to dark energy universe}.
\newblock {\em Phys. Rev.} {\bf 2006}, {\em D74},~086005,
  \href{http://xxx.lanl.gov/abs/hep-th/0608008}{{\normalfont
  [arXiv:hep-th/hep-th/0608008]}}.
\newblock
  doi:{\changeurlcolor{black}\href{https://doi.org/10.1103/PhysRevD.74.086005}{\detokenize{10.1103/PhysRevD.74.086005}}}.

\bibitem[Lozano and Garcia-Compean(2019)]{Lozano:2019gck}
Lozano, L.; Garcia-Compean, H.
\newblock {Emergent Dark Matter and Dark Energy from a Lattice Model} {\bf
  2019}.
\newblock  \href{http://xxx.lanl.gov/abs/1912.11224}{{\normalfont
  [arXiv:hep-th/1912.11224]}}.

\bibitem[Chamings \em{et~al.}(2019)Chamings, Avgoustidis, Copeland, Green, and
  Pourtsidou]{Chamings:2019kcl}
Chamings, F.N.; Avgoustidis, A.; Copeland, E.J.; Green, A.M.; Pourtsidou, A.
\newblock {Understanding the suppression of structure formation from dark
  matter $2013$ dark energy momentum coupling} {\bf 2019}.
\newblock  \href{http://xxx.lanl.gov/abs/1912.09858}{{\normalfont
  [arXiv:astro-ph.CO/1912.09858]}}.

\bibitem[Liu and Xu(2019)]{Liu:2019xhn}
Liu, L.H.; Xu, W.L.
\newblock {The running curvaton} {\bf 2019}.
\newblock  \href{http://xxx.lanl.gov/abs/1911.10542}{{\normalfont
  [arXiv:astro-ph.CO/1911.10542]}}.

\bibitem[Cheng \em{et~al.}(2019)Cheng, Ma, Wu, Zhang, and Chen]{Cheng:2019bkh}
Cheng, G.; Ma, Y.; Wu, F.; Zhang, J.; Chen, X.
\newblock {Testing interacting dark matter and dark energy model with
  cosmological data} {\bf 2019}.
\newblock  \href{http://xxx.lanl.gov/abs/1911.04520}{{\normalfont
  [arXiv:astro-ph.CO/1911.04520]}}.

\bibitem[Cahill(2019)]{Cahill:2019lwc}
Cahill, K.
\newblock {Zero-point energies, dark matter, and dark energy} {\bf 2019}.
\newblock  \href{http://xxx.lanl.gov/abs/1910.09953}{{\normalfont
  [arXiv:physics.gen-ph/1910.09953]}}.

\bibitem[Bandyopadhyay and Chatterjee(2019)]{Bandyopadhyay:2019vdd}
Bandyopadhyay, A.; Chatterjee, A.
\newblock {Time-dependent diffusive interactions between dark matter and dark
  energy in the context of $k-$essence cosmology} {\bf 2019}.
\newblock  \href{http://xxx.lanl.gov/abs/1910.10423}{{\normalfont
  [arXiv:gr-qc/1910.10423]}}.

\bibitem[Kase and Tsujikawa(2019)]{Kase:2019veo}
Kase, R.; Tsujikawa, S.
\newblock {Scalar-Field Dark Energy Nonminimally and Kinetically Coupled to
  Dark Matter} {\bf 2019}.
\newblock  \href{http://xxx.lanl.gov/abs/1910.02699}{{\normalfont
  [arXiv:gr-qc/1910.02699]}}.

\bibitem[Ketov(2019)]{Ketov:2019rzg}
Ketov, S.V.
\newblock {Inflation, Dark Energy and Dark Matter in Supergravity}.
\newblock  {Meeting of the Division of Particles and Fields of the American
  Physical Society (DPF2019) Boston, Massachusetts, July 29-August 2, 2019},
  2019,  \href{http://xxx.lanl.gov/abs/1909.05599}{{\normalfont
  [arXiv:hep-th/1909.05599]}}.

\bibitem[Mukhopadhyay \em{et~al.}(2019)Mukhopadhyay, Paul, and
  Majumdar]{Mukhopadhyay:2019jla}
Mukhopadhyay, U.; Paul, A.; Majumdar, D.
\newblock {Probing Pseudo Nambu Goldstone Boson Dark Energy Models with Dark
  Matter -- Dark Energy Interaction} {\bf 2019}.
\newblock  \href{http://xxx.lanl.gov/abs/1909.03925}{{\normalfont
  [arXiv:astro-ph.CO/1909.03925]}}.

\bibitem[Yang \em{et~al.}(2019)Yang, Pan, Vagnozzi, Di~Valentino, Mota, and
  Capozziello]{Yang:2019nhz}
Yang, W.; Pan, S.; Vagnozzi, S.; Di~Valentino, E.; Mota, D.F.; Capozziello, S.
\newblock {Dawn of the dark: unified dark sectors and the EDGES Cosmic Dawn
  21-cm signal}.
\newblock {\em JCAP} {\bf 2019}, {\em 1911},~044,
  \href{http://xxx.lanl.gov/abs/1907.05344}{{\normalfont
  [arXiv:astro-ph.CO/1907.05344]}}.
\newblock
  doi:{\changeurlcolor{black}\href{https://doi.org/10.1088/1475-7516/2019/11/044}{\detokenize{10.1088/1475-7516/2019/11/044}}}.

\bibitem[Guendelman and Kaganovich(1996)]{Guendelman:1996qy}
Guendelman, E.I.; Kaganovich, A.B.
\newblock {The Principle of nongravitating vacuum energy and some of its
  consequences}.
\newblock {\em Phys. Rev.} {\bf 1996}, {\em D53},~7020--7025,
  \href{http://xxx.lanl.gov/abs/gr-qc/9605026}{{\normalfont
  [arXiv:gr-qc/gr-qc/9605026]}}.
\newblock
  doi:{\changeurlcolor{black}\href{https://doi.org/10.1103/PhysRevD.53.7020}{\detokenize{10.1103/PhysRevD.53.7020}}}.

\bibitem[Gronwald \em{et~al.}(1998)Gronwald, Muench, Macias, and
  Hehl]{Gronwald:1997ei}
Gronwald, F.; Muench, U.; Macias, A.; Hehl, F.W.
\newblock {Volume elements of space-time and a quartet of scalar fields}.
\newblock {\em Phys. Rev.} {\bf 1998}, {\em D58},~084021,
  \href{http://xxx.lanl.gov/abs/gr-qc/9712063}{{\normalfont
  [arXiv:gr-qc/gr-qc/9712063]}}.
\newblock
  doi:{\changeurlcolor{black}\href{https://doi.org/10.1103/PhysRevD.58.084021}{\detokenize{10.1103/PhysRevD.58.084021}}}.

\bibitem[Guendelman and Kaganovich(1999)]{Guendelman:1999tb}
Guendelman, E.I.; Kaganovich, A.B.
\newblock {Dynamical measure and field theory models free of the cosmological
  constant problem}.
\newblock {\em Phys. Rev.} {\bf 1999}, {\em D60},~065004,
  \href{http://xxx.lanl.gov/abs/gr-qc/9905029}{{\normalfont
  [arXiv:gr-qc/gr-qc/9905029]}}.
\newblock
  doi:{\changeurlcolor{black}\href{https://doi.org/10.1103/PhysRevD.60.065004}{\detokenize{10.1103/PhysRevD.60.065004}}}.

\bibitem[Guendelman(1999)]{Guendelman:1999qt}
Guendelman, E.I.
\newblock {Scale invariance, new inflation and decaying lambda terms}.
\newblock {\em Mod. Phys. Lett.} {\bf 1999}, {\em A14},~1043--1052,
  \href{http://xxx.lanl.gov/abs/gr-qc/9901017}{{\normalfont
  [arXiv:gr-qc/gr-qc/9901017]}}.
\newblock
  doi:{\changeurlcolor{black}\href{https://doi.org/10.1142/S0217732399001103}{\detokenize{10.1142/S0217732399001103}}}.

\bibitem[Guendelman and Kaganovich(2008)]{Guendelman:2007ph}
Guendelman, E.I.; Kaganovich, A.B.
\newblock {Absence of the Fifth Force Problem in a Model with Spontaneously
  Broken Dilatation Symmetry}.
\newblock {\em Annals Phys.} {\bf 2008}, {\em 323},~866--882,
  \href{http://xxx.lanl.gov/abs/0704.1998}{{\normalfont
  [arXiv:gr-qc/0704.1998]}}.
\newblock
  doi:{\changeurlcolor{black}\href{https://doi.org/10.1016/j.aop.2007.09.003}{\detokenize{10.1016/j.aop.2007.09.003}}}.

\bibitem[Guendelman \em{et~al.}(2014)Guendelman, Nissimov, Pacheva, and
  Vasihoun]{Vasihoun:2014hpa}
Guendelman, E.; Nissimov, E.; Pacheva, S.; Vasihoun, M.
\newblock {A New Mechanism of Dynamical Spontaneous Breaking of Supersymmetry}.
\newblock {\em Bulg. J. Phys.} {\bf 2014}, {\em 41},~123--129,
  \href{http://xxx.lanl.gov/abs/1404.4733}{{\normalfont
  [arXiv:hep-th/1404.4733]}}.

\bibitem[Guendelman \em{et~al.}(2015)Guendelman, Nissimov, and
  Pacheva]{Guendelman:2015qva}
Guendelman, E.; Nissimov, E.; Pacheva, S.
\newblock {Vacuum structure and gravitational bags produced by
  metric-independent space–time volume-form dynamics}.
\newblock {\em Int. J. Mod. Phys.} {\bf 2015}, {\em A30},~1550133,
  \href{http://xxx.lanl.gov/abs/1504.01031}{{\normalfont
  [arXiv:gr-qc/1504.01031]}}.
\newblock
  doi:{\changeurlcolor{black}\href{https://doi.org/10.1142/S0217751X1550133X}{\detokenize{10.1142/S0217751X1550133X}}}.

\bibitem[Guendelman \em{et~al.}(2016)Guendelman, Nissimov, and
  Pacheva]{Guendelman:2015jii}
Guendelman, E.; Nissimov, E.; Pacheva, S.
\newblock {Unified Dark Energy and Dust Dark Matter Dual to Quadratic Purely
  Kinetic K-Essence}.
\newblock {\em Eur. Phys. J.} {\bf 2016}, {\em C76},~90,
  \href{http://xxx.lanl.gov/abs/1511.07071}{{\normalfont
  [arXiv:gr-qc/1511.07071]}}.
\newblock
  doi:{\changeurlcolor{black}\href{https://doi.org/10.1140/epjc/s10052-016-3938-7}{\detokenize{10.1140/epjc/s10052-016-3938-7}}}.

\bibitem[Guendelman \em{et~al.}(2012)Guendelman, Singleton, and
  Yongram]{Guendelman:2012gg}
Guendelman, E.; Singleton, D.; Yongram, N.
\newblock {A two measure model of dark energy and dark matter}.
\newblock {\em JCAP} {\bf 2012}, {\em 1211},~044,
  \href{http://xxx.lanl.gov/abs/1205.1056}{{\normalfont
  [arXiv:gr-qc/1205.1056]}}.
\newblock
  doi:{\changeurlcolor{black}\href{https://doi.org/10.1088/1475-7516/2012/11/044}{\detokenize{10.1088/1475-7516/2012/11/044}}}.

\bibitem[Guendelman \em{et~al.}(2015)Guendelman, Nissimov, and
  Pacheva]{Guendelman:2015rea}
Guendelman, E.; Nissimov, E.; Pacheva, S.
\newblock {Dark Energy and Dark Matter From Hidden Symmetry of Gravity Model
  with a Non-Riemannian Volume Form}.
\newblock {\em Eur. Phys. J.} {\bf 2015}, {\em C75},~472,
  \href{http://xxx.lanl.gov/abs/1508.02008}{{\normalfont
  [arXiv:gr-qc/1508.02008]}}.
\newblock
  doi:{\changeurlcolor{black}\href{https://doi.org/10.1140/epjc/s10052-015-3699-8}{\detokenize{10.1140/epjc/s10052-015-3699-8}}}.

\bibitem[Guendelman \em{et~al.}(2016)Guendelman, Nissimov, and
  Pacheva]{Guendelman:2016lea}
Guendelman, E.; Nissimov, E.; Pacheva, S.
\newblock {Gravity-Assisted Emergent Higgs Mechanism in the Post-Inflationary
  Epoch}.
\newblock {\em Int. J. Mod. Phys.} {\bf 2016}, {\em D25},~1644008,
  \href{http://xxx.lanl.gov/abs/1603.06231}{{\normalfont
  [arXiv:hep-th/1603.06231]}}.
\newblock
  doi:{\changeurlcolor{black}\href{https://doi.org/10.1142/S0218271816440089}{\detokenize{10.1142/S0218271816440089}}}.

\bibitem[Guendelman \em{et~al.}(2019)Guendelman, Nissimov, and
  Pacheva]{Guendelman:2018zcb}
Guendelman, E.; Nissimov, E.; Pacheva, S.
\newblock {Modified Gravity and Inflaton Assisted Dynamical Generation of
  Charge Confinement and Electroweak Symmetry Breaking in Cosmology}.
\newblock {\em AIP Conf. Proc.} {\bf 2019}, {\em 2075},~090030,
  \href{http://xxx.lanl.gov/abs/1808.03640}{{\normalfont
  [arXiv:hep-th/1808.03640]}}.
\newblock
  doi:{\changeurlcolor{black}\href{https://doi.org/10.1063/1.5091244}{\detokenize{10.1063/1.5091244}}}.

\bibitem[Guendelman \em{et~al.}(2015)Guendelman, Nissimov, and
  Pacheva]{Guendelman:2014waa}
Guendelman, E.; Nissimov, E.; Pacheva, S.
\newblock {Unification of Inflation and Dark Energy from Spontaneous Breaking
  of Scale Invariance}.
\newblock  {Proceedings, 8th Mathematical Physics Meeting, Summer School and
  Conference on Modern Mathematical Physics: Belgrade, Serbia, August 24-31,
  2014},  2015, pp. 93--103,
  \href{http://xxx.lanl.gov/abs/1407.6281}{{\normalfont
  [arXiv:hep-th/1407.6281]}}.

\bibitem[Frieman \em{et~al.}(2008)Frieman, Turner, and Huterer]{Frieman:2008sn}
Frieman, J.; Turner, M.; Huterer, D.
\newblock {Dark Energy and the Accelerating Universe}.
\newblock {\em Ann. Rev. Astron. Astrophys.} {\bf 2008}, {\em 46},~385--432,
  \href{http://xxx.lanl.gov/abs/0803.0982}{{\normalfont
  [arXiv:astro-ph/0803.0982]}}.
\newblock
  doi:{\changeurlcolor{black}\href{https://doi.org/10.1146/annurev.astro.46.060407.145243}{\detokenize{10.1146/annurev.astro.46.060407.145243}}}.

\bibitem[Mathews \em{et~al.}(2017)Mathews, Kusakabe, and
  Kajino]{Mathews:2017xht}
Mathews, G.J.; Kusakabe, M.; Kajino, T.
\newblock {Introduction to Big Bang Nucleosynthesis and Modern Cosmology}.
\newblock {\em Int. J. Mod. Phys.} {\bf 2017}, {\em E26},~1741001,
  \href{http://xxx.lanl.gov/abs/1706.03138}{{\normalfont
  [arXiv:astro-ph.CO/1706.03138]}}.
\newblock
  doi:{\changeurlcolor{black}\href{https://doi.org/10.1142/S0218301317410014}{\detokenize{10.1142/S0218301317410014}}}.

\bibitem[Liddle(2009)]{Liddle:2009zz}
Liddle, A.
\newblock {\em {Einfuehrung in die moderne Kosmologie}};  2009.

\bibitem[Liddle(1998)]{Liddle:1998ew}
Liddle, A.R.
\newblock {\em {An introduction to modern cosmology}};  1998.

\bibitem[Dodelson(2003)]{Dodelson:2003ft}
Dodelson, S.
\newblock {\em {Modern Cosmology}}; Academic Press: Amsterdam,  2003.

\bibitem[Dodelson \em{et~al.}(2009)Dodelson et~al.]{Dodelson:2009kq}
Dodelson, S.; others.
\newblock {The Origin of the Universe as Revealed Through the Polarization of
  the Cosmic Microwave Background} {\bf 2009}.
\newblock  \href{http://xxx.lanl.gov/abs/0902.3796}{{\normalfont
  [arXiv:astro-ph.CO/0902.3796]}}.

\bibitem[Baumann \em{et~al.}(2009)Baumann, Cooray, Dodelson, Dunkley, Fraisse,
  Jackson, Kogut, Krauss, Smith, and Zaldarriaga]{Baumann:2008aj}
Baumann, D.; Cooray, A.; Dodelson, S.; Dunkley, J.; Fraisse, A.A.; Jackson,
  M.G.; Kogut, A.; Krauss, L.M.; Smith, K.M.; Zaldarriaga, M.
\newblock {CMBPol Mission Concept Study: A Mission to Map our Origins}.
\newblock {\em AIP Conf. Proc.} {\bf 2009}, {\em 1141},~3--9,
  \href{http://xxx.lanl.gov/abs/0811.3911}{{\normalfont
  [arXiv:astro-ph/0811.3911]}}.
\newblock
  doi:{\changeurlcolor{black}\href{https://doi.org/10.1063/1.3160890}{\detokenize{10.1063/1.3160890}}}.

\bibitem[Dodelson(2000)]{Dodelson:1999hm}
Dodelson, S.
\newblock {Cosmic microwave background: Past, future, and present}.
\newblock {\em Int. J. Mod. Phys.} {\bf 2000}, {\em A15S1},~765--783,
  \href{http://xxx.lanl.gov/abs/hep-ph/9912470}{{\normalfont
  [arXiv:hep-ph/hep-ph/9912470]}}.
\newblock [,765(1999)],
  doi:{\changeurlcolor{black}\href{https://doi.org/10.1142/S0217751X00005401}{\detokenize{10.1142/S0217751X00005401}}}.

\bibitem[Dabrowski \em{et~al.}(2009)Dabrowski, Garecki, and
  Blaschke]{Dabrowski:2008kx}
Dabrowski, M.P.; Garecki, J.; Blaschke, D.B.
\newblock {Conformal transformations and conformal invariance in gravitation}.
\newblock {\em Annalen Phys.} {\bf 2009}, {\em 18},~13--32,
  \href{http://xxx.lanl.gov/abs/0806.2683}{{\normalfont
  [arXiv:gr-qc/0806.2683]}}.
\newblock
  doi:{\changeurlcolor{black}\href{https://doi.org/10.1002/andp.200810331}{\detokenize{10.1002/andp.200810331}}}.

\bibitem[Angus \em{et~al.}(2019)Angus et~al.]{Angus:2018tko}
Angus, C.R.; others.
\newblock {Superluminous Supernovae from the Dark Energy Survey}.
\newblock {\em Mon. Not. Roy. Astron. Soc.} {\bf 2019}, {\em 487},~2215--2241,
  \href{http://xxx.lanl.gov/abs/1812.04071}{{\normalfont
  [arXiv:astro-ph.HE/1812.04071]}}.
\newblock
  doi:{\changeurlcolor{black}\href{https://doi.org/10.1093/mnras/stz1321}{\detokenize{10.1093/mnras/stz1321}}}.

\bibitem[Zhang \em{et~al.}(2019)Zhang et~al.]{Zhang:2018gbq}
Zhang, Y.; others.
\newblock {Dark Energy Survey Year 1 results: Detection of Intra-cluster Light
  at Redshift $\sim$ 0.25}.
\newblock {\em Astrophys. J.} {\bf 2019}, {\em 874},~165,
  \href{http://xxx.lanl.gov/abs/1812.04004}{{\normalfont
  [arXiv:astro-ph.CO/1812.04004]}}.
\newblock
  doi:{\changeurlcolor{black}\href{https://doi.org/10.3847/1538-4357/ab0dfd}{\detokenize{10.3847/1538-4357/ab0dfd}}}.

\bibitem[Bahamonde \em{et~al.}(2018)Bahamonde, Böhmer, Carloni, Copeland,
  Fang, and Tamanini]{Bahamonde:2017ize}
Bahamonde, S.; Böhmer, C.G.; Carloni, S.; Copeland, E.J.; Fang, W.; Tamanini,
  N.
\newblock {Dynamical systems applied to cosmology: dark energy and modified
  gravity}.
\newblock {\em Phys. Rept.} {\bf 2018}, {\em 775-777},~1--122,
  \href{http://xxx.lanl.gov/abs/1712.03107}{{\normalfont
  [arXiv:gr-qc/1712.03107]}}.
\newblock
  doi:{\changeurlcolor{black}\href{https://doi.org/10.1016/j.physrep.2018.09.001}{\detokenize{10.1016/j.physrep.2018.09.001}}}.

\bibitem[Ade \em{et~al.}(2014)Ade et~al.]{Planck:2013jfk}
Ade, P.A.R.; others.
\newblock {Planck 2013 results. XXII. Constraints on inflation}.
\newblock {\em Astron. Astrophys.} {\bf 2014}, {\em 571},~A22,
  \href{http://xxx.lanl.gov/abs/1303.5082}{{\normalfont
  [arXiv:astro-ph.CO/1303.5082]}}.
\newblock
  doi:{\changeurlcolor{black}\href{https://doi.org/10.1051/0004-6361/201321569}{\detokenize{10.1051/0004-6361/201321569}}}.

\bibitem[Adam \em{et~al.}(2016)Adam et~al.]{Adam:2014bub}
Adam, R.; others.
\newblock {Planck intermediate results. XXX. The angular power spectrum of
  polarized dust emission at intermediate and high Galactic latitudes}.
\newblock {\em Astron. Astrophys.} {\bf 2016}, {\em 586},~A133,
  \href{http://xxx.lanl.gov/abs/1409.5738}{{\normalfont
  [arXiv:astro-ph.CO/1409.5738]}}.
\newblock
  doi:{\changeurlcolor{black}\href{https://doi.org/10.1051/0004-6361/201425034}{\detokenize{10.1051/0004-6361/201425034}}}.

\bibitem[Arkani-Hamed \em{et~al.}(2000)Arkani-Hamed, Hall, Kolda, and
  Murayama]{ArkaniHamed:2000tc}
Arkani-Hamed, N.; Hall, L.J.; Kolda, C.F.; Murayama, H.
\newblock {A New perspective on cosmic coincidence problems}.
\newblock {\em Phys. Rev. Lett.} {\bf 2000}, {\em 85},~4434--4437,
  \href{http://xxx.lanl.gov/abs/astro-ph/0005111}{{\normalfont
  [arXiv:astro-ph/astro-ph/0005111]}}.
\newblock
  doi:{\changeurlcolor{black}\href{https://doi.org/10.1103/PhysRevLett.85.4434}{\detokenize{10.1103/PhysRevLett.85.4434}}}.

\bibitem[Martin \em{et~al.}(2014)Martin, Ringeval, and Vennin]{Martin:2013tda}
Martin, J.; Ringeval, C.; Vennin, V.
\newblock {Encyclopædia Inflationaris}.
\newblock {\em Phys. Dark Univ.} {\bf 2014}, {\em 5-6},~75--235,
  \href{http://xxx.lanl.gov/abs/1303.3787}{{\normalfont
  [arXiv:astro-ph.CO/1303.3787]}}.
\newblock
  doi:{\changeurlcolor{black}\href{https://doi.org/10.1016/j.dark.2014.01.003}{\detokenize{10.1016/j.dark.2014.01.003}}}.

\end{thebibliography}

\end{document}